\documentclass [prb,preprint,superscriptaddress,floatfix,amsmath,amssymb] {revtex4-2}
\usepackage{amsmath}
\usepackage{graphicx}
\usepackage{hyperref}
\usepackage{color}
\usepackage{amssymb}
\usepackage{bm}

\newcommand{\beq} {\begin{equation}}
\newcommand{\eeq} {\end{equation}}
\newcommand{\bea} {\begin{eqnarray}}
\newcommand{\eea} {\end{eqnarray}}
\newcommand{\be} {\begin{equation}}
\newcommand{\ee} {\end{equation}}

\newcommand{\bo}{\bar \omega}

\newcommand{\sign}{\text{sign}}

\DeclareMathOperator{\sgn}{sgn}

\begin{document}
\title {Interplay between superconductivity and non-Fermi liquid  at a quantum-critical point in a metal.\\
 V.  The $\gamma$ model and its phase diagram.  The case $\gamma =2$.}
\author{Yi-Ming Wu}
\affiliation{School of Physics and Astronomy and William I. Fine Theoretical Physics Institute,
University of Minnesota, Minneapolis, MN 55455, USA}
\author{Shang-Shun Zhang}
\affiliation{School of Physics and Astronomy and William I. Fine Theoretical Physics Institute,
University of Minnesota, Minneapolis, MN 55455, USA}
\author{Artem Abanov}
\affiliation{Department of Physics, Texas A\&M University, College Station, 77843 USA}
\author{Andrey V. Chubukov}
\affiliation{School of Physics and Astronomy and William I. Fine Theoretical Physics Institute,
University of Minnesota, Minneapolis, MN 55455, USA}

\date{\today}
\begin{abstract}
  This paper is a continuation and a partial summary of our analysis of the pairing  at a quantum-critical point (QCP) in a metal for a set of quantum-critical systems, whose low-energy physics is described by an effective model with  dynamical electron-electron interaction
 $V(\Omega_m) = ({\bar g}/|\Omega_m|)^\gamma$ (the $\gamma$-model). Examples include pairing at
   the onset of various spin and charge density-wave and nematic orders and pairing in SYK-type models.
  In previous papers, we analyzed the physics for $\gamma <2$. We
  have shown that the onset temperature for the pairing $T_p$ is finite, of order ${\bar g}$,  yet the gap equation at $T=0$ has an infinite set of  solutions within the same spatial symmetry. As the consequence,  the condensation energy $E_c$ has an infinite number of minima.
   The spectrum of $E_c$ is discrete, but becomes more dense as $\gamma$ increases. Here we consider the case
    $\gamma =2$. The $\gamma=2$ model attracted special interest in the past as it describes the pairing by an Einstein phonon  in the limit when the dressed phonon mass $\omega_D$  vanishes.  We show that for $\gamma =2$,  the spectrum of $E_c$ becomes continuous. We argue that the associated gapless "longitudinal" fluctuations destroy superconducting phase coherence at a finite $T$,
     such that at $0<T< T_p$ the system displays pseudogap behavior of preformed pairs. We show that for each gap function from the continuum spectrum,   there is an  infinite array of dynamical vortices in the upper half-plane of frequency.
      For the electron-phonon case,  our   results show that $T_p =0.1827 {\bar g}$, obtained in earlier studies,
    marks the onset of the pseudogap behavior, while the actual superconducting $T_c$ vanishes at $\omega_D \to 0$.
       \end{abstract}
\maketitle

\section{ Introduction.}

 This work presents a continuation and a partial summary of our analysis of the interplay between non-Fermi liquid (NFL) physics and  superconductivity  for a set of quantum-critical (QC) itinerant systems, whose low-energy dynamics can be described  by an effective model of spin-full electrons with dynamical interaction on the Matsubara axis  $V(\Omega_m) =  ({\bar g}/|\Omega_m|)^\gamma$ (the $\gamma$-model).  Examples include pairing near spin-density-wave, charge-density wave, and Ising-nematic  instabilities in isotropic and anisotropic 3D and 2D systems, pairing of fermions at a half-filled Landau level,
 pairing of dispersion-less fermions randomly coupled to phonons, and so on.
 We listed the examples in the first paper in the series, Ref. \cite{paper_1}, and discussed earlier works.  In that and subsequent papers~\cite{paper_1,paper_2,paper_3,paper_4}, hereafter called Papers I-IV, we analyzed the $\gamma$-model with exponents $0<\gamma <2$.
   We rationalized Eliashberg-type approach, solved generalized Eliashberg equations, and found that the solution with a non-zero pairing gap  develops below a finite $T_p$, which for a generic $\gamma = O(1)$ is of order ${\bar g}$. The corresponding  gap function $\Delta (k, \omega_m)$ can be roughly approximated as $\Delta (\omega_m) f(k)$, where a normalized $f(k)$ has a particular
    spatial symmetry ($d-$wave, $s^{+-}$, etc), and $\Delta (\omega_m)$ is a
    sign-preserving function of Matsubara frequency, whose amplitude increases with decreasing $T$.   At $T=0$, $\Delta (0)$ is of order $T_p$ (the ratio $\Delta (0)/T_p$ is a $\gamma-$dependent number~\cite{Kotliar2018,Wu_19}).
    In this respect, the pairing at a QCP is similar to  pairing away from a QCP, when $V(\Omega_m)$ saturates at a finite value at $\Omega_m=0$.   However, on a more careful look, we found  qualitative difference between the two cases. Namely,
     away from a QCP,  Eliashberg gap equation at $T=0$ has at most a finite number of solutions with a given spatial symmetry.  At a
     QCP,  it  has an infinite number of solutions for $\Delta (\omega)$, i.e., the condensation energy, $E_c$, has an infinite number of local minima.  The solutions $\Delta_n (\omega_m)$, labeled by an integer $n$, are topologically distinct in the sense that
     $\Delta_n (\omega_m)$ changes sign $n$ times along the positive Matsubara axis, and each such point is a vortex in the upper half-plane of frequency.

      The ultimate goal of our studies of the $\gamma-$model is to understand how the existence of an infinite set of solutions affects the interplay between pairing (i.e., the appearance of a non-zero $\Delta (\omega)$) and  a true
      superconductivity.  In a conventional Eliashberg theory (ET) of SC out of a non-critical Fermi liquid, phase fluctuations are small in the same parameter that allows one to neglect vertex corrections. In this situation, the onset temperature for the pairing, $T_p$, and the actual superconducting $T_c$ nearly coincide. To address a possible reduction of superconducting $T_c$ by phase fluctuations, one then has to include the effects not incorporated into ET, e.g., a localization of electrons when the interaction exceeds the fermionic bandwidth.  We analyze whether
       the emergence of an infinite number of solutions for the gap  at a QCP gives rise to a substantial reduction of $T_c/T_p$ ratio  already   when the interaction is smaller than the fermionic bandwidth.  For $\gamma <2$, which we analyzed in Papers I-IV,  the spectrum of the condensation energies, $E_{c,n}$,  is discrete, and $E_{c,0}$ is the largest.   In this situation,  the  physics at small $T$ is determined by a single  solution $\Delta_0 (\omega_m)$, and phase fluctuations are weak.  We showed, however, that  the set becomes more dense at $\gamma$ increases:  the ratios $(E_{c,n+1}-E_{c,n})/E_{c,n}$ get progressively smaller.

       In this paper, we analyze the case $\gamma =2$.  We argue that for this $\gamma$, the set of the gap functions and the spectrum of the condensation energies become continuous.  Specifically, at $\gamma =2-\delta$, and $\delta = 0+$, $\Delta_n (\omega_m)$ with $n < 1/\delta$ become equal to $\Delta_0 (\omega_m)$ for all $\omega_m >0$, and
        $E_{c,n}$  become equal to
        $E_{c,0}$, while $\Delta_n (\omega_m)$ and $E_{c,n}$ with infinite $n > 1/\delta$ form a continuous, one-parameter gapless spectra,
         $\Delta_\xi (\omega_m)$  $E_{c,\xi}$.
        Here $\xi$ is a continuous variable, that runs between zero and infinity and depends on how the double limit $n \to \infty$ and $\delta \to 0$ is taken (we define $\xi$ such that
         the minimum of $E_{c,\xi}$ is at $\xi =0$, and $E_{c,\infty} =0$).
        We argue that fluctuation corrections to superconducting order parameter from $E_{c,\xi}$
    destroy long-range superconducting order at any finite $T$. We emphasize that this holds for itinerant fermions, in the limit when the interaction is smaller than the bandwidth, and the ET is rigorously justified.

    We present a corroborative evidence that the $\gamma=2$ model is critical. It comes from the analysis of the gap equation on the real frequency axis and in the upper half-plane of frequency, $z = \omega' + i \omega^{''}$, $\omega^{''} >0$.
          The gap function $\Delta (z)$ generally cannot be obtained from $\Delta (\omega_m)$  by just replacing $i\omega_m$ by
          $z$ as such gap function is not guaranteed to be analytic. To obtain an analytic function, one has to perform a more sophisticated analysis~\cite{Karakozov_91,Marsiglio_91,combescot,Wang_19}. As a consequence,  $\Delta (\omega)$ on the real axis can be quite different from $\Delta (\omega_m)$.   For the $\gamma$-model, some difference is  expected on general grounds, particularly for $\gamma >1$, because while
            the interaction $V(\Omega_m)$ on the Matsubara axis is positive (attractive) for all $\gamma$, the one on the real axis is complex, $V(\Omega) = e^{i\pi \gamma/2 ~{\text {sgn}}\Omega} /|\Omega|^\gamma$,  and
            its real part
            $V'(\Omega) = ({\bar g}/|\Omega|)^\gamma \cos{\pi \gamma/2}$ becomes repulsive for  $\gamma >1$.

            In Paper IV we compared the forms of $\Delta_n (\omega_m)$ and $\Delta_n (\omega)$ for $1<\gamma <2$.
        We found that at small $\omega, \omega_m < {\bar g}$ and at large $\omega, \omega_m > {\bar g} (|\log{(2-\gamma)}|/(2-\gamma))^{1/2}$,
          the two gap functions transfer into each other under a rotation  $i\omega_m \to \omega$.
         However, at intermediate  ${\bar g} <\omega, \omega_m < (|\log{(2-\gamma)}|/(2-\gamma))^{1/2}$,  $\Delta_n (\omega_m) = a_n/|\omega_m|^\gamma$ is a sign-preserving
          function of frequency, while $\Delta_n (\omega) = |\Delta_n (\omega)|e^{i\eta_n (\omega)}$ oscillates, and  its phase $\eta_n (\omega)$ winds up by  $2\pi k_\gamma$, where $k_\gamma$ is an integer, which depends on $\gamma$, but not on $n$.  We extended the analysis to complex $z$  in the upper frequency half-plane and showed that  there exists  an array of $k$  dynamical vortices, centered at some complex $z_i$.

         Here we show that for $\gamma =2$, the gap functions on the real axis form a continuous set,  each
          $\Delta_\xi (\omega)$ oscillates up to an infinite frequency, and its phase winds up by an infinite number of $2\pi$. Accordingly, the number of vortices at $z_i$ becomes infinite, and the array of $z_i$
           extends up to an infinite frequency, where, we argue, each $\Delta_\xi (\omega)$  develops an essential singularity.
          We show that for each gap function from the set, the density of states (DoS) $N_\xi(\omega+i0)$ has an infinite number of maxima and minima, and does not recover the normal state form up to $\omega = \infty$.  For the solution with  $\xi =0$,  which was studied before~~\cite{Karakozov_91,Marsiglio_91,combescot,Schmalian_19,Schmalian_19a},
   $N(\omega+i0)$ reduces to a set of $\delta-$functions at some $\omega_i$.

    We combine the results for $\gamma =2$ and earlier results for $\gamma <2$ (Papers I-IV) and
  present the  phase diagram of the $\gamma-$model for $\gamma \leq 2$, Fig.\ref{fig:phasediagram2}.
   For all $\gamma$,  the ground state is a superconductor with a finite superfluid stiffness $\rho_s$, and
   the onset temperature for the pairing, $T_p$, is finite. However, superconducting $T_c$ decreases with $\gamma$ and
   vanishes for $\gamma =2$.
    In between $T_p$ and $T_c$, the system displays
   a pseudogap (preformed pairs) behavior. One feature of this phase is "gap filling" behavior, as $T$  increases towards $T_p$.
 In the next paper we consider the case $\gamma >2$. We show that the behavior at a finite $T$ remains largely the same as for $\gamma = 2$, however new physics emerges at $T=0$ and gives rise to a reduction and eventual vanishing of $\rho_s$ even in the ground state.

The model with the pairing interaction $V(\Omega_m) = ({\bar g}/|\Omega_m|)^2$ attracted a substantial attention
 on its own as it describes the pairing, mediated by an Einstein boson,  in the limit  where the effective (dressed) Debye frequency $\omega_D$ vanishes
   \footnote{The model also describes  strong coupling limit  of the interaction between dispersion-less fermions and phonons (SYK-Yukawa model), Refs.~\cite{Schmalian_19,Schmalian_19a,Wang_19,Chubukov_2020b,Classen_21}}.
  Electron-phonon model at $\omega_D \to 0$ has been studied before by a large number of
authors~\cite{Scalapino_66,*Scalapino_69,ad,Bergmann,Rainer_86,Karakozov_91,Marsiglio_91,combescot}.
  We use the results of these studies, particularly the works by Karakozov, Maksimov, and Mikhailovsky~\cite{Karakozov_91},  Marsiglio and Carbotte~\cite{Marsiglio_91}, and Combescot~\cite{combescot}  as the input for some of our calculations.
   This limit is often termed strong coupling as the dimensionless coupling constant $\lambda = ({\bar g}/\omega_D)^2$ diverges at $\omega_D \to 0$.  However, the interaction ${\bar g}$ is still assumed to be  smaller than the Fermi energy, $E_F$. Indeed,  ET includes contributions
  to all orders in $\lambda$ within the ladder approximation,  but neglects vertex corrections to ladder series. The latter  hold in powers of Migdal-Eliashberg parameter $\lambda_E  = {\bar g}^2 N_0/\omega_D = \lambda (N_0 \omega_D)$, where $N_0 \sim 1/E_F$ in the
   DoS per unit volume.  For  small enough ${\bar g}/E_F$, $\lambda_E$ remains small even when $\lambda$ is large.  From this perspective, the  strong coupling limit
    of the ET is the double limit in which $\omega_D$ and ${\bar g}/E_F$ tend to zero simultaneously, such that  $\lambda_E$ remains small.
    In physical terms,  the smallness of $\lambda_E \ll \lambda$ comes about because
    in a process that gives rise to a vertex correction, fermions are forced to vibrate at a phonon frequency, far away from their own resonance, while in the processes, which form series in $\lambda$,   fermions are vibrating near their resonance
   frequencies.   The smallness of $\lambda_E$ also allows one to neglect the renormalization of the bosonic propagator by fermions,  both in the normal and in the superconducting  state.

Previous studies have found that a non-zero gap function emerges at  $T_p \approx 0.25 \omega_D e^{-1/\lambda}$ at weak coupling (Refs. \cite{Hertel_1971,Geilikman_1972,Karakozov_1976,Dolgov_1995,Wang_2013,Marsiglio_2018,Mirabi_2020})
  and at  $T_p = 0.1827 {\bar g}$ at strong coupling~\cite{ad,Marsiglio_91,combescot,Chubukov_2020b}.
   \footnote{This formula was originally obtained  semi-analytically by Allen and Dynes~\cite{ad}. They expressed it as $T_p \sim \omega_D \sqrt{\lambda}$ to emphasize that at strong coupling $T_p$ becomes  larger than $\omega_D$.
  Given that $\lambda = ({\bar g}/\omega_D)^2$, their formula reduces to $T_p \sim {\bar g}$.}.
 To understand the interplay between the onset of pairing and $T_c$, one has to also compute superfluid stiffness, $\rho_s$.    At weak coupling, $\rho_s \sim E_F \gg T_p$~\cite{Randeria_1993,Scalapino_1993,Benfatto_2001,randeria_1,Sharapov_2002,cee_2}. In this situation,  $T_c$ and $T_p$ almost coincide.  At strong coupling, the situation is more complex.  At $T=0$, the $\rho_s \sim T_p/\lambda_E$ (see below).  Within the validity of ET,  this stiffness  exceeds $T_p$.  If we were to neglect  the continuum spectrum of the condensation  energy, we would obtain that $T_p$ and $T_c$ again also coincide, as thermal  corrections to SC order parameter are of order $T/\rho_s$ and hence remain small for all $T < T_p$.  Including the additional corrections from the continuum of $E_{c,\xi}$, we find that thermal corrections
  actually hold in powers of $T/(\omega_D \lambda_E)$ and become of order one at
  $T  \sim  \omega_D \lambda_E$, which we identify with the actual $T_c$. At small $\omega_D/{\bar g}$, this $T_c$ is much smaller than $T_p$, even if we set $\lambda_E = O(1)$.
 In between $T=T_p$  and $T_c$ the system  displays a preformed pairs behavior.
 When $\omega_D$ increases and becomes of order ${\bar g}$,
  the pseudogap region shrinks and the system gradually recovers BCS-like behavior (Fig.\ref{fig:phasediagram1}).

The structure of the paper is the following.  In
Sec.\ref{sec:Eli}
we present the Eliashberg gap equations that we use in this paper. In Sec. \ref{sec:matsubara} we discuss the solution of the gap equation along  the Matsubara axis at $T=0$ and $\gamma \to 2$.  We first obtain, in Secs. \ref{sec:Mats_linearized} and \ref{sec:Mats_sign_preserving}, the   exact solution of the linearized gap equation, $\Delta_{\infty} (\omega_m)$, which
 changes sign an infinite number of times between $\omega_m =0$ and  $\omega_m  \sim {\bar g}$, and sign-preserving solution $\Delta_0(\omega_m)$, which tends to
  a finite value at $\omega_m \to 0$.  At larger $\omega_m > {\bar g}$, both $\Delta_{\infty} (\omega_m)$ and $\Delta_{0} (\omega_m)$ scale
   as $1/|\omega_m|^2$.
   In Sec.  \ref{sec:Mats_expansion}, we obtain the solutions of the non-linear gap equation in the order-by-order expansion in the gap magnitude and show
     that  they form a one-parameter continuum set
      $\Delta_\xi (\omega)$, for which $\Delta_{\infty} (\omega_m)$ and $\Delta_0 (\omega_m)$ are the two limiting cases.
      In Sec. \ref{sec:real}, we analyze the properties of the  gap function  $\Delta (\omega)$ along the real frequency axis. We first obtain, in Sec. \ref{sec:real_1}, the  exact solution of the linearized gap equation on the real axis, $\Delta_\infty (\omega)$,  and show that it oscillates not only at $\omega < {\bar g}$, but also at  $\omega > {\bar g}$, with a different period.  In Sec. \ref{sec:real_2} we consider
       the real-frequency form of $\Delta_0 (\omega)$, which does not change sign on the Matsubara axis.
    We use as an input the results from earlier works~\cite{Karakozov_91,Marsiglio_91,combescot}, which demonstrated that $\Delta_0 (\omega) = |\Delta_0 (\omega)| e^{i\eta (\omega)}$ oscillates at $\omega > {\bar g}$, and argue that the phase $\eta (\omega)$  winds up by an infinite number of $2\pi$ between $\omega =O({\bar g})$
    and $\omega = \infty$. In Sec. \ref{sec:real_3} we present a one-parameter continuum set of $\Delta_\xi (\omega)$, which in the two limits reduces to
    $\Delta_{\infty} (\omega)$ and $\Delta_0 (\omega)$.
     In Sec. \ref{sec:complex} we extend $\Delta_\xi (\omega)$
  into the upper frequency half-plane ($\omega \to z$) and show that for each $\xi$, there is an infinite array of vortices in the upper frequency half-plane and an  essential singularity at $|z| = \infty$.
  In Sec. \ref{sec:omega_D} we consider the gap equation at a  finite $\omega_D$. We argue that the number of vortices becomes finite and  the high-frequency behavior of the gap function becomes regular; however, this holds only above a frequency, which scales inversely with $\omega_D$.
  In Sec. \ref{sec:stiffness} we consider fluctuation corrections to superconducting order parameter $\Delta \langle e^{i\eta} \rangle$
    We argue that  the ground state is a superconductor,
    however corrections to $\langle e^{i\eta} \rangle$ become $O(1)$ already at
     $T \leq \omega_D$.  We identify this scale with the actual superconducting $T_c$ and discus pseudogap behavior in
      between $T_p \sim {\bar g}$ and $T_c$.
          In Sec. \ref{sec:gamma} we combine the results for $\gamma =2$  and for $\gamma <2$ from Papers I-IV and obtain the full phase diagram
   of the $\gamma$ model for $\gamma \leq 2$.  We present our conclusions in Sec. \ref{sec:conclusions}.
   Some technical aspects are discussed in the Appendices.

\section{Eliashberg equations}
\label{sec:Eli}

  The Eliashberg gap equation  for the $\gamma$-model is obtained by combining the equations for the pairing vertex $\Phi$ and the self-energy $\Sigma$.
   The two  equations are obtained in a standard way, by summing up ladder series and neglecting vertex corrections (see Paper I and the text below for justification).
   On the Matsubara axis we have   ($\Phi = \Phi (\omega_m),  \Sigma = \Sigma (\omega_m)$):
   \bea
   \Phi (\omega_m) = {\bar g}^\gamma \pi T \sum_{m'} \frac{\Phi (\omega_{m'})}{\sqrt{(\omega_{m'} + \Sigma (\omega_{m'}))^2 +\Phi^2 (\omega_{m'})}}
    ~\frac{1}{(|\omega_m - \omega_{m'}|^2 + \omega^2_D)^{\gamma/2}} \nonumber \\
  \Sigma (\omega_m) = {\bar g}^\gamma \pi T \sum_{m'} \frac{\omega_{m'} + \Sigma (\omega_{m'})}{\sqrt{(\omega_{m'} + \Sigma (\omega_{m'}))^2 +\Phi^2 (\omega_{m'})}}
    ~\frac{1}{(|\omega_m - \omega_{m'}|^2 + \omega^2_D)^{\gamma/2}}
  \label{r_1}
    \eea
    Introducing $\Delta (\omega_m) = \Phi (\omega_m) \omega_m /(\omega_{m} + \Sigma (\omega_{m}))$ and substituting into (\ref{r_1}), we obtain after a simple algebra the equation that contains only $\Delta (\omega_m)$:
  \beq
   \Delta (\omega_m) = {\bar g}^\gamma \pi T \sum_{m'} \frac{\Delta (\omega_{m'}) - \Delta (\omega_m) \frac{\omega_{m'}}{\omega_m}}{\sqrt{(\omega_{m'})^2 +\Delta^2 (\omega_{m'})}}
    ~\frac{1}{(|\omega_m - \omega_{m'}|^2 + \omega^2_D)^{\gamma/2}}.
     \label{ss_11_01}
  \eeq
  For $\gamma =2$ this reduces to 
  \beq
   \Delta (\omega_m) = {\bar g}^2 \pi T \sum_{m'} \frac{\Delta (\omega_{m'}) - \Delta (\omega_m) \frac{\omega_{m'}}{\omega_m}}{\sqrt{(\omega_{m'})^2 +\Delta^2 (\omega_{m'})}}
    ~\frac{1}{|\omega_m - \omega_{m'}|^2 + \omega^2_D}.
     \label{ss_11_0}
  \eeq
This is the same equation as for the interaction with an Einstein phonon
\cite{Scalapino_66,*Scalapino_69,maki4,maki3,ad,Bergmann,Rainer_86,Marsiglio_88,Karakozov_91,Marsiglio_91,combescot,Schmalian_19,
Schmalian_19a,Chubukov_2020b}. 
 As we said, we consider the limit $\omega_D \to 0$. 
 The self-action term with $m'=m$ in the r.h.s. of 
 (\ref{ss_11_0}) can be safely eliminated  because the numerator vanishes at $m=m'$.
 Setting then $\omega_D =0$, we
     obtain the gap equation at a QCP:
 \beq
   \Delta (\omega_m) = {\bar g}^2 \pi T \sum_{m' \neq m} \frac{\Delta (\omega_{m'}) - \Delta (\omega_m) \frac{\omega_{m'}}{\omega_m}}{\sqrt{(\omega_{m'})^2 +\Delta^2 (\omega_{m'})}}
    ~\frac{1}{(\omega_m - \omega_{m'})^2}.
     \label{ss_11}
  \eeq
At $T=0$,  $\pi T \sum_{m' \neq m} \to (1/2) \int d \omega'_m$.

The gap equation on the real axis is more conveniently expressed in terms of
\begin{equation}
D(\omega) = \Delta (\omega)/\omega.
\label{el7}
\end{equation}
The equation has the form~\cite{combescot}
\begin{equation}
D(\omega) \omega B(\omega) = A(\omega) + C(\omega)
\label{el8}
\end{equation}
where
\begin{eqnarray}
&&A(\omega) = -\frac{{\bar g}^2}{2}
\int_0^{\infty} d \omega^\prime
 \Re \frac{D(\omega^\prime)}{\sqrt{1 - D^2(\omega^\prime)}} A_T \nonumber \\
&& A_T = \frac{\tanh{\frac{\omega^\prime}{2T}} + \tanh{\frac{\omega}{2T}}}{(\omega^\prime + \omega)^2} +  \frac{\tanh{\frac{\omega^\prime}{2T}} - \tanh{\frac{\omega}{2T}}}{(\omega^\prime - \omega)^2} -
\frac{1}{T \cosh^2{\frac{\omega}{2T}}} ~\frac{\omega^\prime}{(\omega^\prime)^2 - \omega^2} \nonumber \\
&&B(\omega) = 1 + \frac{{\bar g}^2}{2\omega}
\int_0^{\infty} d \omega^\prime
 \left[\Re \frac{1}{\sqrt{1 - D^2(\omega^\prime)}}\right] B_T \nonumber\\
&& B_T = \frac{\tanh{\frac{\omega^\prime}{2T}} +
\tanh{\frac{\omega}{2T}}}{(\omega^\prime + \omega)^2} -
 \frac{\tanh{\frac{\omega^\prime}{2T}} - \tanh{\frac{\omega}{2T}}}{
(\omega^\prime - \omega)^2} + \frac{1}{T \cosh^2{\frac{\omega}{2T}}}
~\frac{\omega}{(\omega^\prime)^2 - \omega^2} \nonumber \\
&&C(\omega) = - i \frac{{\bar g}^2 \pi}{2 \sqrt{1 -
D^2 (\omega)}}~ ~\left[ \frac{dD (\omega)}{d\omega}
\tanh{\frac{\omega}{2T}} -T
~\left( \frac{dD^2 (\omega)}{d\omega^2} + \left(\frac{dD (\omega)}{d\omega}\right)^2~\frac{D(\omega)}{1 - D^2(\omega)}\right)\right]
\label{nt1}
\end{eqnarray}
where  the integrals are principal values.
At $T=0$, the expressions  simplify to
\begin{eqnarray}
A(\omega) &=& - {\bar g}^2  \int_0^{\infty} \frac{d \omega^\prime}{(|\omega|
 + \omega^\prime)^2}
\Re \frac{D(\omega^\prime)}{\sqrt{1 -D^2(\omega^\prime)}} \nonumber \\
B(\omega) &=& 1 +
\frac{{\bar g}^2}{|\omega|} \int_0^{\infty}
 \frac{d \omega^\prime}{(|\omega| + \omega^\prime)^2}
\Re \frac{1}{\sqrt{1 -D^2(\omega^\prime)}} \nonumber \\
C(\omega) &=& - i \frac{\pi{\bar g}^2}{2\sqrt{1 -D^2(\omega)}}~\frac{d D(\omega)}{d\omega}
~\mbox{sign}\omega,
\label{e124}
\end{eqnarray}
and the gap equation becomes
\begin{eqnarray}
&&- i \frac{\pi {\bar g}^2}{2} ~\frac{\frac{dD (\omega)}{d\omega}}{\sqrt{1 -
D^2 ( \omega)}}
~\mbox{sign} \omega
=
 \label{e125}
   \\
&& D(\omega) \omega ~\left(1  + \frac{{\bar g}^2}{|\omega|}
 \int_0^{\infty} \frac{d \omega^\prime}{(|\omega| + \omega^\prime)^2}
\Re \frac{1}{\sqrt{1 -D^2(\omega^\prime)}}\right) +
\int_0^{\infty} \frac{d \omega^\prime}{(|\omega| + \omega^\prime)^2}
\Re \frac{D(\omega^\prime)}{\sqrt{1 -D^2(\omega^\prime)}} \nonumber
\end{eqnarray}

The functions $A(\omega)$ and $B(\omega)$ can be equivalently expressed in terms of the solution of the gap equation on the Matsubara axis~\cite{Marsiglio_88,Marsiglio_91,combescot}:
\bea
A(\omega) &=&  2\pi T \sum_{m=0}^\infty \frac{\Delta (\omega_m)}{\sqrt{\Omega^2_m + \Delta^2 (\omega_m)}} \frac{\omega^2_m - \omega^2}{(\omega^2_m + \omega^2)^2} \nonumber \\
B(\omega) &=& 1+ 4\pi T \sum_{m=0}^\infty \frac{1}{\sqrt{\Omega^2_m + \Delta^2 (\omega_m)}} \frac{\omega^2_m}{(\omega^2_m + \omega^2)^2}.
\label{nt1a}
\eea
 This simplifies numerical calculations: the recipe is to  first solve for the gap at the Matsubata points $\omega_m = \pi T(2m+1)$ and then use Eqs. (\ref{nt1a})
  as an input for the calculation of $D(\omega)$ on the real axis.

\section{Gap equation along the Matsubara axis at $T=0$}
\label{sec:matsubara}

In Papers I-IV we analyzed the gap equation for $\gamma <2$ and found that  at $T=0$
  it has an infinite, discrete set of  solutions at $\Delta(\omega_m)=\Delta_n (\omega_m)$. A gap function $\Delta_n (\omega_m)$ changes sign $n$ times between $\omega_m=0$ and $\omega_m = O({\bar g})$ and decays as $1/|\omega_m|^\gamma$ at larger frequencies. The two end points of the set are the sign-preserving solution $\Delta_0 (\omega_m)$ and the solution of the linearized gap equation $\Delta_{\infty} (\omega_m)$, which changes sign an infinite number of times.   The existence of this infinite set is a distinct feature  of the pairing at a QCP. Away from a QCP,  the number of solutions  becomes finite $(n=0,1.. n_{max}$), and far away from a QCP  only the $n=0$ solution remains, like in a conventional BCS/Eliashberg theory.

Here we extend this analysis to $\gamma =2$. We show that for this $\gamma$, the set of gap functions
  becomes $\Delta_\xi (\omega_m)$, where $0 \leq\xi \leq \infty$ is a continuous variable.
    We first analyze the two end points, $\Delta_{\infty} (\omega_m)$ and $\Delta_0 (\omega_m)$, and then obtain the gap function for arbitrary $\xi$.

 \subsection{Linearized gap equation}
\label{sec:Mats_linearized}

 The linearized gap equation at $T=0$ is obtained from (\ref{ss_11}) by  assuming that
 the gap function is infinitesimally small, $\Delta(\omega_m)=\Delta_\infty(\omega_m)$. In terms of $D_\infty(\omega_m) = \Delta_\infty (\omega_m)/\omega_m$ we have
 \beq
    D_\infty(\omega_m) = \frac{{\bar g}^2}{2 \omega_m} \int d\omega'_m \frac{D_\infty (\omega'_{m}) - D_\infty (\omega_m)}{(\omega_m - \omega_{m'})^2}
    ~{\text{sign}} \omega'_m.
     \label{ss_11_2}
  \eeq

 One can verify that the  leading term in $D_\infty(\omega_m)$ at small $\omega_m \ll {\bar g}$  is obtained by neglecting the l.h.s. of
 (\ref{ss_11_2}), i.e., by solving
  \beq
     \int_0^\infty d\omega'_m \left[ \frac{D_\infty (\omega'_{m}) - D_\infty (\omega_m)}{(\omega_m - \omega_{m'})^2} + \frac{D_\infty (\omega'_{m}) + D_\infty (\omega_m)}{(\omega_m + \omega_{m'})^2}\right]
     =0
     \label{ss_11_3}
  \eeq
   This approximation is equivalent to neglecting the bare $\omega$ in the fermionic propagator in comparison with the NFL fermionic self-energy without the self-action term, $\Sigma (\omega_m) = - {\bar g}^2/\omega_m$.

 The solution of (\ref{ss_11_3}) is
 \beq
 D_\infty (\omega_m) =2\epsilon~{\text{Re}} \left[e^{i(\beta \log{\left(\frac{|\omega_m|}{{\bar g}}\right)^2} + \phi)}\right] ~\mbox{sign}\omega = 2\epsilon \cos{\left(\beta \log{\left(\frac{|\omega_m|}{{\bar g}}\right)^2} + \phi\right)} ~\mbox{sign}\omega
 \label{5_1}
 \eeq
  where
  $\epsilon$ is an infinitesimally small real overall factor, $\phi$ is a phase factor, which is arbitrary at this stage, and
   $\beta = 0.38187$ satisfies $\pi \beta \tanh(\pi \beta) =1$ and is the solution of
   \beq
   \int_{-\infty}^\infty dx  \frac{|x|^{2i\beta}-{\text{sign}} x}{(x-1)^2} =0,
    \label{5_8}
    \eeq
    The function  $D_\infty (\omega_m)$ is scale-invariant (an arbitrary phase factor $\phi$ can be absorbed into the prefactor for $\omega$ under the logarithm). This is the consequence of the fact that ${\bar g}$ falls off from the gap equation (\ref{ss_11_2}), once we neglect  $D_\infty (\omega_m)$ in the l.h.s.

     We now analyze the full gap equation. By power counting, the r.h.s of (\ref{ss_11_2}) is of order
      $D_\infty (\omega_m)  ({\bar g}/|\omega_m|)$. This justifies neglecting $D_\infty (\omega_m)$ in the l.h.s. for $|\omega_m| < O({\bar g})$, but for larger frequencies it must be kept.

  We  obtained the exact solution of Eq. (\ref{ss_11_2}). The derivation parallels the one
   for smaller $\gamma$ in Papers I and IV. We skip the details and present
    the final result:
 \beq
 D_\infty (\omega_m) = \epsilon
  \frac{{\bar g}}{\omega_m} \int_{-\infty}^\infty dk b_k e^{-ik \log{(\omega_m/{\bar g})^2}},
 \label{nn_2}
 \eeq
where
  \beq
  b_k = \frac{e^{-i I_k }}{\left[\cosh(\pi (k-\beta))\cosh(\pi (k+\beta))\right]^{1/2}},
  \label{nn_2_1}
  \eeq
  and
  \beq
  I_k = \frac{1}{2} \int_{-\infty}^\infty dk' \log{|\epsilon_{k'} -1|} \tanh{\pi (k'-k + i \delta)},
  \label{nn_2_2}
  \eeq
  \beq
\epsilon_{k'} = \pi k' \tanh (\pi k'),
\label{nn_3}
\eeq
Here $\beta\simeq 0.38187$ is the same as in Eq. (\ref{5_1}).

\begin{figure}
    \includegraphics[width=10cm]{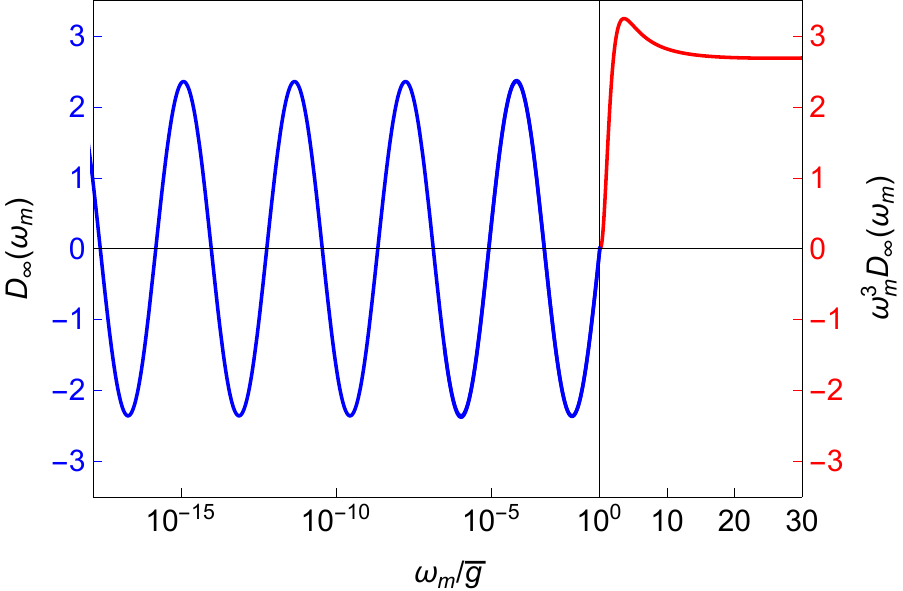}
     \caption{$D_\infty(\omega_m)$ as a function of $\omega_m/{\bar g}$. The scale is logarithmic for $\omega_m < {\bar g}$ and linear
       at $\omega_m > {\bar g}$.  }\label{fig:Dinfty}
  \end{figure}

At  $\omega_m \ll {\bar g}$, the exact  $D_\infty (\omega_m)$  has the form of Eq. (\ref{5_1})  with some particular $\phi$.
At $\omega_m \gg {\bar g}$,
$D_\infty (\omega_m)$ does not oscillate and decreases as $1/(\omega_m)^{3}$ ($\Delta_\infty (\omega_m)$ decreases as $1/(\omega_m)^{2}$).  We plot the exact $D_\infty (\omega_m)$ in Fig. \ref{fig:Dinfty}. The crossover between the two forms occurs at $\omega_m \sim {\bar g}$, as expected.

The corrections to  Eq. (\ref{5_1}) at small $\omega_m$  hold in powers of $|\omega_m|/{\bar g}$;
the leading correction scales as $(|\omega_m|/{\bar g})^{2.78}$.
 The corrections to $1/(\omega_m)^{3}$  at large $\omega_m$ hold in powers of ${\bar g}/|\omega_m|$; the leading correction scales as
$({\bar g}/\rvert \omega_m \rvert)^5 \log (\rvert \omega_m \rvert/{\bar g})$. We present the details of the analysis in  Appendix \ref{sec:app_exact}. There, we also show that at $\omega_m \gg {\bar g}$
 there exists an exponentially small, oscillating component $D_{\infty;u} (\omega_m)$  in the form
 \beq
  D_{\infty;u} (\omega_m) \propto  2\sqrt{2} \epsilon e^{-(|\omega_m|/{\bar g})^{2} }  \cos{\left[   \frac{(\pi^2-2)}{2\pi}
   \left(\frac{|\omega_m|}{{\bar g}}\right)^{2}  +\frac{\pi}{4}\right]}.
  \label{nn_3_6_1}
 \eeq
This term is the contribution to $D_\infty$ from large $k$ and $k'$ in Eqs. (\ref{nn_2_1}) and (\ref{nn_2_2}),
It is completely irrelevant on the Matsubara axis, but we will see that it gives the  dominant contribution to $D_\infty (\omega)$ on the real axis.

\subsection{Non-linear gap equation. Sign-preserving solution.}
\label{sec:Mats_sign_preserving}

We now analyse the full non-linear gap equation, Eq. (\ref{ss_11}).
We first search for a  ``conventional'' sign-preserving solution  $\Delta_0 (\omega_m)$

The analytical analysis uses the same computational steps as in Paper IV and we will be brief.
 We use the identity
\beq
 \int_{-\infty}^\infty d\omega'_m \frac{1 - \frac{\omega'_m}{\omega_m}}{|\omega_m-\omega'_m|^\gamma} =0,
 \label{new_5}
 \eeq
  valid for $\gamma >1$, and re-express Eq. (\ref{ss_11}) as
 \bea
 &&\Delta_0 (\omega_m) \left[1 -
  \frac{{\bar g}^2}{2} \int_{-\infty}^\infty d \omega'_m \frac{1 - \frac{\omega'_m}{\omega_m}}{|\omega_m - \omega'_m|^2} \left(\frac{1}{\sqrt{\Delta^2_0 (\omega'_m)+ (\omega'_m)^2}}- \frac{1}{\Delta_0 (\omega_m)} \right)\right] \nonumber \\
&& =  \frac{{\bar g}^2}{2} \int_{-\infty}^\infty d \omega'_m  \frac{\Delta_0(\omega'_m)-\Delta_0 (\omega_m)}{|\omega_m - \omega'_m|^2 \sqrt{\Delta^2_0 (\omega'_m)+ (\omega'_m)^2}}
\label{new_6}
\eea
Both integrals in (\ref{new_6}) are infra-red convergent and are determined by $\omega_m' \leq  \Delta (\omega'_m)$.  In the limit $\omega_m \to 0$, Eq. (\ref{new_6}) reduces to
  \bea
&&\Delta_0 (0) \left[1 - {\bar g}^2 \int_{0}^\infty d \omega'_m  \frac{\sqrt{\Delta^2_0 (\omega'_m)+ (\omega'_m)^2} - \Delta_0 (0)}{\Delta_0 (0) \sqrt{\Delta^2_0 (\omega'_m)+ (\omega'_m)^2}|\omega'_m|^2}\right] \nonumber \\
&& = {\bar g}^2 \int_{0}^\infty \frac{d \omega'_m ~ \left(\Delta_0(\omega'_m)-\Delta_0 (0)\right)}{\sqrt{\Delta^2_0
 (\omega'_m)+ (\omega'_m)^2}|\omega'_m|^2}
\label{new_7}
\eea
We assume and then verify that   $\Delta_0 (\omega'_m) \approx \Delta_0 (0)$ for
$\omega'_m \leq \Delta _0 (\omega'_m)$, relevant for both integrals in (\ref{new_7}).
Substituting  into (\ref{new_7}), we find
  \beq
  1\approx {\bar g}^2 \int_{0}^\infty d \omega'_m  \frac{\sqrt{\Delta^2_0 (0) + (\omega'_m)^2} - \Delta_0(0)}{\Delta_0 (0) \sqrt{\Delta^2_0 (0) + (\omega'_m)^2}|\omega'_m|^2}
\label{new_8}
\eeq
The integral can be evaluated analytically and yields $\Delta_0 (0) = {\bar g}$.   Substituting further $\Delta_0 (\omega'_m) = {\bar g}$ into the r.h.s. of (\ref{new_6}), we find
that $\Delta_0 (\omega_m)$ varies quadratically with $\omega_m$ at small $\omega_m$ and for $\omega_m \leq {\bar g}$ remains
 comparable to $\Delta_0 (0)$.
\begin{figure}
  \includegraphics[width=10cm]{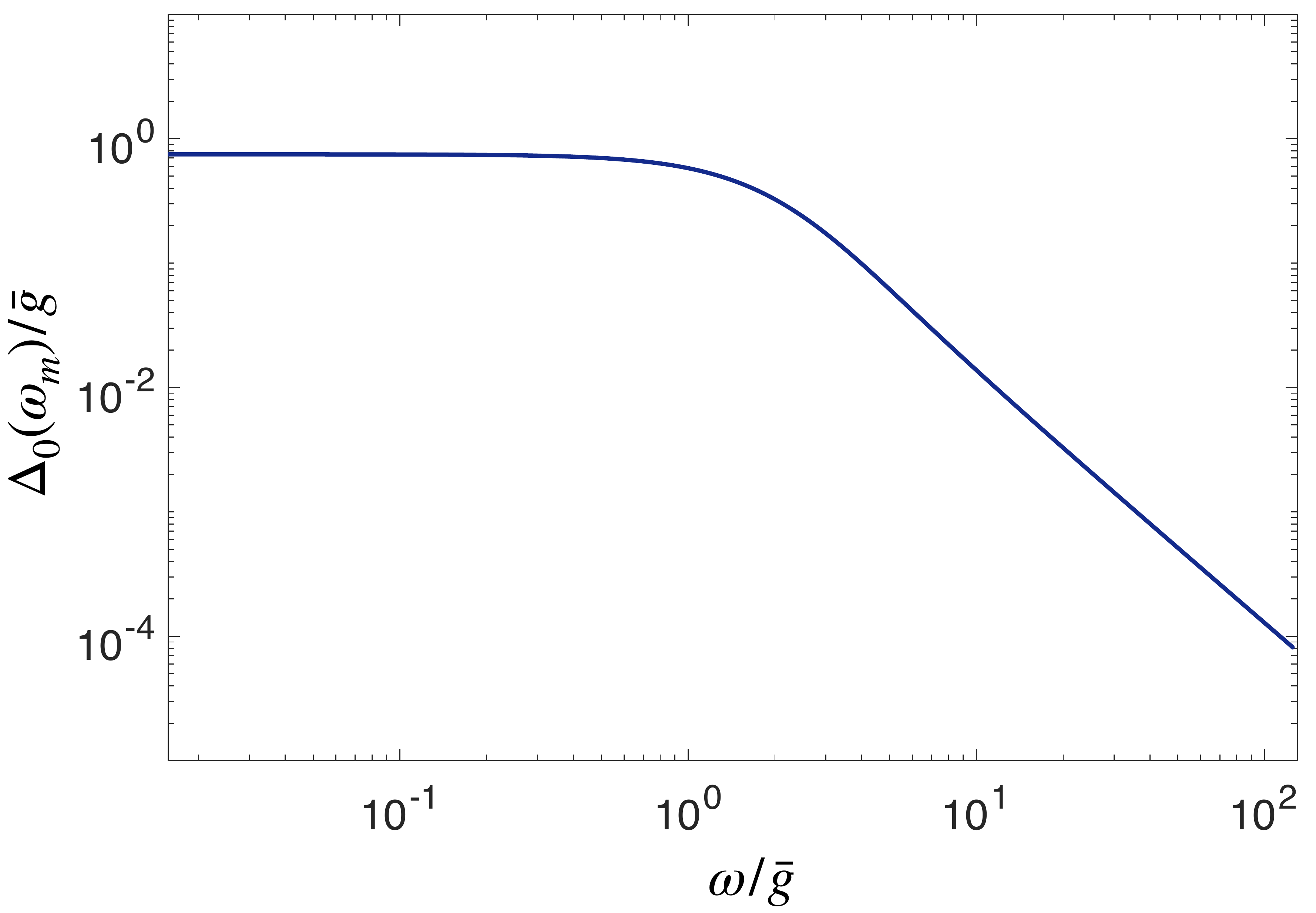}
  \caption{Sign-preserving solution $\Delta_0(\omega_m)$ of the nonlinear gap equation along the Matsubara axis.
  We obtained $\Delta_0(\omega_m)$ by solving the non-linear gap equation numerically and taking the limit $T \to 0$.
   At $\omega_m<\bar g$, $\Delta_0(\omega_m)$
   remains comparable to $\bar g$;
   at larger frequencies it decays as $1/\omega_m^2$.}\label{fig:Delta0}
\end{figure}
 In the opposite limit of large $\omega_m$, the prefactor for $\Delta_0 (\omega)$ in the l.h.s. of (\ref{new_6}) is approximately $1$, and in the r.h.s. of this equation  $1/|\omega_m|^2$ can be pulled out from the integral.  This yields
  \beq
 \Delta_0 (\omega_m) \approx Q \left(\frac{{\bar g}}{|\omega_m|}\right)^2
 \label{4_4}
 \eeq
  where
  \beq
  Q = \int_{0}^\infty \frac{d \omega'_m \Delta_0  (\omega'_m)} {\sqrt{\Delta^2_0 (\omega'_m)+ (\omega'_m)^2}}
  \eeq
The integral is determined by $\omega_m' \sim {\bar g}$ and is of order ${\bar g}$. Then
 $ \Delta_0 (\omega_m) \sim {\bar g}^3/|\omega_m|^2$  at high frequencies.
The full gap function is sign-preserving.
 We show the numerical result for $ \Delta_0 (\omega_m)$ in Fig.\ref{fig:Delta0}.
 At small $\omega_m$ we find
  $\Delta_0 (0) \approx 0.75 {\bar g}$. This fully agrees with the earlier result, Ref. \cite{Marsiglio_91}.  We note in passing that the first numerical evidence that $\Delta_0 (0)$ scales with ${\bar g}$ has been obtained in Ref. \cite{frank_1}.

 \subsection{Continuous set of solutions.  Expansion in the gap magnitude}
\label{sec:Mats_expansion}

 The solutions $\Delta_\infty (\omega_m)$ and $\Delta_0 (\omega_m)$ (or, equivalently, $D_\infty (\omega_m)$ and $D_0 (\omega_m)$) also
 exist
  for $\gamma <2$. For such $\gamma$, these two solutions  are the end points of a discrete set of topologically distinct solutions $\Delta_n (\omega)$.
  We argue below that the set becomes continuous for $\gamma =2$.
  For a continuous set there is no one-to-one correspondence between a particular member of the set and integer $n$, and we will show how this correspondence gets lost at $\gamma = 2-0$.

Comparing $D_\infty (\omega_m)$ and $D_0 (\omega_m)$, we see that they have the same form $1/(\omega_m)^3$
 for $\omega_m > {\bar g}$, but are very different for $\omega_m < {\bar g}$.  We therefore focus on the range $\omega_m < {\bar g}$ and use to our advantage the fact that we know the analytic form of $D_\infty (\omega_m)$ in this range, Eq. (\ref{5_1}).  We use this $D_\infty (\omega_m)$ as an input and expand
 it
 in powers of $D^2 (\omega'_m)$ in the r.h.s. of  the
 gap equation (\ref{ss_11}).
   Specifically, we will be searching for the solution of (\ref{ss_11}) in the form
   \begin{equation}
D(\omega_m)=\sum_{j=0}^\infty D^{(2j+1)}(\omega_m)
\label{exx1}
\end{equation}
 where
 \beq
  D^{(1)} (\omega_m) = D_\infty (\omega_m) = 2\epsilon\cos{f(\omega_m)}~{\text {\sign}} \omega_m
 \label{5_11a}
  \eeq
 with
  \beq
  f(\omega_m) = \beta \log{\left(\frac{|\omega_m|}{\bar g}\right)^2}  +\phi
  \label{5_11b}
  \eeq
  We will see that $D^{(2j+1)}\sim \epsilon^{2j+1}$.

  Substituting
    $D(\omega_m)$ from (\ref{exx1}) into (\ref{ss_11}) and expanding in $D^2(\omega'_m)$ in the r.h.s. of (\ref{ss_11}), we obtain the set of equations, which express $D^{(2j+1)}$ for a given $j$ in terms of
    $D^{(2j+1)}$ with smaller $j$.  For $j=1$, we have
   \beq
   D^{(3)} (\omega_m) \omega_m - \frac{{\bar g}^2}{2} \int d \omega'_m \left( D^{(3)} (\omega'_m) - D^{(3)} (\omega_m)\right) \frac{\sign{\omega'_m}}{(\omega_m-\omega'_m)^2} =  K_3 (\omega_m)
   \label{5_12a}
   \eeq
    where the source term is
     \beq
     K_3 (\omega_m) = -
      \frac{{\bar g}^2}{4}
      \int d \omega'_m \left(D^{(1)} (\omega'_m) - D^{(1)} (\omega_m)\right) [D^{(1)} (\omega'_m)]^2\frac{\sign{\omega'_m}}{(\omega_m-\omega'_m)^2}
\label{5_12}
\eeq
The source term is of order $\epsilon^3$, hence $D^{(3)} \propto \epsilon^3$ ($D^{(5)} \propto \epsilon^5$ and so on).
Substituting $D^{(1)} (\omega_m)$ from Eq. \eqref{5_11a} and evaluating the integrals, we find the source term for $D^{(3)}$ as the sum of the two terms,
 $K_3 = K_{3a} + K_{3b}$, where
  \beq
  K_{3a} (\omega_m) = - \epsilon^3 \frac{\bar g}{\omega_m}  \cos\left(3 f(\omega)\right)
    \left(2\pi \beta \coth(2\pi \beta) - 3 \pi \beta \tanh(3\pi \beta)\right)~{\text {\sign}} \omega_m
  \label{5_14}
 \eeq
 and
  \beq
  K_{3b} (\omega_m) =  -  \epsilon^3 \frac{\bar g}{\omega_m}  \cos\left(f(\omega) \right)
   \frac{1+ \sinh^2 (\pi \beta)}{\sinh^2 (\pi \beta)} ~{\text {\sign}} \omega_m
 \label{5_15}
 \eeq
 Solving for $D^{(3)}$ we find that the first term gives rise to
 $\epsilon^3 \cos \left( 3 f(\omega)\right)$, while  the second term accounts for the renormalization of the prefactor for $\log(\omega^2_m)$ in $f(\omega_m)$ in (\ref{5_11b})
     To order $\epsilon^2$, the dressed $f(\omega_m)$, which we label $f_\epsilon (\omega_m)$, becomes
  \beq
  f_\epsilon (\omega_m) = \beta_\epsilon \log{\left(\frac{|\omega_m|}{\bar g}\right)^2} +\phi_\epsilon
  \label{5_16}
  \eeq
  where
  \beq
  \beta_\epsilon   =\beta \left(1-\epsilon^2/2\right) \approx \beta (1- \epsilon^2)^{1/2}
  \label{5_17}
  \eeq
 The full $D(\omega_m)$ to order $\epsilon^3$ is
  \beq
  D(\omega_m) = 2\left(\epsilon \cos{f_\epsilon(\omega_m)} + Q_3 \epsilon^3 \cos{3 f (\omega_m)}\right) ~{\text {\sign}} \omega_m
  \label{5_18}
  \eeq
   where
   \beq
   Q_3 = \frac{2\pi \beta \coth{(2\pi \beta)} - 3 \pi \beta \tanh{(3 \pi \beta)}}{2(1 -3\pi \beta \tanh{(3\pi \beta)})} = \frac{5- (\pi \beta)^2}{16} \approx 0.222
   \label{5_19}
   \eeq
   Expanding to next order, we find (i) $\epsilon^5 \cos{5 f (\omega_m)}$ term with the prefactor $Q_5 = 0.043$, (ii)
   $O(\epsilon^4)$
   corrections to
   $\beta_{\epsilon}$
   in (\ref{5_17})  ($\beta_\epsilon = 1-0.5 \epsilon^2 + 0.806 \epsilon^4)$), and (iii) $O(\epsilon^2)$ corrections to $Q_3$ ($Q_3 \to Q_{3,\epsilon}$) and to the argument of $\cos{3 f (\omega_m)}$ in (\ref{5_18}). We verified that the last correction changes $\cos{3 f (\omega_m)}$ to
     $\cos{3 f_\epsilon (\omega_m)}$ with the same $f_\epsilon$ as in (\ref{5_16}).  This is the strong indication that the series contain the same fully renormalized $f_\epsilon (\omega_m)$ in each term.
    Combining the results, we obtain, for $\omega_m \ll {\bar g}$, $ D(\omega_m) =D_\epsilon (\omega_m)$, where
    \beq
  D_\epsilon (\omega_m) = 2\epsilon\left(\cos{f_\epsilon (\omega_m)} +  Q_{3,\epsilon} \epsilon^3 \cos{3 f_\epsilon (\omega_m)} + Q_{5,\epsilon} \epsilon^5 \cos{5 f_\epsilon(\omega_m)} + ...\right) ~{\text {\sign}} \omega_m
  \label{5_20}
  \eeq
  We emphasize that a continuous set of solutions exists only for $\gamma =2$.  Applying the same perturbative analysis for
   $\gamma <2$, we find that the expansion  holds in $\epsilon^2 ({\bar g}/|\omega_m|)^{2-\gamma}$ and breaks at a finite
  $\omega_{min} \sim {\bar g} \epsilon^{2/(2-\gamma)}$ (see  Appendix \ref{app:no_omega} for more detail).
  At smaller $\omega_m$, $\Delta (\omega_m)$ saturates, and $D(\omega_m) \propto 1/\omega_m$.  The forms of $D(\omega_m)$  at $\omega_m < \omega_{min}$  and $\omega_m > \omega_{min}$ match  only for a discrete set of $\epsilon = \epsilon_n$, which implies that for $\gamma <2$ the solutions of the full non-linear gap equation form a discrete set.

  Because  $f_\epsilon (\omega_m)$ contains $\log {\omega^2_m}$,
      each $D_\epsilon (\omega_m)$ from (\ref{5_20})  changes sign an infinite number of times down to  $\omega_m=0$,
        i.e., in our original classification
        the gap functions from the set are  different  realizations of $n=\infty$.
        At $\omega_m =0$, each $D_\epsilon (\omega_m)$  has an essential singularity as neither $\lim_{\omega_m \to 0} D_\epsilon (\omega_m)$
         nor $\lim_{\omega_m \to 0} 1/D_\epsilon (\omega_m)$ exist.

For a generic $\epsilon$, Eq. (\ref{5_20}) is valid for $\omega_m < \bar g$.  At larger $\omega_m$, $D_\epsilon (\omega_m) = D_\epsilon/|\omega_m|^2$.  We expect that for every $\epsilon$, the crossover to proper high-frequency behavior
 can be achieved by fixing the phase factor $\phi_\epsilon$ in (\ref{5_16})  (see Paper I for a similar analysis for the linearized gap equation for $\gamma <1$).

\begin{figure}
  \includegraphics[width=15cm]{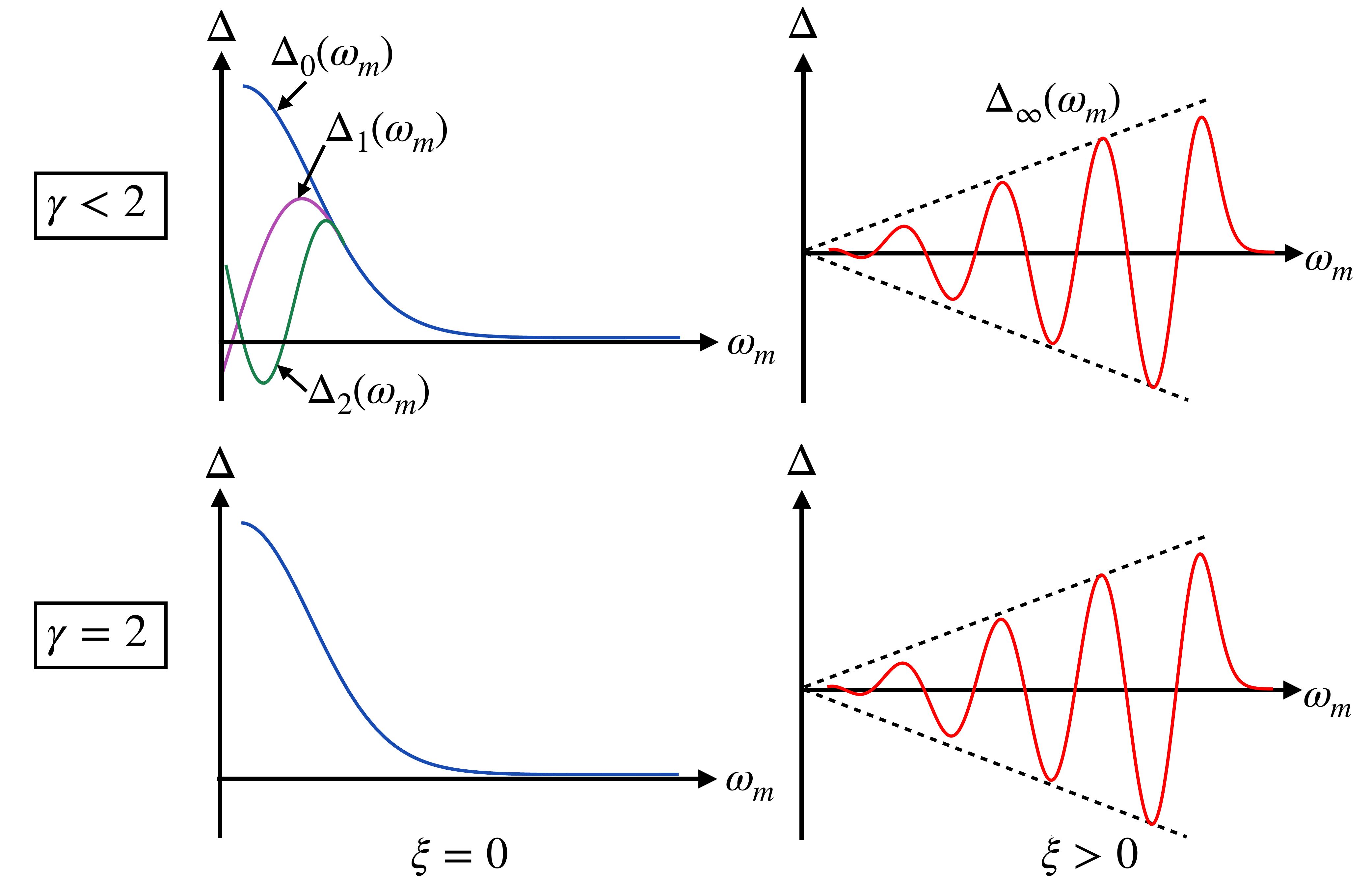}
  \caption{The gap function $\Delta_n(\omega_m)$ for $\gamma<2$ and $\gamma=2$. For $\gamma<2$, $\Delta_n(\omega_m)$ changes sign $n$ times. As $\gamma$ gets close to $2$, the frequency region where these $n$ sign changes happen, shrinks to progressively smaller $\omega_m=0$, and
    at $\gamma =2-0$,  $\Delta_n (\omega_m)$  with finite $n$ collapse into  $\Delta_0(\omega_m)$ at all  $\omega_m >0$.
     The continuum set of $\Delta_\xi (\omega_m)$ at $\gamma =2-0$ emerges from $\Delta_n (\omega_m)$ with $n \to \infty$, and the continuous
   parameter $\xi$ is determined by how the double limit $n \to \infty$ and $\gamma \to 2$ is taken.  As the consequence, all $\Delta_\xi (\omega_m)$ with $\xi >0$ change sign infinite number of times between $\omega_m =0$ and $\omega_m \sim {\bar g}$.  The solution of the linearized gap equation is the $\xi \to \infty$ limit of this set.}\label{fig:Delta_sketch}
\end{figure}

 Next, we see from Eq. (\ref{5_17})
  that
 $\beta^2_\epsilon$
  decreases with increasing $\epsilon$. It is natural to expect that
   it vanishes at some $\epsilon_{cr} = O(1)$.
   The expansion in (\ref{5_20}) holds only as long as $\beta_{\epsilon}$ is real, as  there is no solution of the nonlinear gap equation for imaginary $\beta_\epsilon$ (see Paper I for detailed discussion on this).
    For $\epsilon \leq \epsilon_{cr}$, $\beta_\epsilon$ is small, and
   the range, where $D(\omega_m)$ oscillates, is confined to small $\omega_m \leq {\bar g} e^{-\pi/\beta_\epsilon}$.
   By properly taking the double limit $\epsilon \to \epsilon_{cr}$ and $\omega_m \to 0$, one can
      obtain  an infinite set of gap functions, which change sign a given number of times  in the  immediate vicinity of  $\omega_m = 0$. At $\epsilon = \epsilon_{cr}$ all these gap functions coincide with $\Delta_0 (\omega_m)$ at any $\omega_m >0$.
       This agrees with the observation in Paper IV that as $\gamma$ increases towards $2$, the region, where $\Delta_n (\omega_m)$ with  finite $n$ change sign, gets confined to progressively smaller $\omega_m$, while at larger $\omega_m$, all $\Delta_n (\omega_m)$ with $n =0,1,2...$  nearly  coincide.
        We illustrate this in Fig.\ref{fig:Delta_sketch}.
        For consistency with the notations in previous sections, it is convenient to introduce $\xi = (\epsilon_{cr} - \epsilon)/\epsilon$ and label the continuum set of the gap functions by
        $\Delta_\xi (\omega_m)$. Then the end point solutions $\epsilon \to 0$ and $\epsilon = \epsilon_{cr}$ are $\Delta_\infty (\omega_m)$ and $\Delta_0 (\omega_m)$.

 It is beyond the ability of the order-by-order expansion to determine the  form of  $\Delta_\xi(\omega_m)$ near $\xi =0$.
 On general grounds, we expect that  corrections to $f_\epsilon (\omega_m) \to f_\xi (\omega_m)$ in (\ref{5_16})
 become relevant starting already from  small frequencies, and that at $\xi =0$,
 the gap function coincides with $\Delta_0 (\omega_m)$, which we found in the previous section.
  A way to reproduce this behavior is to assume that at $\xi \to 0$, the series for $D (\omega_m)$ in (\ref{5_20})
    become  geometrical $[Q_{2n+1} \epsilon_{cr}^{2n+1} \approx (-1)^n]$. In this case,
\beq
 D_\xi (\omega_m) \sim \frac{\cos{ f_\xi (\omega_m)}}{\alpha^2 + \cos^2 {f_\xi (\omega_m)}} ~{\text {\sign}} \omega_m,
 \label{con_14}
  \eeq
  where $\alpha \sim \xi^2$
  and $f_\xi (\omega_m) \sim \sqrt{\alpha}\log{{\bar g}/|\omega_m|} + f^* (\omega_m)$, where
   $f^* (\omega_m)$ is a regular function of $\omega_m$, which at low frequencies reduces to
  $\pi/2 + O(\omega_m/{\bar g})$. For any $\xi >0$, this $D_\xi (\omega_m)$ changes sign an infinite number of times, but at
   $\xi =0$, $D_{\xi =0}(\omega_m) \sim {\bar g}/\omega_m$, as we expect. We also note that between the nodes (the vortex points), $D_\xi(\omega_m)$ from (\ref{con_14}) is large, of order $1/\xi$. Extending this $D(\omega_m)$ to complex frequencies, $z = \omega' + i \omega^{''}$, we find that there exist anti-vortices at small $z$ in the lower frequency half-plane.
   At $\xi=0$, vortices and anti-vortices annihilate at $z=0$, leaving a regular gap function $\Delta_0 (\omega_m)$.

In Appendix \ref{sec:AppA} we consider the extended $\gamma-$ model with non-equal interactions in the particle-particle and particle-hole channels and introduce $M \neq 1$ as a measure of the difference of the two interactions. For the extended model, there is a critical $M_{max}$, below which the ground state is a non-Fermi liquid with $\Delta =0$.  For $\gamma =2$, $M_{max}=0$. We obtain the set of $\Delta_\xi (\omega_m)$ at small $\omega_m$ at $M =0+$ and show that all gap functions from the continuous set appear simultaneously with the overall magnitude $M^{1/2}$.

We next analyze the condensation energy $E_{c}$. We define $E_{c}$ as the difference between the actual ground state energy $E_\Delta$ at a finite $\Delta (\omega_m)$ and the would be ground state energy of the normal state, $E_{\Delta=0}$.
 The expression for $E_c$ for $\gamma =2$ has been obtained before:
 ~\cite{maki3,Bardeen,haslinger,Emil2020} and we just copy it here:
\begin{eqnarray}
&& E_{c} = -N_0 \int_0^\infty d \omega_m \omega_m
\frac{(\sqrt{1 + D^2 (\omega_m)}-1)^2}{\sqrt{1 + D^2(\omega_m)}} \nonumber \\
&& - N_0 {\bar g}^2\int_0^\infty d \omega_m d \omega'_m
\frac{\left(\sqrt{1 + D^2(\omega_m)} -{\sqrt{1 + D^2(\omega'_m)}}\right)^2}{\sqrt{1 + D^2(\omega_m)} \sqrt{1 + D^2 (\omega'_m)}}
~\frac{\omega_m \omega'_m}{(\omega^2_m - (\omega'_m)^2)^2}
\label{5_22}
\end{eqnarray}
This formula has been derived with the use of  (\ref{ss_11}) and is therefore valid only for the solutions of the gap equation. Both terms in (\ref{5_22}) are negative, i.e. any solution of the gap equation lowers the  ground state energy compared to the normal state.

\begin{figure}
  \includegraphics[width=15cm]{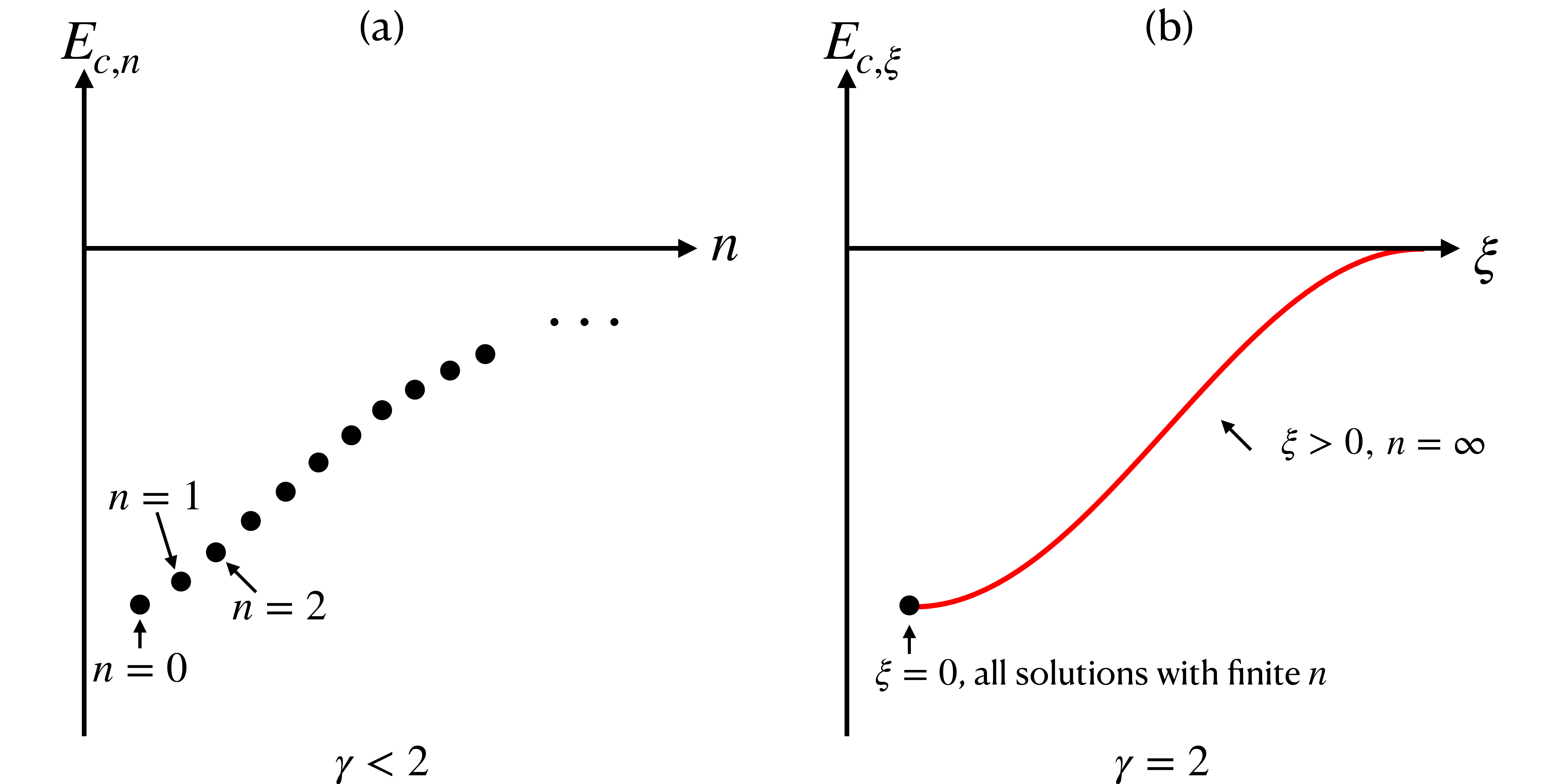}
  \caption{(a)  The condensation energy $E_c$ the solutions of the Eliashberg gap equation for $\gamma <2$.  $E_c = E_{c,n}$ is a discrete function of a number of a solution, $n$.
      The largest condensation energy  is for $n=0$.
   (b)
  The condensation energy $E_{c,\xi}$ for  $\gamma =2$.  $E_{c,\xi}$
  is a continuous function of the parameter $\xi$. The condensation energy at $\xi =0$ is the accumulation point of
  all $E_{c,n}$ from $\gamma <2$ with finite  $n =0,1..$.
     Every other point on the curve $E_{c,\xi}$ comes from the limit $n \to \infty$, and different $\xi$  correspond to different ways how the double limit $n\to \infty$
      and $\gamma \to 2$ is taken.
       In the limit $\xi\to \infty$, $E_c$ is the condensation energy
    for infinitesimally small gap function $\Delta_\infty (\omega_m)$.}\label{fig:Ec}
\end{figure}

  Substituting (\ref{5_20}) into (\ref{5_22}), we find that
  $E_c = E_{c,\xi}$  is a continuous function of $\xi$.  At $\xi \gg 1$,
    \beq
    E_{c,\xi} = - a N_0 \frac{{\bar g}^2}{\xi^4}
    \label{con_1}
    \eeq
    where $a = O(1)$.
    It is natural to expect that  $|E_{c,\xi}|$  increases with decreasing $\xi$ and reaches a maximum at $\xi =0$,
    see Fig. \ref{fig:Ec}.
    \footnote{The second term in (\ref{5_22}) diverges logarithmically at $\xi=0$ if we use $D_0(\omega) \approx \Delta (0)/\omega$ at small frequencies.  This divergence comes from  the putative normal state energy $E_{\Delta =0}$  while the
  ground  state energy $E_\Delta$ remains finite. For $\xi >0$, both $E_{\Delta=0}$ and $E_\Delta$ have logarithmic singularities, which cancel out in $E_c = E_{\Delta} - E_{\Delta=0}$.}

In the next  two Sections we present corroborative evidence for the special, critical  behavior
 of the $\gamma$-model with $\gamma =2$  from the analysis of the gap function on the real frequency axis and in the upper half-plane of frequency.

\section{Gap equation along the real frequency axis}
\label{sec:real}

  As we said in the Introduction,  the analysis of the gap equation for the $\gamma$ model along real frequency axis  should generally be more revealing than the analysis along the Matsubara axis, because the pairing interaction on the real axis
$V(\Omega) = (\cos (\pi\gamma/2) + i \sign{(\Omega)}\sin (\pi \gamma/2) )  ({\bar g}/|\Omega|)^\gamma$
is complex. The real part of the interaction becomes repulsive for $\gamma >1$,  and the imaginary
  part vanishes at $\gamma =2$ for any non-zero $\Omega$.  This makes the $\gamma =2$ case special.

We present the results for $\Delta (\omega)$ on the real axis in the same order as in previous section: we first obtain the solution of the linearized gap equation, which we label $\Delta_\infty (\omega)$, then analyze the solution $\Delta_0 (\omega)$, and then show that there is a one-parameter continuous set of solutions $\Delta_\epsilon (\omega)$  in between $\Delta_\infty (\omega)$ and $\Delta_0 (\omega)$.

\subsection{Linearized gap equation in real frequencies}
\label{sec:real_1}

The linearized gap equation in real frequencies
is obtained by taking the limit $\Delta (\omega) \rightarrow 0$ in (\ref{e125}). We again
introduce
 $D_{\infty}(\omega) = \Delta_{\infty}(\omega)/\omega$ and re-write the gap equation as
\begin{equation}
D_{\infty}({\omega})  = - \frac{{\bar g}^2}{\omega}  \left[i \frac{\pi}{2}~\frac{d D_\infty (\omega)}{d\omega} \sign \omega  + \frac{D_{\infty}(\omega)}{\omega}  + \int_0^{\infty} \frac{d { \omega}^\prime}{(|\omega| + \omega^{\prime})^2}~\Re D_{\infty}(\omega^\prime)\right],
\label{139}
\end{equation}
 where $\Re$ stands for the real part.
 The $D_{\infty}({\omega}) $ term in the l.h.s. of (\ref{139}) is the analog of $D_{\infty} (\omega_m)$ in the l.h.s. of the gap equation (\ref{ss_11_2}) on the Matsubara axis, and, like there, it originates from the bare $\omega$ term in the fermionic Green's function.  Neglecting this term, we find that the solution
  of (\ref{139}) is
\beq
D_{\infty}(\omega) = -2 i \epsilon \cos{\left[\beta \left(\log{\left(\frac{\omega}{{\bar g}}\right)^2}  -i\pi  \sign (\omega)\right) + \phi\right]},
\label{5_23}
\eeq
where $\beta =0.38187$  the same as in (\ref{5_1}), and $\epsilon$ is infinitesimally small.
We note that this $D_{\infty}(\omega)$ can be obtained from  $D_\infty (\omega_m)$, Eq. (\ref{5_1}),
 by  rotating from $i\omega_m$ to $\omega + i0$.
In explicit form,
\bea
&& D'_{\infty} (\omega) = 2 \epsilon   \sin{\left(\beta \log{\left(\frac{\omega}{{\bar g}}\right)^2} + \phi\right)} \sinh(\pi \beta)  \sign{\omega} \nonumber \\
&& D^{''}_{\infty} (\omega) = -2 \epsilon   \cos{\left(\beta \log{\left(\frac{\omega}{{\bar g}}\right)^2} + \phi\right)} \cosh(\pi \beta)  \nonumber \\
\label{5_24}
\eea
Observe that $D'_{\infty} (-\omega) =-D'_{\infty} (\omega)$ and $D^{''}_{\infty} (-\omega) = D^{''}_{\infty} (\omega)$, as it should be.
The relation
\begin{equation}
\int_0^{\infty} dx~\frac{x^{2i \beta}}{(x+1)^2} = \frac{2\pi \beta}{\sinh(2\pi \beta)} = \frac{1}{\sinh^2 (\pi \beta)},
\label{l42}
\end{equation}
is useful for the verification that $D_{\infty}(\omega)$ satisfies Eq. (\ref{139})
 without $D_{\infty}({\omega})$ in the l.h.s.
  Using another relation
\begin{equation}
\int_0^{\infty} dx~\frac{x^{i \beta}}{x-1} = i \pi \coth(\pi \beta),
\label{l42_1}
\end{equation}
one can verify that $D'_{\infty}$ and $D^{''}_{\infty}$ satisfy KK relations:
\beq
\frac{2}{\pi} \int dx \frac{D'_{\infty} (x) x}{x^2-\omega^2} =-D^{''}_{\infty} (\omega),
\quad \frac{2 \omega}{\pi} \int dx\frac{D^{''}_{\infty } (x)}{x^2-\omega^2} =D^{'}_{\infty } (\omega),
\label{5_25}
\eeq
 where the integrals are principle values.
\begin{figure}
     \includegraphics[width=12cm]{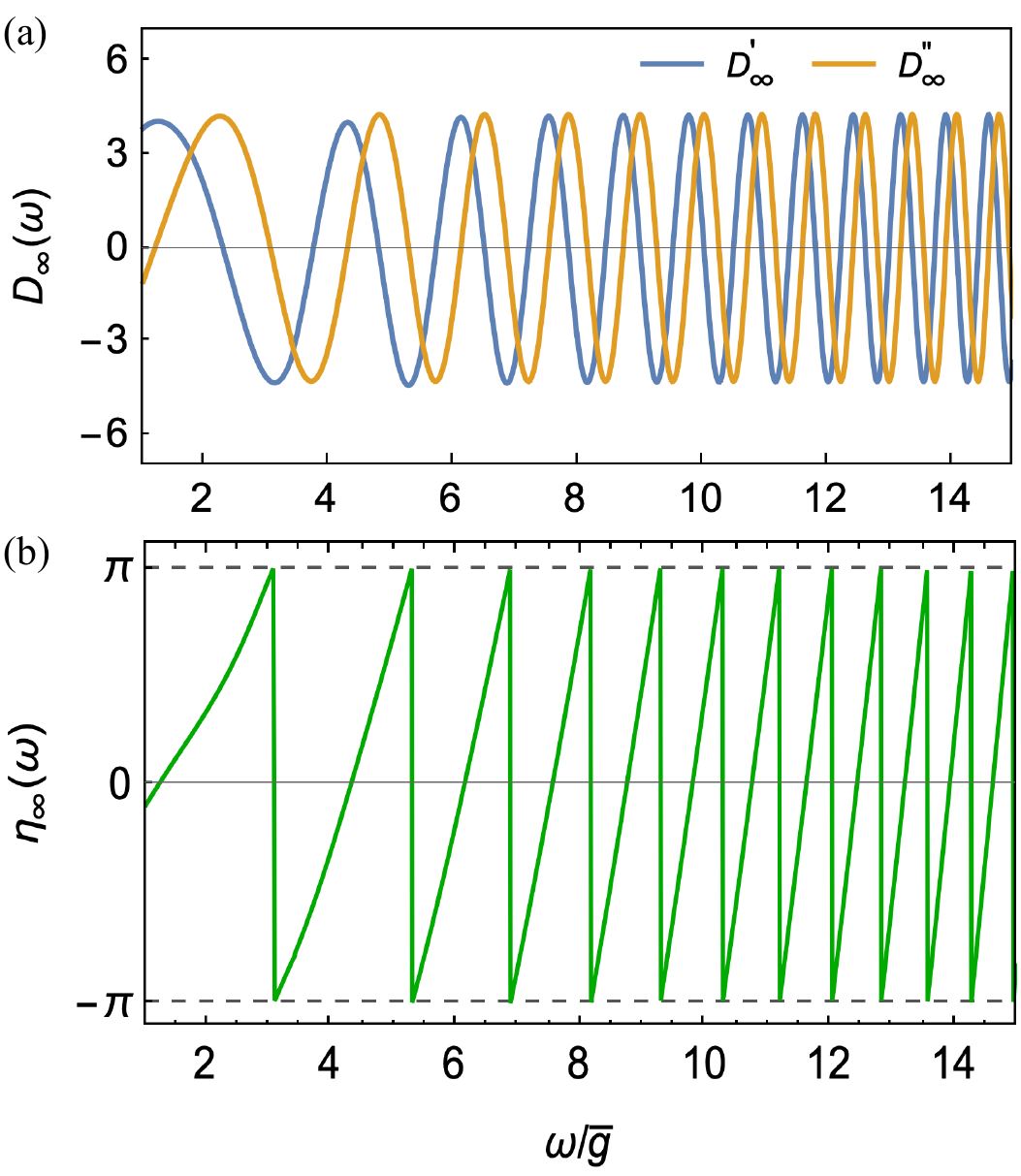}
     \caption{Real and imaginary part (a) and the phase $\eta_{\infty} (\omega)$ (b) of $D_{\infty}(\omega)$. The periodicity of oscillation is set by $[(\omega/{\bar g})^2 + \log(\omega/{\bar g}  )^2]/\pi$.}
     \label{fig:Dinfty_real}
   \end{figure}

 We next consider $\rvert \omega \rvert > {\bar g}$.  To obtain
 $D_\infty (\omega)$ in this region, we take as an input the exact solution on the Matsubara axis and analytically
  continue it to the real axis.  By construction, this can be done by replacing $\omega_m$ by $(-i z)$ - the function
   $\Delta_\infty (z)$ is guaranteed to be analytic in the upper half-plane of frequency.
   However, because we don't have the exact analytical expression for $D_\infty (\omega_m)$ for an arbitrary
    $\omega_m$,
     we have to replace $\omega_m$ by $-i (\omega + i0)$ in Eq. (\ref{nn_2}) and obtain $\Delta_\infty (\omega)$ by  integrating over $k$.
   For small $\omega < {\bar g}$, we find, after this integration, series of corrections to (\ref{5_23}) in powers of $\omega/{\bar g}$.
    For large $\omega > {\bar g}$, the largest contribution to $\Delta_\infty (\omega)$ comes from the continuation of the
     universal
      oscillating term $\Delta_{\infty;u}$,  Eq. (\ref{nn_3_6_1}).  Upon rotation to the real axis, this term splits into two. One remains exponentially small,  but in the other the exponential factor cancels out. As a result, on the real axis we have
      (see Appendix \ref{sec:app_exact} for details)
   \beq
 D_{\infty;u}(\omega) \sim \sqrt{2} \epsilon  e^{{i \over \pi} \left[ \left({\omega \over {\bar g}}  \right)^2 + \log \left({\omega \over {\bar g}}  \right)^2  \right]}.
\label{5_25_aa}
\eeq
 Other contributions contain powers of ${\bar g}/|\omega|$ and are smaller. Neglecting them, we obtain
  $D_\infty (\omega) = D_{\infty;u} (\omega)$  at $\omega \gg {\bar g}$.

Comparing this form with (\ref{5_23}), we see that both $D'_{\infty }(\omega)$ and $D^{''}_{\infty }(\omega)$ continue oscillating at $\omega > {\bar g}$, but with the period set predominantly by  $(\omega/{\bar g})^2$ rather than by $\log{(\omega/{\bar g})^2}$.
  In Fig.~\ref{fig:Dinfty_real},
   we plot real and imaginary parts of $D_{\infty }(\omega) $ and the phase of the gap, $\eta_{\infty} (\omega)$, defined via
   $D_{\infty }(\omega) = |D_{\infty }(\omega)| e^{i\eta_\infty (\omega)}$,
   or, equivalently, via
   $\eta_\infty (\omega) = {\text{Im}} \log{D_\infty (\omega)} = {\text{Im}} \log{\Delta_\infty (\omega)}$. We see that the phase winds up an infinite number of times between $\omega =0$ and $O(\bar g)$ and
  between $O(\bar g)$ and $\infty$.  Oscillations  at $\omega < {\bar g}$ are directly related to oscillations of $\Delta_\infty (\omega_m)$ on the Matsubara axis, and there is one-to-one correspondence between each  phase winding by $2\pi$ on a real axis and a  vortex on the Matsubara axis.
   Oscillations and phase winding at $\omega > {\bar g}$ are present  on the real axis, but not on the Matsubara axis.  It is natural to relate this discrepancy to the fact that the pairing interaction is attractive on the Matsubara axis, but
    on the real axis, $V' (\Omega)$ is repulsive, and a non-zero $D_\infty (\omega)$ comes from $V^{''} (\Omega) \propto \delta (\Omega^2)$ (see more on this below).

\subsection{The function $D_0 (\omega)$}
\label{sec:real_2}
We now consider the opposite limit of the real-axis partner of  sign-preserving $D_0 (\omega_m)$.
At $\omega_m < {\bar g}$, $D_0 (\omega_m)  \approx \Delta_0 (0)/\omega_m$, and $D_0 (\omega)$
 on the real axis must also be close to $D_0 (0)/\omega$ at $\omega < {\bar g}$.  At larger $\omega_m > {\bar g}$, we will see that $D_0 (\omega_m)$  and $D_0 (\omega)$ are very different: $D_0 (\omega_m)$ decays as $1/\omega^3_m$, while $D_0 (\omega)$  does not decay and oscillates in sign.

 The solution of the gap equation along the real axis  for $\omega > {\bar g}$ has been found by Combescot~\cite{combescot}, who build his analysis on  earlier results by Karakozov, Maksimov, and Mikhailovsky~\cite{Karakozov_91} and by Marsiglio and Carbotte~\cite{Marsiglio_91}. We follow Ref. ~\cite{combescot} below.

It is convenient to introduce $\phi_0 (\omega)$ via $D_0(\omega) = 1/\sin{\phi_0 (\omega)}$ and re-express the
  gap equation (\ref{e125}) at $T=0$ as the equation on $\phi (\omega)$. The equation is
 \beq
  \frac{d\phi_0 (\omega)}{d\omega}  = \frac{2}{\pi {\bar g}^2}
 \left[\omega B(\omega) - A(\omega) \sin{\phi_0}\right]
 \label{5_30}
 \eeq
 where $ A(\omega)$ and $B(\omega)$ are given by Eq. (\ref{e124}).  The initial condition  for $\phi_0$  is $\phi_0 ({\bar g}) \approx  {\bar g}/\Delta_0 (0) = O(1)$, consistent with
  $\phi_0 (\omega) \approx\omega/\Delta_0 (0)$ at $\omega < {\bar g}$.

At $\omega \geq {\bar g}$,  $B(\omega) \approx 1 + {\bar g}^2/\omega^2$ and $A(\omega) \approx -\alpha {\bar g}^3 /\omega^2$, where $\alpha \approx 1.27 $ (Ref. \cite{combescot}) . The $A(\omega)$ term can then be neglected if
$\phi_0 (\omega)$ is real, as we will assume and then verify.
 Without this term, Eq. (\ref{5_30}) can be solved easily, and the result is
\beq
 \phi_0 (\omega) \approx \frac{1}{\pi} \left(\log{\left(\frac{\omega}{{\bar g}}\right)^2} +
\left(\frac{\omega}{{\bar g}}\right)^2  + C\right)
\label{5_31}
\eeq
where $C = {\bar g}/\Delta_0 (0) -1/\pi$. We
  see that $\phi_0 (\omega)$ is real, as we anticipated.  We note that this $\phi_0 (\omega)$ coincides with
  the argument of the exponent for $D_\infty (\omega)$ in (\ref{5_25_aa})

\begin{figure}
     \includegraphics[width=12cm]{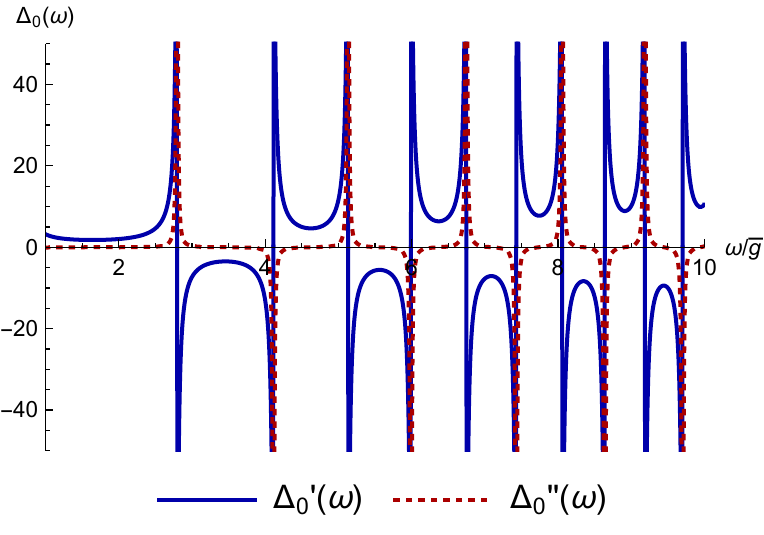}
     \caption{$\Delta_0(\omega) = \omega/\sin{\phi_0 (\omega)}$ for $\phi_0(\omega)$ given by \eqref{5_31}. The real part of the gap $\Delta'_0 (\omega)$ diverges at the set of points where $\phi_0(\omega) = p\pi$, $p=1,2...$. The imaginary part $\Delta^{''}_0 (\omega)$ is a set of $\delta-$functions at these points.  The behavior of $\Delta' _0(\omega)$ has been obtained in Refs. \cite{Karakozov_91,Marsiglio_91,combescot,Schmalian_19}. }\label{fig:Delta_real}
   \end{figure}

 The function
 \beq
 D_0(\omega) = \frac{1}{\sin{\phi_0 (\omega+i0)}}
 \label{5_31a}
 \eeq
  is a sign-changing function of $\omega$, whose real part almost diverges at a set of frequencies where
  $\phi_0 (\omega) = p\pi$, and $p=1,2..$ is an integer. The imaginary component $D^{''}_0 (\omega)$ is a set of $\delta-$functions at these frequencies. We plot the gap function $\Delta_0(\omega) = \omega D_0(\omega)$  in Fig.\ref{fig:Delta_real}.

 To analyze the phase winding, we again introduce the phase factor  via $D_0 (\omega) = |D_0 (\omega)|e^{i\eta_0 (\omega)}$ and
     consider how $\eta_0 (\omega)$ varies at $\omega \geq {\bar g}$.
      The imaginary component
     $D^{''}_0 (\omega)$ in (\ref{5_31a}) is infinitesimally small, except in the vicinity of $\omega_p$, where $\phi_0 (\omega_p) = p\pi$.
     We use Eq. (\ref{5_31}) for $\phi_0 (\omega)$   and
      express $D_0 (\omega)$ near each such point as
     \beq
      D_0(\omega) \approx \frac{\pi {\bar g}^2}{2} \frac{(-1)^p \omega_p}{\omega^2_p + {\bar g}^2} \frac{1}{\omega - \omega_p + i\delta}.
     \label{5_34}
     \eeq
     Then
     \beq
     e^{i\eta_0 (\omega)} = (-1)^p \frac{\omega - \omega_p - i\delta}{\sqrt{(\omega - \omega_p)^2 + \delta^2}}
     \label{5_34_a}
     \eeq

 \begin{figure}
  \includegraphics[width=9cm]{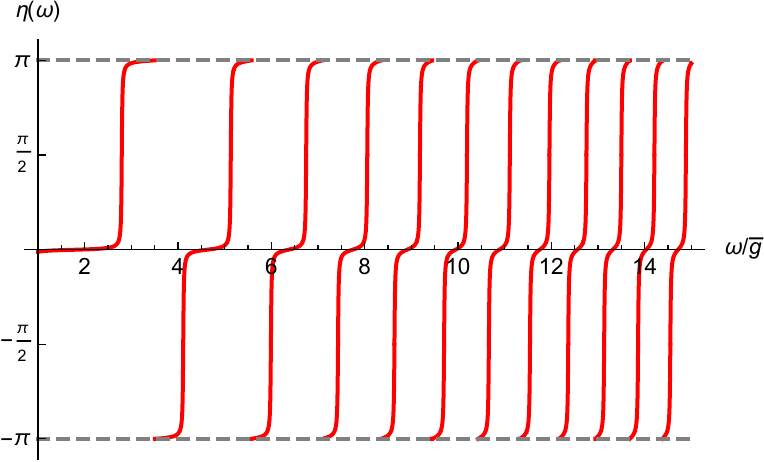}
  \caption{Variation of the phase of the gap $\eta_0(\omega)$ ($\Delta_0 (\omega) = |\Delta_0 (\omega)| e^{i\eta_0 (\omega)}$). We restrict $\eta_0 (\omega)$ to $(-\pi, \pi)$. Phase slips of $\eta_0 (\omega)$ continue up to infinite frequency. }\label{fig:eta}
\end{figure}
    We plot $\eta_0 (\omega)$ in Fig.\ref{fig:eta}.
    We see that the phase rapidly changes by $\pi$  around each $\omega_p$. If we restrict $\eta_0 (\omega)$ to $(-\pi, \pi)$, we find that the phase jumps by $2\pi$ in between $\omega_p$ and $\omega_{p+1}$.  The number of $\omega_p$ is infinite, hence the number $2\pi$ jumps is also infinite.
   We reiterate that behavior has no analog  the Matsubara axis, where $D_0 (i\omega_m)$
     is real and positive for $\omega_m >0$, hence $\eta_0 =0$.
    \footnote{For a generic $\gamma$, $\Delta_0 (\omega_m)$ and $\Delta^{''}_0 (\omega)$ are related by Cauchy formula: $\Delta_0 (\omega_m) = (2/\pi) \int_0^\infty d \omega \Delta^{''}_0 (\omega) \omega/(\omega^2 + \omega^2_m)$.  For $\gamma <2$, typical $\omega$ are of order $\omega_m$, and to reproduce $\Delta_0 (\omega_m) \propto 1/|\omega_m|^\gamma$
one needs $\Delta^{''}_0(\omega) = \sin (\pi \gamma/2) \text{ sign} \omega/ |\omega|^\gamma$.
 The
        case $\gamma =2$ is an exception here because $1/\omega^2_m$ dependence of $\Delta_0 (\omega_m)$ is
       obtained by pulling $1/\omega^2_m$ out of the denominator in the Cauchy formula.
        The remaining integral is determined by  $\omega = O({\bar g})$ rather than $O(\omega_m)$.
     Because of this, the fact that $\Delta _0(\omega_m) \propto 1/\omega^2_m$  at large frequencies does not imply  that $\Delta^{''} (\omega)$ must behave as $1/\omega^2$. }

\subsection{The one-parameter set of gap functions}
\label{sec:real_3}

We follow the same strategy as in the analysis on the Matsubara axis and expand the non-linear gap equation (\ref{e125})  in powers of $D^2$. We search for the solution in the form
\begin{equation}
D(\omega) = \sum^{\infty}_{j=0} D^{(2j+1)}  (\omega)
\label{exx2}
\end{equation}
where $D^{(1)} (\omega) = D_{\infty}(\omega)$
  and higher-order terms are obtained by solving  Eq. (\ref{e125}) iteratively.
  For $\omega < {\bar g}$, we use Eq. (\ref{5_23}) for $D_{\infty}(\omega)$.
  The computational steps are the same  as in Sec. \ref{sec:Mats_expansion}, and we obtain
 \beq
  D_\epsilon(\omega) = -2i \epsilon \left(\cos{{\tilde f}_\epsilon (\omega)} + \epsilon^2 Q_{3,\epsilon} \cos{3 {\tilde f}_\epsilon (\omega)} + \epsilon^4 Q_{5,\epsilon} \cos{5 {\tilde f}_\epsilon (\omega)} + ...\right)
  \label{5_20a}
  \eeq
where $Q_{i,\epsilon}$ are the same as in (\ref{5_20}) and
\beq
{\tilde f}_\epsilon (\omega) = \beta_\epsilon \left(\log{\left(\frac{\omega}{{\bar g}}\right)^2} - i\pi \sign{\omega}\right)  +  \phi_\epsilon
\label{5_29}
\eeq
 This $D_\epsilon (\omega)$  could also be  obtained directly from (\ref{5_20})
   by replacing  $\log{\omega^2_m}$ by
   $\log {\omega^2} - i \pi$
    in each term in (\ref{5_20}).

We recall that the  continuous  set  exists for $\epsilon \leq  \epsilon_{cr}$. For any $\epsilon <\epsilon_{cr}$, $D(\omega)$ oscillates an infinite number of times down to $\omega =0$. As $\epsilon$ approaches $\epsilon_{cr}$, log-oscillations shift to  progressively smaller frequencies. At  $\epsilon=\epsilon_{cr}$, $\beta_\epsilon$ vanishes and  log-oscillations disappear.
   The behavior of $D (\omega)$ at $\omega \to 0$  at $\epsilon \to \epsilon_{cr}$ depends on how the double limit $\omega \to 0$ and $\epsilon \to \epsilon_{cr}$ is taken.

   Like we did in Sec. \ref{sec:Mats_expansion}, we introduce $\xi = (\epsilon_{cr} - \epsilon)/\epsilon$ and re-express
 $\Delta_\epsilon (\omega)$ as $\Delta_\xi (\omega)$. The two limits $\epsilon =0$
 and $\epsilon=\epsilon_{cr}$ now correspond to $\xi = \infty$ and $\xi =0$, respectively.  This brings the notations in line with the ones we used in Secs. \ref{sec:real_1} and \ref{sec:real_2}.

 On the Matsubara frequency, all $\Delta_\xi (\omega_m)$ behave in the same way at $\omega_m > {\bar g}$:
 $\Delta_\xi \propto 1/\omega^2_m$.
 On the real axis, the dependence on $\xi$ is more complex. To see this,
  we use the solution of the linearized gap equation $D^{(1)} \propto  i e^{i\phi_0 (\omega)}$
with $\phi_0 (\omega)$,  given by (\ref{5_31}),
and evaluate $D^{(2n+1)}$
in order-by-order
  expansion of the non-liner gap equation in $D^2$.
   Collecting the series, we obtain
  the closed form expression
   \beq
  D_\xi (\omega)  =   \frac{-2ie^{i\phi_0 (\omega)}}{1 +\xi -  e^{2i \phi_0 (\omega)}/(1+\xi)}=\frac{1}{\sin[\phi_0(\omega)+i\log(1+\xi)]}
  \label{5_20b}
  \eeq
This expression can be equivalently obtained by solving
  Eq. (\ref{5_30}) for $\phi (\omega)$ with the initial condition
   $\phi ({\bar g}) = {\bar g}/\Delta_0 (0) +i \log{(1+\xi)}$.

  The parameter $\xi$ runs between $0$ and $\infty$.
  For $\xi=0$, Eq. (\ref{5_20b}) yields $D_0(\omega)= 1/\sin{\phi_0 (\omega)}$, which agrees  with (\ref{5_31a}) (one should add $i0$ to $\omega$ in this case). At  $\xi \to \infty$ we recover, by construction,  the solution of the linearized gap equation.
    For any $\xi$, including $\xi =0$, $D_\xi(\omega)$ oscillates up to an infinite frequency,
     and its phase $\eta_{\xi} (\omega)$  winds up by an infinite number of $2\pi$ between $\omega \sim {\bar g}$ and $\omega = \infty$.

We see therefore that in both limits $\omega \ll {\bar g}$ and $\omega \gg {\bar g}$, the solutions of the non-linear gap equation form a continuous one-parameter set, Eqs. (\ref{5_20a}) and  (\ref{5_20b}).
  We conjecture that for any value of $\xi$, one can use a free phase factor $\phi_\xi$ in (\ref{5_29}) to merge small-$\omega$ and large-$\omega$ expressions into a single $D_\xi (\omega)$.

\subsection{density of states}

The fermionic  DoS is defined as  $N(\omega) = (-N_0/\pi) {\text{Im}} G_l (\omega)$, where $N_0$ is the DoS in the normal state and
 \beq
 G_l (\omega) = -i \pi \sqrt{\frac{1}{1 - D^2 (\omega)}}
\label{nn_6_6_2}
  \eeq
 is a retarded
 Green's function, integrated over the dispersion.

\begin{figure}
   \includegraphics[width=16cm]{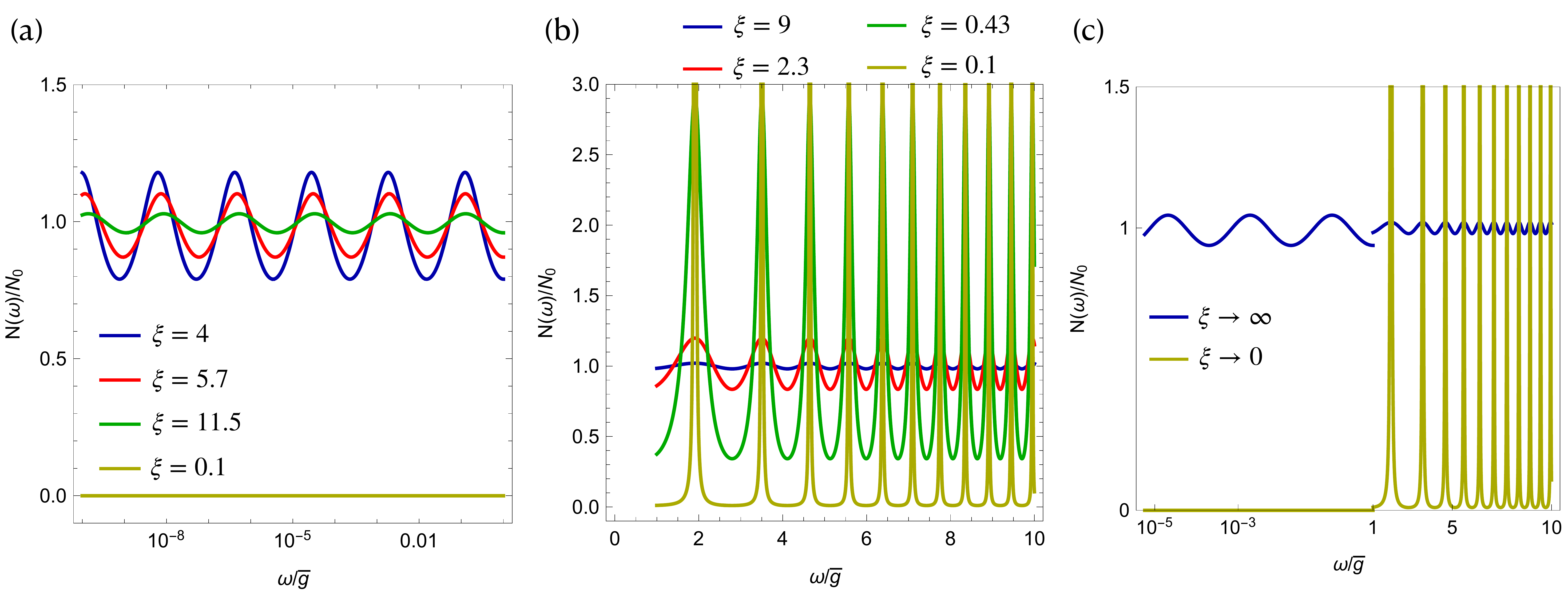}
   \caption{DoS $N_\xi(\omega)$ for (a) $\omega<\bar{g}$ and (b) $\omega>\bar{g}$ and for different $\xi$.
    For all $\xi>0$, $N_\xi (\omega)$ remains finite down to $\omega=0$ (a gapless superconductivity).
   For $\xi=0$, the DoS $N_0 (\omega)$ vanishes at small $\omega$ and has a set of $\delta-$functional peaks at $\omega > {\bar g}$.
    In panel (c) we present the schematic plot of the DoS at all frequencies.}\label{fig:DOS}
 \end{figure}

 In a BCS superconductor,
$N(\omega) \propto \text{Re} \omega/\sqrt{\omega^2 -\Delta^2}$ vanishes at $\omega < \Delta$,  has an integrable
 singularity at $\omega = \Delta +0$, and is non-zero for all $\omega >\Delta$
  because quasiparticle states  in a BCS superconductor form a continuum $\omega = \pm \sqrt{\Delta^2 + (\epsilon_k-\mu)^2}$.  In our case, the form of $N(\omega) = N_\xi (\omega)$ strongly depends on $\xi$.
   At small $\omega < {\bar g}$, $N_\xi(\omega)$  remains finite down to $\omega =0$ for all $\xi >0$.
  In this respect, all such solutions describe gapless superconductivity.  The gap function $\Delta_{0} (\omega)$ tends to a finite $\Delta_0 (0)$ at small $\omega$, and the corresponding $N_0(\omega)$
    vanishes, like in BCS superconductor. We show this in Fig.\ref{fig:DOS}(a).

At $\omega > {\bar g}$, $\Delta_{\xi}(\omega)$ is given by (\ref{5_20b}), and
 $N_\xi(\omega) = N_0 \Im \tan[\phi_0(\omega) +i \log{(1+\xi)}]$.
 For $\xi >0$, $N_\xi(\omega)$  oscillates around $N_0$ up to $\omega = \infty$. The amplitude of the oscillations increases
  with decreasing $\xi$.   For $\xi =0$,
  $N_0(\omega) = N_0 \delta/(\cos^2 \phi_0 (\omega) + \delta^2)$, where $\delta = 0+$. This DoS  consists of
 a set of $\delta-$functions at the points $\omega_k$, for which
    $\phi_0 (\omega_k) = \pi/2 + k\pi$ ($k$ is an integer) (Refs \cite{combescot,Schmalian_19,Schmalian_19a}).  We show this in Fig.\ref{fig:DOS}(b).
   In Fig.\ref{fig:DOS}(c) we  show
   $N_\xi (\omega)$ in the whole range of frequencies.

   The function $N_0(\omega)$  is the true DoS  at $T=0$, as the $\xi=0$  solution has the lowest condensation energy.
    It is different from the DoS in a BCS-type superconductor, which is  non-zero at all $\omega > \Delta$ and approaches $N_0$  at $\omega \to \infty$.  We emphasize that a qualitative distinction holds only for $\gamma =2$. For smaller $\gamma$, the DoS for the $n=0$ solution evolves as a function of frequency, but still remains non-zero at all $\omega > \Delta$ and approaches
    $N_0$ at infinite $\omega$ (see Paper IV).

In a BCS superconductor, a continuous $N(\omega)$ at $\omega >\Delta$ is the consequence of the fact that fermionic energy
 $E_k$ is a  continuous function of the normal state dispersion $\epsilon_k$,
 $E_k = \sqrt{\epsilon^2_k + \Delta^2}$.  The form of $N_0 (\omega)$ as a set of $\delta$-functional peaks raises the issue whether fermionic energies get quantized at $\gamma =2$.  To address this issue, we compute the total weight of each level:  $N_k = (1/2\pi) \int N_0 (\omega)/N_0$, where the integration is confined to the vicinity of $\omega_k$.
Using $\phi_0 (\omega) \approx \omega^2/\pi$,
we obtain $N_k = 1/\sqrt{8(1+2k)}$.  We see that  $N_k <1$ for all $k$. Because of this,   $\omega_k$ cannot be viewed as true quantized fermionic energy levels, as a fermion is necessary distributed between $\omega_k$ with different $k$.

\section{Gap function in the upper frequency half-plane.}
\label{sec:complex}

  Comparing $D_\xi (\omega)$  and $D_\xi (\omega_m)$, we see that they are similar at
 small frequencies, but very different at $\omega, \omega_m > {\bar g}$. Indeed,
    on the real axis, the phase $\eta_\xi (\omega)$ winds up by an infinite number of $2\pi$ between $\omega = O({\bar g})$ and $\omega = \infty$, while near  the Matsubara axis, $\eta_\xi (\omega_m) =0$ in this frequency range. The discrepancy implies that phase winding must end somewhere between the real and the Matsubara axis. We now argue that
    there is a set of vortices
  in the upper frequency half-plane, at $|z| \geq {\bar g}$ and the phase winding drops by $2\pi$ each time
   the axes of $z$ passes through a vortex upon rotation away from the real axis.

We use the Cauchy relation
 \beq
 \Delta (z) = \frac{2}{\pi} \int_0^\infty dx \frac{x \Delta^{''} (x)}{x^2 -z^2}
 \label{5_35}
 \eeq
 to  extend the gap function $\Delta (x) = x D(x)$ from the real axis to complex $z = \omega' +i \omega^{''}$ with $\omega^{''} >0$.   We use Eq. (\ref{5_20b}) for the gap function as we expect vortices to be present at $|z| > {\bar g}$. Like before, we first consider the  cases $\xi=0$ and $\xi \to \infty$,  and then extend the analysis to arbitrary $\xi$.

\subsection{case $\xi=0$}

Using the expansion near $\phi (\omega_p) = p \pi$, Eq. (\ref{5_34}), we approximate $\Delta^{''}_0 (\omega)$ as
 \beq
 \Delta^{''}_0 (\omega) \approx \frac{\pi^2 {\bar g}^2}{2} \sum_{p=1}^\infty \frac{(-1)^{p+1} \omega^2_p}{\omega^2_p + {\bar g}^2} \delta (\omega - \omega_p).
     \label{5_36}
     \eeq
Substituting into (\ref{5_35}), we obtain
\beq
 \Delta_0 (z) = \pi {\bar g}^2 \sum_{p=1}^\infty \frac{(-1)^{p+1} \omega^3_p}{\omega^2_p + {\bar g}^2} \frac{1}{\omega^2_p -z^2}.
 \label{5_37}
 \eeq
   Here $\omega_p$ is a  solution of $\phi_0 (\omega_p) = \pi p$, where  $\phi_0 (\omega)$ is given by (\ref{5_31}).
  We verified numerically that KK relations on the real axis are satisfied, i.e., if we use (\ref{5_36}), we reproduce with high accuracy $\Delta' (\omega)$.
   On the Matsubara axis,
$z=i\omega_{m}$,
Eq. (\ref{5_37}) yields, at $\omega_m \gg {\bar g}$,
  \beq
  \Delta_0 (\omega_m) = a \frac{{\bar g}^3}{\omega^2_m}
  \label{5_38}
  \eeq
  where
   $a = \pi \sum_{p=1}^\infty \left[(-1)^{p+1} \omega^3_p/((\omega^2_p + {\bar g}^2){\bar g})\right]$.
   Approximating $\omega_p$  by $ {\bar g} \pi \sqrt{p}$,
    we find $a =2.56$.   The number  is somewhat larger than
   $1.27$, obtained by solving the gap equation on the Matsubara axis
  (Ref. \cite{combescot} and Sec. \ref{sec:Mats_sign_preserving}). The difference likely comes from subleading terms in $\phi (\omega)$.

\begin{figure}
  \includegraphics[width=12cm]{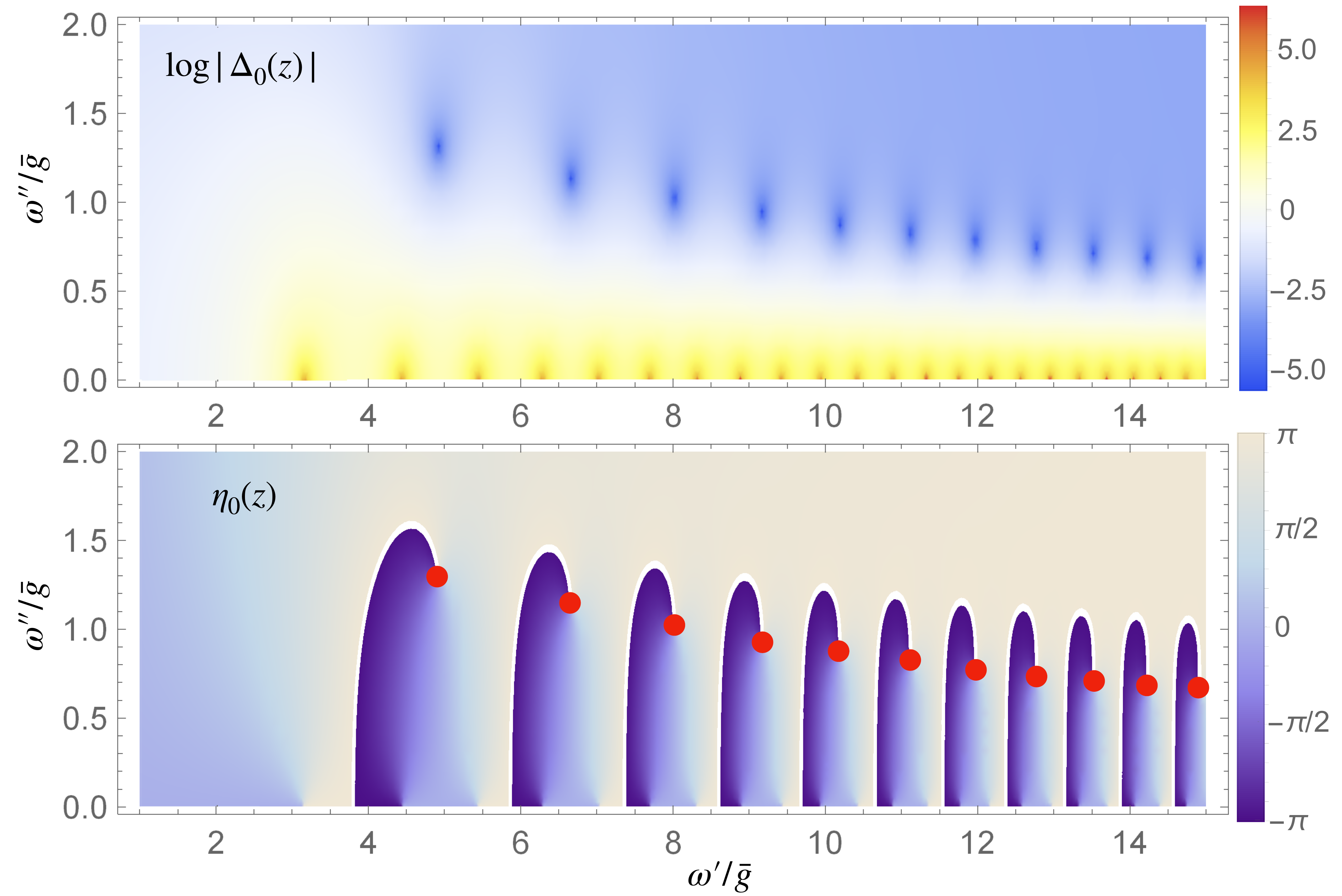}
  \caption{$\Delta_0(z)$ in the upper half plane. Upper panel: $\log{|\Delta_0(z)|}$.
  Blue spots mark the locations of dynamical vortices, where $|\Delta_0(z)|=0$. Lower panel: the phase of the gap $\eta_0(z)$ defined via $\Delta_0(z) = |\Delta_0(z)| e^{i\eta_0 (z)}$. The phase slips by $2\pi$ upon crossing a white line in the direction from   near-white to  dark-blue. The white lines are the locations of points where $\Delta^{''}_0 (z) =0$ and $\Delta'_0 (z) <0$. }\label{fig:DeltaZ}
\end{figure}

  We plot $\Delta_0 (z)$ for a generic $z$ in the upper half-plane in Fig.\ref{fig:DeltaZ}
  We clearly see
  that there is a set of points, where $\Delta'_0 (z) = \Delta^{''}_0 (z) =0$.  These points are the centra of dynamical vortices with anti-clockwise circulation $2\pi$. The vortices are located along a particular line
  in the complex plane. The set extends to an infinite frequency, i.e., the number of vortices is infinite. This is consistent with an infinite phase winding along the real axis.  We verified that
  if we use a  more accurate form of $\omega_p$, the positions of the vortices shift a bit, but their number  remains infinite.

\begin{figure}
  \includegraphics[width=12cm]{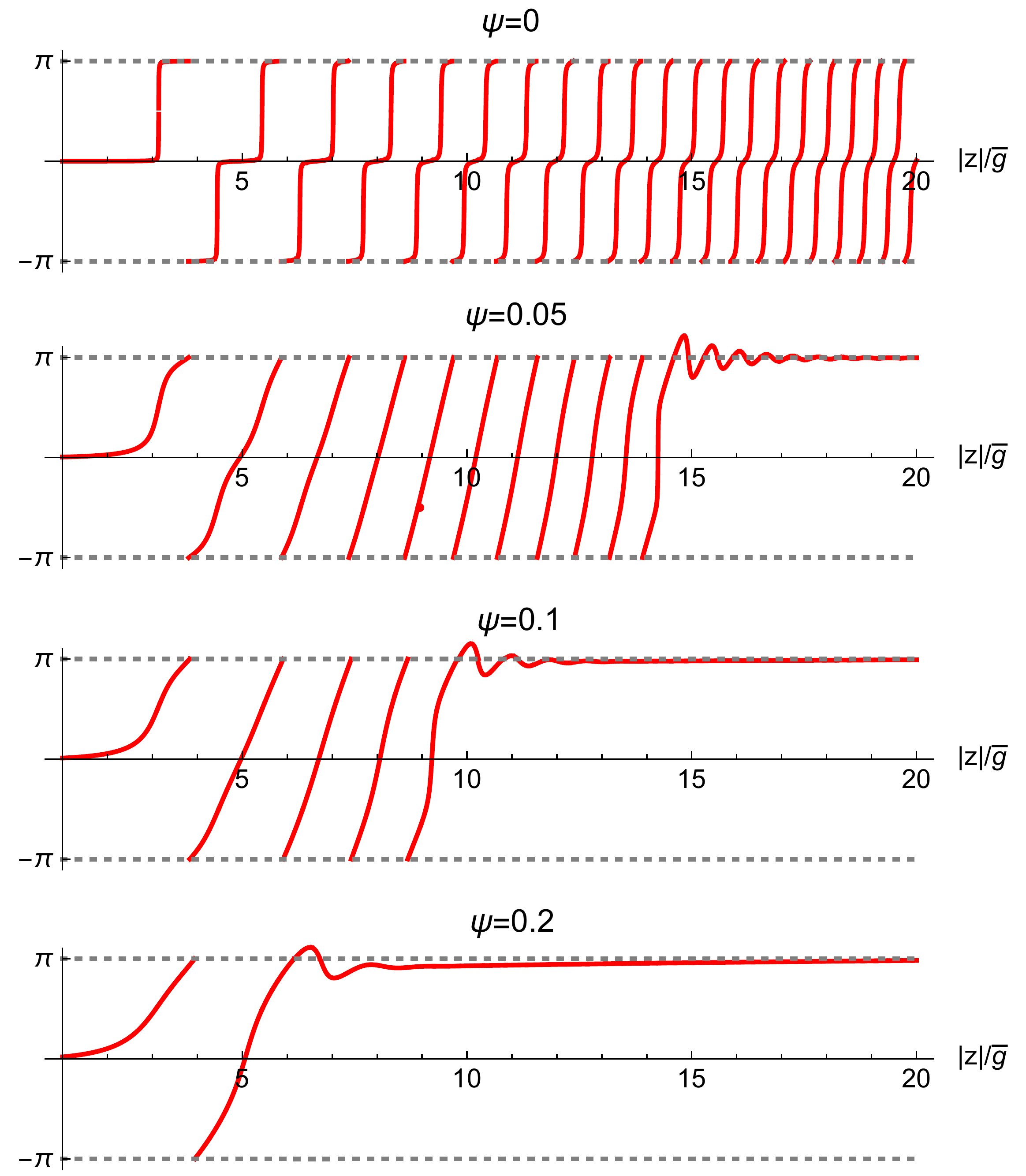}
  \caption{Phase variation $\eta_0(|z|,\psi)$
  along different paths specified by $\psi$, defined via $z = |z| e^{i\psi}$.
   Along real axis, $\psi =0$; along Matsubara axis, $\psi = \pi/2$.
    Along the real axis the phase $\eta_0(\omega)$ winds up an infinite number of times, i.e., the winding number (the number of $2\pi$ phase slips) is infinite. For a finite $\psi$, phase winding ends at some finite $|z|$, and the winding number becomes finite.
    }\label{fig:eta_z_psi}
\end{figure}

To see how the winding
number changes once we rotate from the real to the Matsubara axis, we introduce $z= |z| e^{i\psi}$ ($\psi =0$ along the positive real semi-axis and $\pi/2$   along the Matsubara axis) and check the winding of the phase of $\Delta_0 (z) = |\Delta_0 (z)| e^{i\eta_0 (z)}$
between  $|z| \sim {\bar g}$ and $|z| \to \infty$ along the directions in the upper frequency half-plane, specified by  $\psi$. We show the results in Fig.\ref{fig:eta_z_psi}.
We see that for any  $\psi >0$, the phase $\eta_0 (z)$ winds for $|z|$ below a certain value, and then saturates.  At larger $|z|$, both $\Delta'_0 (z)$ and $\Delta^{''}_0 (z)$ scale as $1/|z|^2$ with no oscillations.   Counting the total phase winding $\delta \eta_0$ between $|z| = O({\bar g})$ and $|z|=\infty$, we see that $\delta \eta_0 = 2\pi s$, where $s$ is an integer. It  decreases by one every time the direction set by $\psi$ passes through a vortex.
  The winding vanishes at some $\psi \leq \pi/4$.

\subsection{case $\xi =\infty$}

\begin{figure}
  \includegraphics[width=12cm]{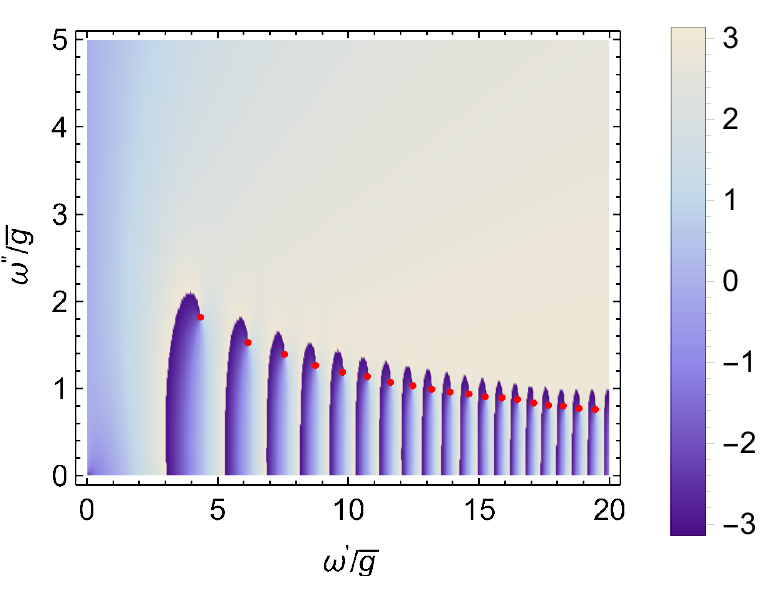}
  \caption{The phase $\eta_\infty(z)$ of $\Delta_{\infty}(z)$ in the first quarter of the complex plane of frequency ($\omega'>0$, $\omega^{''} >0$). }\label{fig:eta_infty}
\end{figure}

\begin{figure}
  \includegraphics[width=10cm]{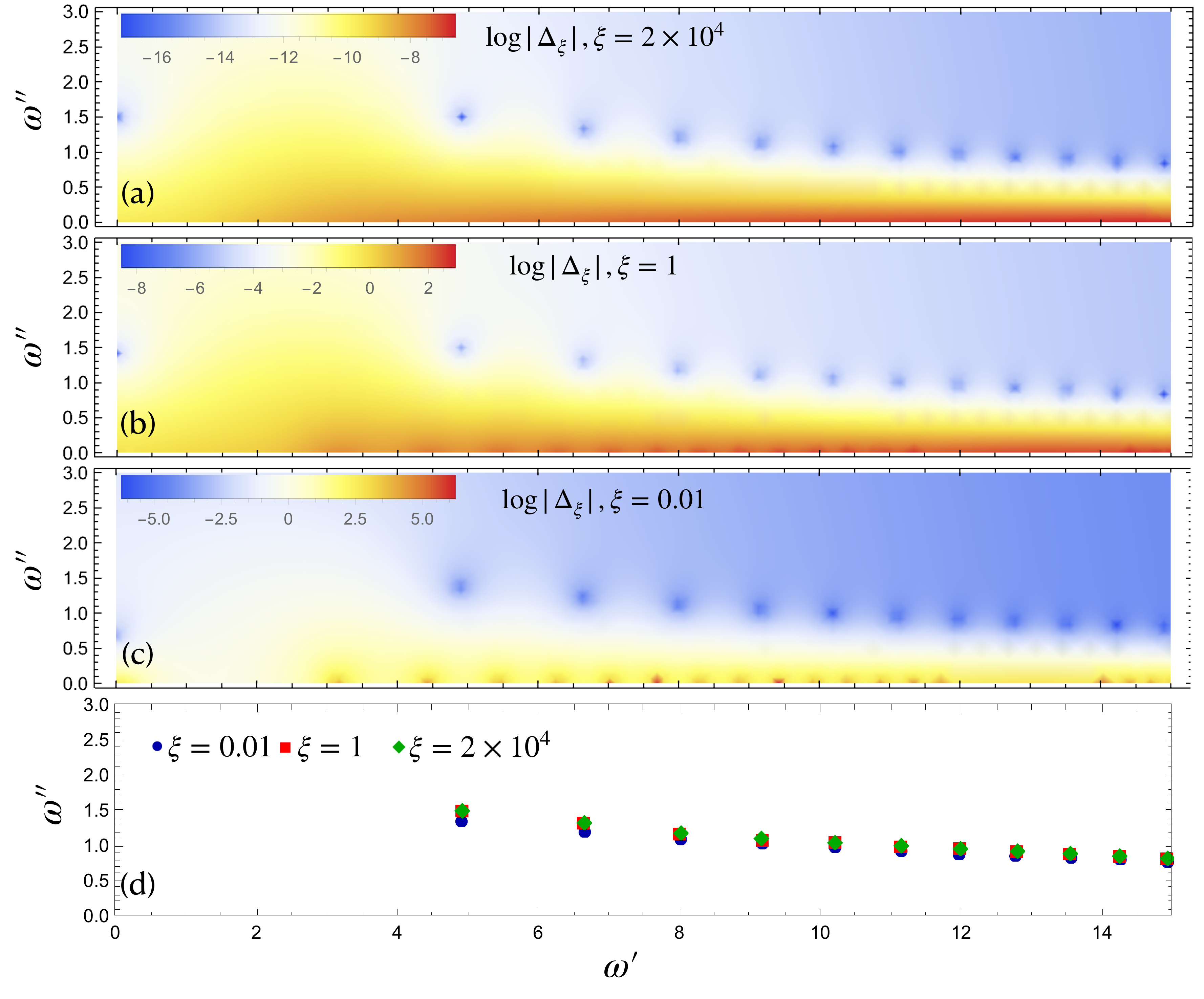}
  \caption{(a)-(c)Gap functions $\Delta_{\xi}$ defined in \eqref{5_20b} for different $\xi$ in the frequency upper half plane. Here we take $\phi(\omega)\approx(\omega/\bar{g})^2/\pi$ (d) Comparison of the vortex positions for different $\xi$ obtained by approximating $\phi(\omega)$ by $(\omega/\bar{g})^2/\pi$. The results suggest that the positions of the vortices almost do not depend on the value of $\xi$.}\label{fig:VortexPosition}
\end{figure}

We next consider the opposite limit $\xi = \infty$. The form of $\Delta_\infty ( z )$ can be obtained starting from (\ref{nn_2}) and replacing  $\omega^2_m$ by
$|z|^2  e^{i(2\psi-\pi)}$. This gives
\beq
 \Delta_\infty ( z ) \propto  \int_{0}^\infty dk \left( b_k e^{-ik \log{|z|^2/{\bar g}^2}+ (2\psi-\pi)k} + b_{-k} e^{i k \log{|z|^2/{\bar g}^2}-(2\psi-\pi)k} \right),
 \label{delta_z}
 \eeq
 where $b_k$ is defined in (\ref{nn_2_1}).
  We obtain $\Delta_\infty ( z )$ by numerical integration. We plot its  phase $\eta_\infty(z)$  in Fig.~\ref{fig:eta_infty}.
 We again see that there is an infinite array of vortices. The array extends to an infinite frequency, where it approaches the real axis.  The vortex arrangement in Fig.~\ref{fig:eta_infty} is remarkable similar to that
   in Fig. \ref{fig:DeltaZ} for $\xi=0$.  Moreover, if we approximate $\phi (\omega)$ by the leading term $(\omega/{\bar g})^2/\pi$, we find that the positions of the vortices are at the same $z_i$ in both cases. We can see this by comparing Fig.\ref{fig:VortexPosition}(a) where $\xi \to \infty$ and Fig.\ref{fig:VortexPosition}(c) where $\xi \to 0$. The gap function $\Delta_{\xi}(z)$ are very similar in these two cases, despite that the overall factors are different. The vortex positions for these two cases are almost identical, as can be seen from Fig.\ref{fig:VortexPosition}(d).

\subsection{arbitrary $\xi$}

The same infinite array of vortices exists for all $0<\xi<\infty$. As an example, in Fig.\ref{fig:VortexPosition}(b) we show the gap function for $\xi =1$.
 We clearly see that there is an infinite array of vortices, similar to the ones for $\xi=\infty$ and $\xi=0$,   and the positions of vortices are almost indistinguishable, see Fig.\ref{fig:VortexPosition}(d). Analytically, if we use Eqs. (56) and (58), we find that the positions of vortices are independent on $\xi$.

\subsection{Essential singularity}

There is another  consequence of the existence of an infinite array of vortices -- each the gap function $\Delta_\xi (z)$ has an essential singularity
 at $|z|= \infty$. Indeed, one can reach $|z| = \infty$ from the set of vortex points, where $\Delta_\xi (z) =0$, or from the real axis, where $\Delta_\xi (\omega)$ oscillates, and the amplitude of the oscillations does not vanish at $\omega \to \infty$,
 hence neither $\lim_{|z| \to \infty} \Delta_\xi (|z|)$
         nor $\lim_{|z|\to \infty} 1/\Delta_\xi (|z|)$ exist.
         We emphasize that an essential
   singularity is only present  for $\gamma=2$. For smaller $\gamma$, phase winding and  associated vortices exist only at $|z|$ smaller than a certain, $\gamma-$dependent value. At larger $|z|$, $\Delta (z)$ scales as $1/|z|^\gamma$ and vanishes at $|z| = \infty$ no matter how this limit is reached.

Further,  for $\gamma =2$, the very existence of a non-zero  $\Delta_\xi (z)$  for  a generic $z$ away from vortex points, is ultimately related to an essential singularity at $|z| = \infty$. The argument is that  the set of vortex points is complete, hence one can analytically continue the gap function from this set to the upper half-plane of frequency in the same way as $\Delta (z)$
  is obtained from a discrete set of Matsubara points $\omega_m = \pi T(2m+1)$  in standard diagrammatic calculations for interacting fermions.  If this analytical continuation was unique,  we would obtain
  $\Delta (z) \equiv 0$,  because $\Delta (z)=0 $ at the vortex points.  For a non-zero $\Delta_\xi (z)$, the extension must be multi-valued. A  rigorous mathematical argument is that this is the case when
 the end point of the set, $|z| = \infty$, goes outside the domain of analyticity.  This is exactly what we have because of an essential singularity at $|z| = \infty$.

 We conjecture that the multi-value nature of the extension is the reason why
    the set of $\Delta (\omega)$ is a continuous one at $\gamma =2$.  This is plausible, particularly if the vortices are at the same $z_i$ for all $\xi$,  as Fig.\ref{fig:VortexPosition} seems to indicate. However,  at the moment, we cannot prove this.

\section{Finite $\omega_D$}
\label{sec:omega_D}

\subsection{Gap equation at a finite $\omega_D$}

We now consider the case when the bosonic mass is small but finite. By analogy with the phonon case we call this mass $\omega_D$.  On the Matsubara axis,  $\Delta_0 (\omega_m)$ changes little compared to the case $\omega_D =0$.  The set of $\Delta_n (\omega_m)$ still exists at small $\omega_D$, but becomes  discrete and holds up to a finite $n_{max}$. In particular, there is no solution of the linearized gap equation at $T=0$ for any non-zero $\omega_D$.   The value of $n_{max}$ can be estimated by
  noticing that if we,  e.g., depart from the  solution  on the Matsubara axis at $\omega_D=0$  and compute  corrections due to finite $\omega_D$, these corrections  increase at small $\omega_m$ and become $O(1)$ at  $\omega_m \sim \omega_D$.  A simple experimentation shows that this sets $n_{max}$ at
    \beq
   n_{max} \sim  \frac{{\bar g}}{\omega_D}
   \label{con_3}
   \eeq
 On the real axis, the gap equation still has the form  $D(\omega) \omega B(\omega) = A(\omega) + C(\omega)$, and  $A(\omega)$ and $B(\omega)$ remain the same as in (\ref{e124}), up to irrelevant small corrections. However, $C(\omega)$ changes to
\beq
C(\omega) = - i \frac{\pi{\bar g}^2}{2 \omega_D} \frac{D(\omega - \omega_D) - D(\omega)}{\sqrt{1 -D^2(\omega - \omega_D)}} ~ \sign \omega
\label{eee_7}
\eeq
Expanding to first order in $\omega_D$ and introducing, as before, $D(\omega) =1/\sin{\phi (\omega)}$, we obtain after straightforward algebra  that the gap equation reduces to
 \bea
 && {\dot{\phi}}  - \frac{\omega_D}{2} \left((\dot \phi)^2 \tan{\phi (\omega)} + \ddot {\phi}\right)=
\frac{2}{\pi {\bar g}^2}
 \left[\omega B(\omega) - A(\omega) \sin{\phi (\omega)}\right] + ...
 \label{3_43a}
 \eea
  where  dots  stand for the
 terms  with higher powers of $\omega_D$.
  A similar equation at a finite $T$ instead of finite $\omega_D$ has been obtained by Combescot~\cite{combescot}

\begin{figure}
   \includegraphics[width=8cm]{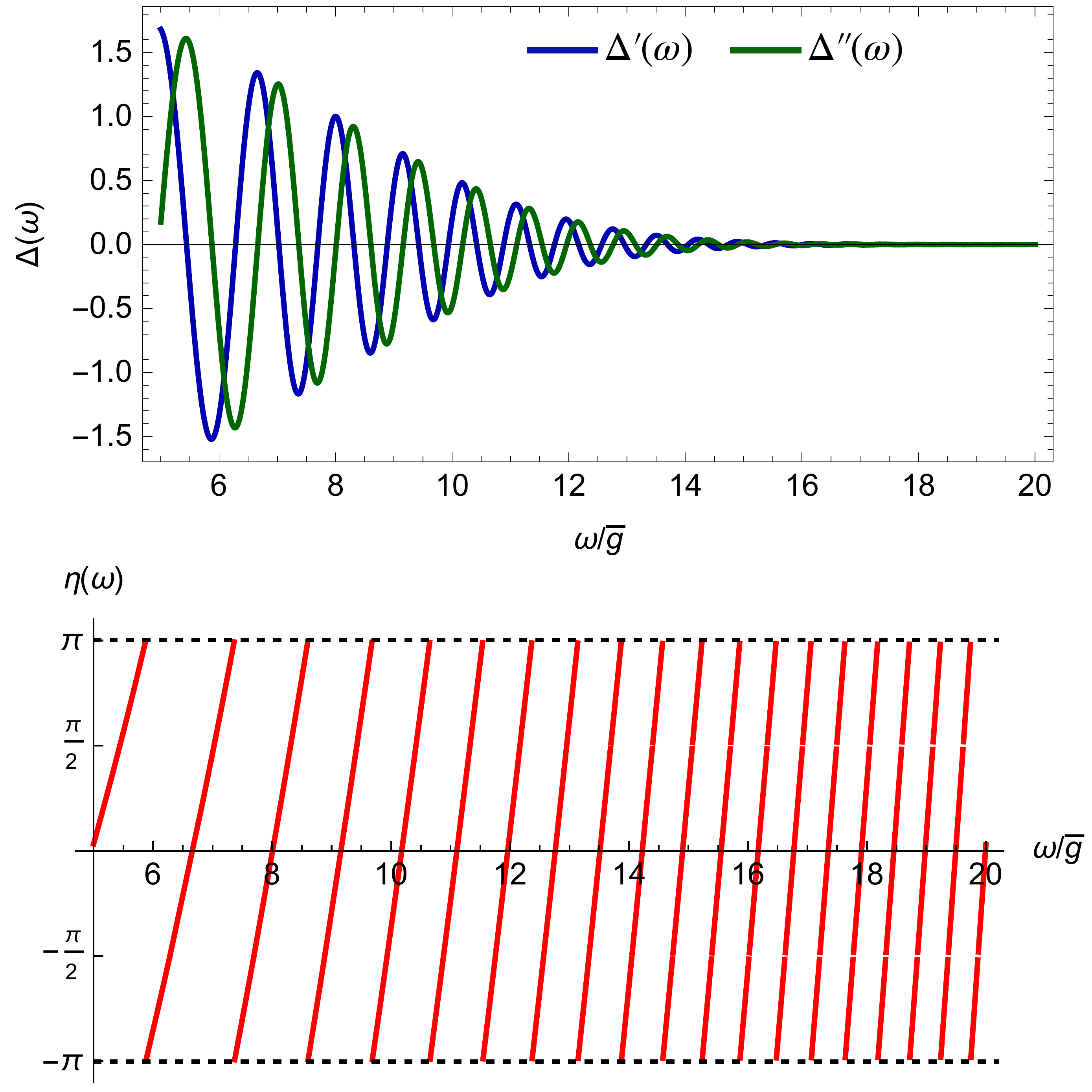}
   \caption{The gap function $\Delta(\omega)$ (a) and the variation of its phase $\eta(\omega)$ at $\omega_D=0.02\bar g$.
    From \eqref{eq:DeltafiniteT}.}\label{fig:Delta_eta_finiteT}
 \end{figure}

\begin{figure}
  \includegraphics[width=15cm]{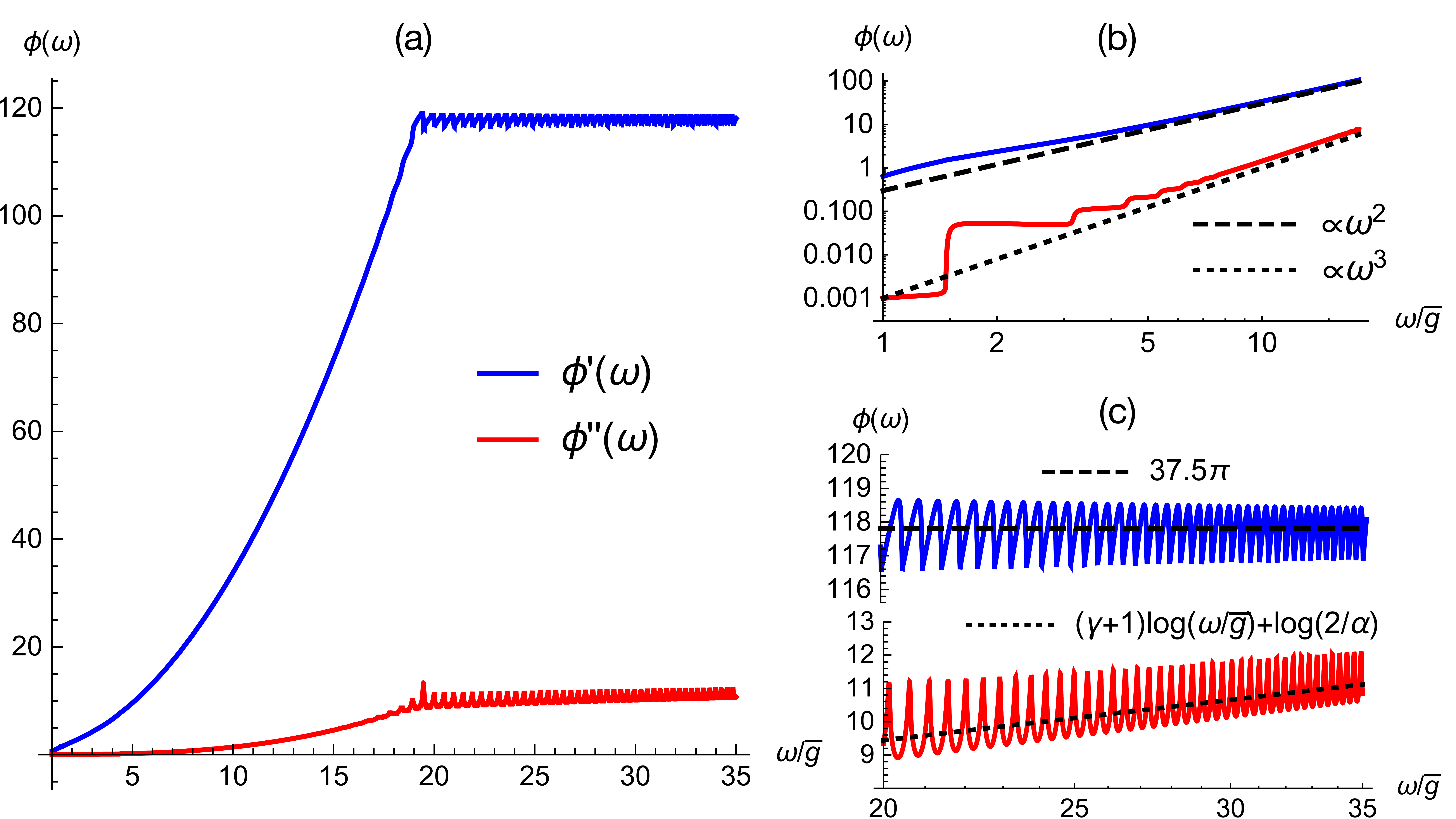}
  \caption{(a) Numerical solution of Eq. \eqref{3_43a} for a complex $\phi (\omega)$ at $\omega_D=0.02\bar g$. (b) and (c):
   Asymptotic forms of $\phi' (\omega)$ and $\phi^{''} (\omega)$.  At $\omega<\omega_{max}$, $\phi'(\omega)$ increases as $\omega^2$, while $\phi''(\omega)$  first displays a step-like behavior and then increases  as $\omega^3$, with $\omega_D$ in the prefactor. At $\omega>\omega_{max}$, $\phi'(\omega)$ saturates at$(2m-1/2)\pi$, where $m=19$ for our chosen $\omega_D$, and $\phi''(\omega)$ increases logarithmically.
   The corrections to asymptotic values oscillate with the period set by $\omega^2/\pi$.
  In the numerical solution, we neglected $\ddot\phi$ term in \eqref{3_43a} compared to $(\dot\phi)^2$ and verified that this is a valid approximation.}\label{fig:Diff_solve_finiteT}
\end{figure}

  For definiteness, let's consider the case $\xi =0$. 
  At $\omega \geq {\bar g}$,   $B(\omega)$ and $A(\omega)$ from (\ref{e124}) can be approximated by $B(\omega) \approx 1 +{\bar g}^2/\omega^2$ and $A(\omega) \approx -1.27 {\bar g}^3/\omega^2$.  To understand the effect of $\omega_D$  we use as an input the solution at $\omega_D$ =0, $\phi (\omega) \approx \omega^2/(\pi {\bar g}^2)+i\delta $. Substituting this input into
 (\ref{3_43a}), expanding near $\omega = \pi {\bar g}/\sqrt{2}$, where $\phi (\omega) = \pi/2$, expressing $\phi (\omega) = \phi^{\prime} (\omega) +i \phi^{\prime \prime} (\omega)$, and solving for  $\phi^{''} (\omega)$, we find that it jumps to $O(\omega_D)$  once $\omega$ exceeds $\pi {\bar g}/\sqrt{2}$. The same happens at all $\omega_n =  \pi {\bar g}/\sqrt{2} (2n+1)^{1/2}$, where $\tan{\phi' (\omega)} =0$.
   After $n$ jumps,
 $\phi^{''} (\omega)$ becomes
 \beq
 \phi^{''} (\omega) =
\frac{\pi \omega_D}{ \sqrt{2} \bar g} \sum_{m=1}^{n} \sqrt{2m+1}
\approx \frac{2 \pi \omega_D}{3{\bar g}}  n^{3/2} \approx   \frac{2 \omega_D}{3 \pi^2} \frac{\omega^3}{{\bar g}^{4}}.
 \label{3_44}
 \eeq
  A more accurate, non-perturbative analysis of (\ref{3_43a}) shows that $\phi^{''} (\omega)$ appears slightly before $\phi' (\omega)$ reaches $\pi/2$. This smoothes up the jumps, but the functional form of $\phi^{''} (\omega)$ in (\ref{3_44})  remains intact.
       When both $\phi'$ and $\phi^{''}$ are nonzero, $D(\omega)$ is a complex function of frequency:
       \beq
   D'_0 (\omega) = \omega \frac{\sin{\phi' (\omega)} \cosh{\phi^{''} (\omega)}}
  {\sin^2{\phi' (\omega)} + \sinh^2{\phi^{''} (\omega)}}~ {\text {sign}} \omega
   \eeq
   and
       \beq
   D''_0 (\omega) = - \omega \frac{ \cos{\phi' (\omega)} \sinh{\phi^{''} (\omega)}}
  {\sin^2{\phi' (\omega)} + \sinh^2{\phi^{''} (\omega)}}
 \label{eee_4a}
   \eeq
At  $\omega > 3\pi^2 {\bar g}^4/(2\omega_D)$, $\phi^{''} (\omega)$ becomes larger than one.   At such frequencies, both $D' (\omega)$ and
$D'' (\omega)$ oscillate with progressively decreasing magnitudes, approximately as the real and the imaginary parts of
   \beq
   -2i e^{i\omega^2/(\pi\bar{g}^2)} e^{-\frac{2 \omega_D}{3\pi^2\bar g} (\omega/{\bar g})^{3}},\label{eq:DeltafiniteT}
   \eeq
 and the phase  $\eta_0 (\omega)$ gradually winds up as $\omega$ increases.
  We show this in Fig.\ref{fig:Delta_eta_finiteT}.
 This behavior holds as long as $|A(\omega)| \ll \omega$, i.e.,  $\omega < \omega_{max}$, where
    \beq
\omega_{max} \sim {\bar g} \left(\frac{\bar g}{\omega_D} \log{\frac{\bar g}{\omega_D}}\right)^{1/3}
\label{eee_1}
\eeq
 At even larger frequencies, the $A(\omega)$ term cannot be neglected, and
    the forms of $\phi' (\omega)$ and $\phi^{''} (\omega)$ change.
     We show the full numerical solution of Eq. (\ref{3_43a}) in Fig.\ref{fig:Diff_solve_finiteT}.
 We see that at $\omega > \omega_{max}$, $\phi^{''} (\omega)$ keeps  increasing, while $\phi' (\omega)$ saturates.
   A simple analysis shows that Eq. (\ref{3_43a}) is satisfied, up to corrections of order $\omega_D$, if
\beq
\phi^{''} (\omega) = 3 \log{\frac{\omega}{{\bar g}}} + 0.45,~~
\phi' (\omega) = -\frac{\pi}{2} + 2 m \pi
\label{5_32}
\eeq
 where $m$ is an  integer.
   Substituting  this complex $\phi (\omega)$  into $\Delta_0(\omega) = \omega/\sin{\phi (\omega)} \approx -2i \omega e^{i \phi (\omega)}$, we see that at $\omega > \omega_{max}$, the real part of the gap function gradually decreases as
     \beq
    \Delta'_0 (\omega) = \frac{ 1.27 {\bar g}^3}{\omega^2}
  \label{5_33}
  \eeq
To obtain $\Delta^{''}_0 (\omega)$ at these frequencies, we need to keep  the $\omega_D$ term in the l.h.s. of
 (\ref{3_43a}) and obtain the correction to
(\ref{5_32}), which we label as $\tilde \phi$.
Solving perturbatively for  ${\tilde \phi} (\omega)$ we obtain
\beq
{\tilde \phi} (\omega) = f\left(\frac{\omega}{\omega_{max}}\right)~e^{i\omega^2/(\pi {\bar g}^2)}
\label{eee_4}
\eeq
 where $f(...)$ is a decreasing function of the argument.
The $\omega^2$ oscillations of ${\tilde \phi} (\omega)$ are clearly visible in the numerical results for $\phi'$ and $\phi^{''}$
 in Fig. \ref{fig:Diff_solve_finiteT}.
 Substituting (\ref{eee_4}) into (\ref{eee_4a}), we obtain
     \beq
   \Delta^{''}_0 (\omega) \sim  \frac{{\bar g}^3}{\omega^2} f\left(\frac{\omega}{\omega_{max}}\right) ~\cos{\frac{\omega^2}{\pi {\bar g}^2}}
  \label{5_33_a}
  \eeq
One can verify that an integer $m$ in (\ref{5_32}) determines the number of $2\pi$ variations of $\eta_0 (\omega)$ on the real axis and, equivalently, the  number of vortices at complex $z_i$.   The value of $m$ decreases one-by-one as $\omega_D$ increases and $\omega_{max}$ decreases. That $m$ is finite implies that there is no essential singularity at $|z| = \infty$.  Indeed, at the  largest frequencies $\Delta (\omega) \propto 1/\omega^2$.

  For completeness, we verified that higher-order terms in $\omega_D$, which we neglected in the l.h.s. of  (\ref{3_43a}),  become important at frequencies $\omega \sim {\bar g}^2/\omega_D$, which well exceed $\omega_{max}$ and  are therefore irrelevant to our purposes.

\section{Dressed superfluid stiffness}
\label{sec:stiffness}

In this section we analyze superfluid stiffness and thermal corrections to a superconducting order parameter.
 As we discussed in the Introduction, we consider the $\gamma =2$-model as the double limit $\omega_D \to 0$, $E_F \to \infty$, such that
  Migdal-Eliashberg parameter $\lambda_E = {\bar g}^2 N_0/\omega_D$ remains small ($N_0 \sim 1/E_F$ is the DoS per unit volume in the normal state). Accordingly, in the analysis below we keep $\omega_D$ small, but finite.

\subsection{Bare stiffness}

A superfluid stiffness is the ratio of the excess energy $E_\eta$ due to inhomogeneous variation of the phase of a superconducting order
parameter $\Delta (r) = \Delta e^{i\eta (r)}$ and
$\int dr\left(\nabla \eta(r)\right)^2$: $E_\eta = \rho_s \int dr  \left(\nabla \eta(r)\right)^2$.
  In the momentum space,
 \beq
 E_\eta = \rho_s ~\sum_q q^2 \eta_q^2
 \label{feb12_2_aa}
\eeq
A way to compute $\rho_s$ is to choose $\eta_q = \delta_{q,q_0}$ and extract $\rho_s$ as the prefactor for $q^2_0$ term  in the particle-particle bubble (the sum of GG and FF terms) at zero frequency and finite $q$ (see Refs. \cite{randeria_1,cee_2,dima_m}

At $\omega_D/{\bar g}>1$ ,
the system is in a weak coupling limit, and
 superfluid  stiffness at $T=0$ is a fraction of the Fermi energy, $\rho_s  (T=0) = E_F/(4\pi)$ (Refs. \cite{randeria_1,cee_2}).
  This stiffness is much larger than $T_c$~\cite{Schriefferbook,*Emery1995}. At $T >0$, $\rho_s (T)$ drops and vanishes at $T_c$,
   but at weak coupling a drop of $\rho_s$ occurs only in the immediate vicinity of $T_c$.

 At small $\omega_D/{\bar g}$, strong mass renormalization $m^*/m = 1 + {\bar g}^2/\omega^2_D$ changes the stiffness to
 \beq
 \rho_s (T=0) \sim E_F \frac{\omega_D \Delta (0)}{{\bar g}^2}  \sim \frac{T_p}{\lambda_E}
  \label{con_4}
  \eeq
   where $T_p \sim \Delta (0)$ is the onset temperature of the pairing.  As long as $\lambda_E \leq 1$,  $\rho_{s} (T=0) > T_p$.

We now relate the stiffness to the strength of thermal phase fluctuations of $\Delta (r) = \Delta e^{i\eta (r)}$.
 For this, consider the correlator
	\begin{equation}
			\langle \eta(r)\eta(0)\rangle=\frac{\int D[\eta]\eta(r)\eta(0)e^{-\rho_s \int dr  \left(\nabla \eta(r)\right)^2/T}}{\int D[\eta]e^{-\rho_s \int dr  \left(\nabla \eta(r)\right)^2/T}}
\label{new_a}
	\end{equation}
 We assume that in equilibrium $\eta (r) =0$ and
 expand $\langle e^{i\eta (r)}\rangle$   as $1 - \langle\eta^2 (r)\rangle/2$.
Transforming (\ref{new_a})
to the momentum space, we obtain
$\langle \Delta (r)\rangle = \Delta (1- \langle \eta^2\rangle)$,
 where
\begin{equation}
\langle\eta^2\rangle = \frac{1}{N} \sum_q \frac{\prod_{q'}\int d \eta_{q'} \eta^2_q ~e^{-\rho_{s} q^2 \eta^2_{q'} /T}}{\prod_{q'}\int d \eta_{q'} ~e^{-\rho_{s} q^2 \eta^2_{q'} /T}}
\label{feb12_4}
\end{equation}
 where $N$ is the number of particles in the system.
Evaluating the integrals, we obtain the conventional result~\cite{Pokrovsky_1979}
\begin{equation}
\langle\eta^2\rangle = \frac{T}{\rho_{s}} \frac{1}{N} \sum_q \frac{1}{q^2}.
\label{feb12_5}
\end{equation}
 We assume for simplicity that spatial dimension $D$ is larger than $2$, in which case the sum converges. By order of magnitude we then have
  \begin{equation}
\langle\eta^2\rangle ~ \sim \frac{T}{\rho_{s} (T)}
\label{feb12_6}
\end{equation}
 As long as $\rho_s (T) > T$, fluctuation corrections to the order parameter are small.
  This does not hold in the immediate vicinity of the onset temperature of the pairing, $T_p$, but as long as $\rho_s (0) \gg T_c$,
   the $T$ range, where fluctuations are strong and destroy phase coherence, is quite narrow, i.e., superconducting $T_c$ remains close to $T_p$. We see that this holds even when $\omega_D$ is small and the reduction of $\rho_s$ by mass renormalization is strong.

 \subsection{Dressing of $\rho_s$ by soft longitudinal fluctuations}

We now argue that in our case the expression for $\langle\eta^2\rangle$ is different due to the presence of a continuum gapless spectrum of condensation energy, $E_{c,\xi}$, where, we remind,   $\xi$ runs between $0$ and $\infty$, and the bottom of the spectrum is at $\xi=0$. We will need states near the bottom of the continuum, at $\xi \ll 1$. For such states,
 we assume
\beq
E_{c,\xi}  = E_{c,0} + b_1  N_0 N \bar{g}^2 \xi^2,
 \label{con_2}
 \eeq
  where $b_1 = O(1)$ and  $N$ is the total number of particles.
 We will also need superfluid  stiffness $\rho_{s,\xi}$ for the states near the bottom of the continuum.
  Evaluating the particle-particle susceptibility for a generic $\Delta_\xi (\omega_m)$  and extracting the $q^2$ term we obtain
  \begin{equation}
\rho_{s,\xi} \sim E_F \frac{\omega_D}{{\bar g}^2}  \int d\omega_m \frac{D^2_\xi (\omega_m)}{1 + D^2_\xi (\omega_m)}
\label{feb12_8}
\end{equation}

 For $\xi= 0$, the integral is determined by $\omega_m \sim {\bar g}$, where $D_0(\omega_m) \sim 1$. This yields $\rho_{s,0}  \sim E_F \omega_D/{\bar g} \sim T_p/\lambda_E$, as in (\ref{con_4}).  For states with $\xi >0$, the magnitude of $\Delta_\xi (\omega_m)$ is reduced, and the stiffness gets smaller.  We assume that for the states near the bottom of the continuum,
   the stiffness is obtained by expanding to first order in $\xi$:
  \beq
  \rho_{s, \xi} =\rho_{s, 0} \left(1 - b_2 \xi\right)
 \label{con_5}
 \eeq
 where $b_2 =O(1)$ is  positive.
  The extra energy of a given state $\xi$ due to phase variation is
  \beq
  E_{\eta,\xi} =  \rho_{s, \xi} \sum_q q^2 \eta_q^2
  \label{con_9}
  \eeq

 We assume (see reasoning below) that all states near the bottom of a continuum  contribute to the variation of the phase, i.e.,
   the averaging in $\langle(\eta_q)^2\rangle$ is over both $\eta_q$ and $\xi$ with the  weight factor $e^{-E_\xi/T}$, where
 \beq
E_\xi = E_{c,\xi} +E_{\eta,\xi} = E_{0} + \delta E_\xi,
\label{aaaa}
\eeq
and
\bea
&&E_{0} = E_{c,0}  + \rho_{s,0} \sum_q q^2 \eta_q^2 \nonumber \\
&& \delta E_\epsilon =  b_1  N_0 g^2 \xi^2 - b_2 \rho_{s,0} \xi \sum_q q^2 \eta_q^2
\label{con_10}
\eea
If we neglected $\delta E_\xi$,  we would obtain the same result as before:
 \beq
\langle\eta^2\rangle = \frac{T}{\rho_{s,0}} \frac{1}{N} \sum_q \frac{1}{q^2}
\label{ee_4_1}
\eeq
Keeping $\delta E_\xi$ we find that $\langle\eta^2\rangle$ has an additional overall factor, which we label as $I_T$. Dropping for simplicity numerical prefactors $b_1$ and $b_2$, we obtain after integrating over $\eta_{q'}$
\beq
I_T  = \frac{\int d \xi e^{- N f(\xi) \frac{1}{1-\xi}}}{\int d {\xi}   e^{-N f(\xi)}}
\label{ee_5}
\eeq
 where
\beq
  f(\xi) = \frac{N_0 {\bar g}^2}{T} {\xi}^2 - \frac{\xi}{2}
\label{ee_7}
\eeq
We assume that the measure of the integration over $\xi$ is non-singular.
The linear in $\xi$ term in $ f(\xi)$ comes from integration over $\eta_{q'}$ with $q' \neq q$ (see Eq. \ref{feb12_4}).   Each integration over $q'$ yields $1/\sqrt{1- \xi}$, and the product of the integrals over all $q'$ yields
$1/(1- \xi)^{N/2} = e^{-(N/2) \log{(1-\xi)}} \approx  e^{(N/2) \xi}$.

 At small $T$, the  function $f(\xi)$ in (\ref{ee_7}) has a minimum at ${\xi} = T/(4 N_0 {\bar g}^2) \sim T/(4\omega_D \lambda_E)$. Then $I_T = 1/(1- T/(4\omega_D \lambda_E))$ and
  \beq
\langle\eta^2\rangle  \sim \frac{T \lambda_E}{T_p} \frac{1}{1- \frac{T}{4 \omega_D \lambda_E}}
\label{ee_4_1_a}
\eeq
 We see that the renormalizations coming from the low-energy states of the continuum spectrum of the condensation energy hold in powers of $T/\omega_D$. With these renormalizations, the fully dressed stiffness is
 \beq
 \rho_s (T) = \frac{T_p}{\lambda_E} \left(1 - \frac{T}{4 \omega_D \lambda_E}\right)
   \label{con_12}
   \eeq
  We  see from  that the value of $\rho_s (T)$ at  $T \to 0$ and $\omega_D \to 0$ depends on the order of limits.
   At $T=0$, $\rho_s (0) = T_p/\lambda_E$ is finite and exceeds $T_p$.  At $\omega_D \to 0$ the corrections to stiffness
   rapidly increase with $T$, and  $\rho_s (T)$ becomes comparable to  $T$ at
   $T \sim \omega_D \lambda_E$.  For the largest $\lambda_E \sim 1$, at which our theory is valid, this holds at $T \sim \omega_D$. It is tempting to associate this temperature with  the actual $T_c$ above which the system looses long-range phase coherence.

Further, there is an analogy between finite $\omega_D$ and finite $2-\gamma$, as the two have similar effect on the gap function
(see Paper IV).  Replacing $\omega_D$ by ${\bar g} (2-\gamma)$, we find that at $\omega_D =0$ and $\gamma <2$,  superconducting  $T_c \sim {\bar g} (2-\gamma)$.

Before concluding this Section, we elaborate on our assumption that the averaging over phase fluctuations should include  low-energy states from the continuum spectra of the condensation energy.
Consider the case $\gamma < 2$, when the spectrum is still discrete and the $n=0$ solution has the lowest condensation energy $E_{c,0}$.  The energies $E_{c,n\geq1}$ are close to
 $E_{c,0}$, yet the solutions with different $n$ are topologically distinct as $\Delta_n (\omega_m)$ has $n$ vortices.
 These other states
   contribute to the renormalization of the phase of $\Delta_0 (\omega_m)$ only if the tunneling amplitude between
  the states $n=0$ and $n >0$ is non-zero, which requires the barrier between $E_{c,0}$ and $E_{c,n}$ to be small. The height
  of the barrier depends on the path along which
   a state without a vortex transforms into a state  with a vortex
  at some small $\omega_m$.  A vortex can either come from $\omega_m = \infty$, in which case the barrier is high, or via a creation of a vortex-antivortex pair at $\omega_m =0$, in which case it is low.
  For a generic $\gamma <2$,  $\Delta_n (z)$ are regular at small $z$ in the complex plane, hence one should not expect an anti-vortex nearby. However, for $\gamma \to2$, our candidate $\Delta_\xi (z)$,
  Eq. (\ref{con_14}), possess  anti-vortices at small $z$ in the lower frequency half-plane.
   In this situation, it is natural to expect that ,
     the barriers between $E_{c,0}$ and  $E_{c,n}$ with $n >0$ are low, hence our reasoning is justified.

\section{Phase diagram of the $\gamma$ model}
\label{sec:gamma}

\subsection{$\gamma =2$, finite $\omega_D$}

\begin{figure}
  \includegraphics[width=10cm]{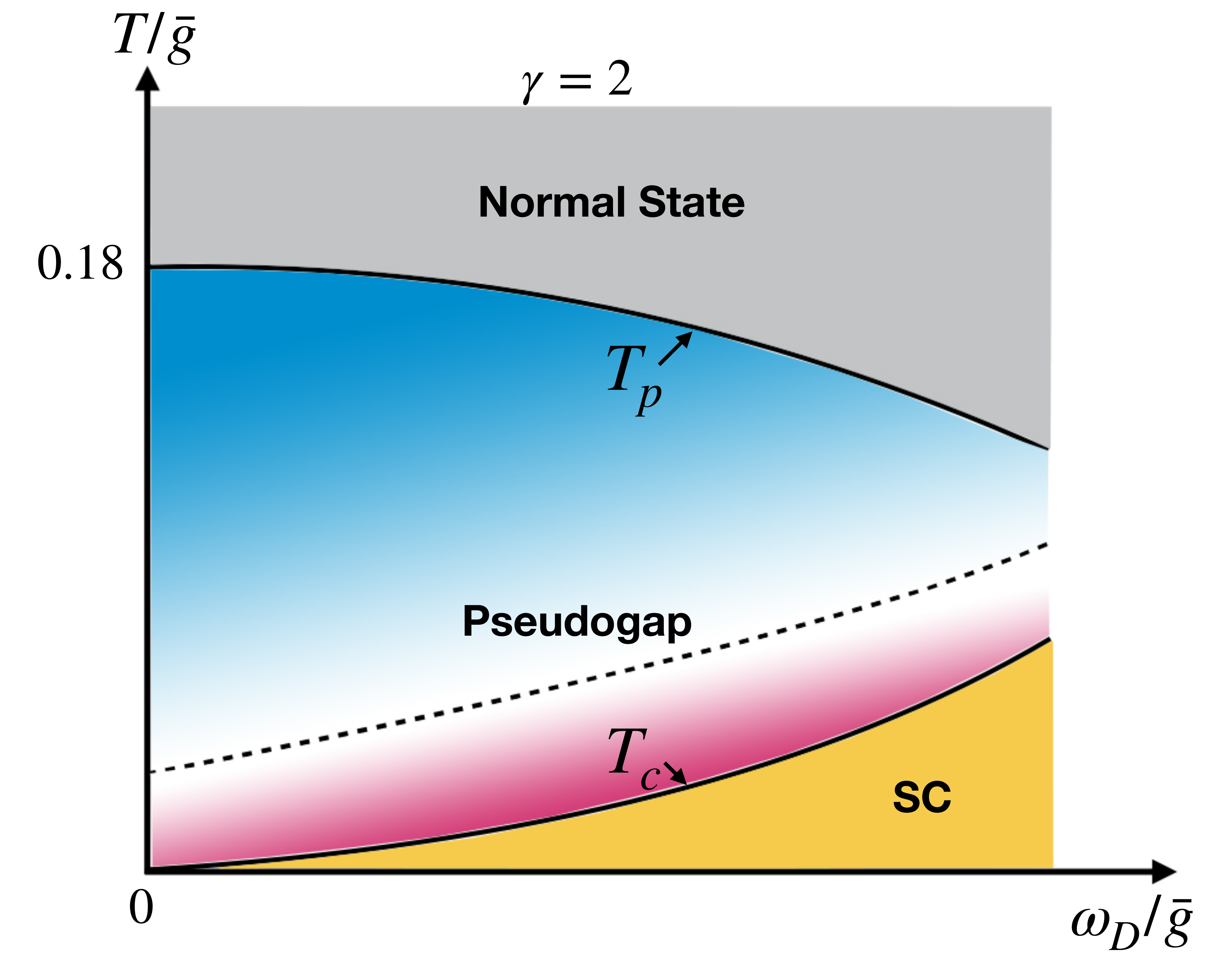}
  \caption{The phase diagram of the $\gamma$ model for $\gamma=2$ in variables $(T/{\bar g}, \omega_D/{\bar g})$, where $\omega_D$ is the mass of a pairing boson.  $T_p$ is the onset temperature of the pairing, and $T_c$ is the actual superconducting transition temperature, below which the system establishes phase coherence.  In between the system displays  pseudogap behavior, in which fermionic pairs are formed, but there is no macroscopic phase coherence.
    The dashed line
    separates the two regimes within the pseudogap phase -- the one at higher $T$, where the system behavior is chiefly
     determined by fermions with the two lowest Matsubara frequencies $\pm\pi T$, and
     the one at lower $T$,  when fermions with all Matsubara frequencies contribute to the pairing. In these two regimes the system displays gap filling and gap closing behavior, respectively.}\label{fig:phasediagram1}
\end{figure}

 In Fig. \ref{fig:phasediagram1}, we present the phase diagram for $\gamma =2$ for  non-zero $\omega_D$ and $T$.
  At $\omega_D=0$, the true transition temperature into a SC state is zero, although the onset temperature for the pairing, $T_p$ is finite. At finite $\omega_D$, $T_c$ is finite but much smaller than $T_p$, at least for small $\omega_D$. In between $T_c$ and $T_p$ the system displays pseudogap behavior:  the spectral function and the DoS display a peak at a finite frequency, but the spectral weight below the peak remains finite.  Close to $T_p$, the pairing is mainly induced by fermions with the two lowest Matsubara frequencies $\pm\pi T$\cite{Wu_19_1,Abanov_19,Wang2016}. In this situation, the position of the  peak in the spectral function and the DoS increases linearly with $T$, and the gap
   fills in as $T$ approaches $T_p$.  In the $T$ range near $T_c$, fermions with all Matsubara frequencies contribute to the pairing, and the positions of the maxima in the spectral function and the DoS move to smaller frequencies as $T$ increases (gap closing behavior). We show the DoS in the two  regimes in Fig.\ref{fig:gap_closing_filling} below.

\begin{figure}
  \includegraphics[width=10cm]{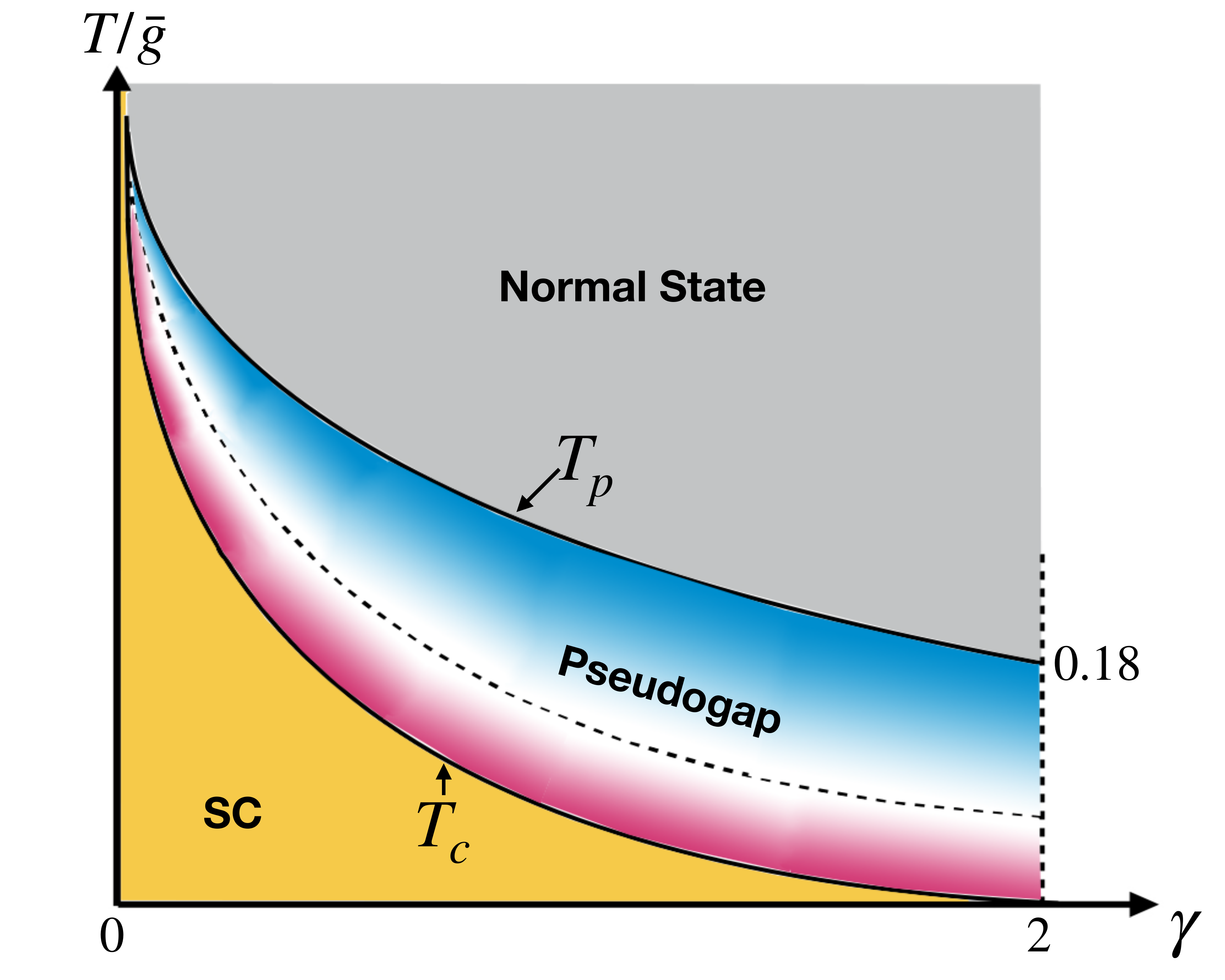}
  \caption{The phase diagram of the $\gamma$ model for a generic $\gamma<2$ at a finite $T$ and vanishing $\omega_D$. For any $\gamma<2$, the true SC transition temperature $T_c$ is finite, but is smaller than the onset temperature for the paring, $T_p$. In between $T_c$ and $T_p$,  the system displays a pseudogap behavior. There are two distinct behaviors in the pseudogap regime, like in Fig. \ref{fig:phasediagram1}: close to $T_p$, the spectral function and the DoS display gap filling behavior, while close to $T_c$, the behavior becomes more conventional and the gap frequency shifts to a smaller value as $T$ increases. }
  \label{fig:phasediagram2}
\end{figure}

\subsection{$\omega_D=0$, $0<\gamma \leq 2$}
In Fig. \ref{fig:phasediagram2} we show the phase diagram for the $\gamma$ model  with $0 < \gamma \leq 2$, at $\omega_D =0$ and finite $T$.  This phase diagram is based on the results of this work and previous  works (Papers I-IV).
 For $\gamma <2$, we found earlier the largest condensation energy is for sign-preserving solution of the gap equation ($n=0$ in our classification).  Still, for any $\gamma >0$, there exists an infinite set of topologically distinct solutions for the gap (all with the same symmetry), labeled by integer $n$. This generates a discrete spectrum of the condensation energy $E_{c,n}$.  The spectrum is sparse near the bottom at small $\gamma$, but becomes dense and flattens up at the bottom as $\gamma$ approaches $2$.
 At $\gamma \leq 2$, the corrections to superconducting order parameter  from the states with $n \neq 0$ are small at low $T$, but
 rapidly increase with increasing $T$ and destroy phase coherence at $T_c \sim {\bar g} (2-\gamma)/\lambda_E$.
  For $\gamma \leq 2$, $T_c \ll T_p$, and there exists a wide intermediate  temperature range  where the system displays a pseudogap behavior. By continuity, we expect that the pseudogap region to exist for all $\gamma >0$ albeit with a smaller width.

\subsection{Properties of the pseudogap phase}

\subsubsection{toy model for $\gamma =2$}

\begin{figure}
  \includegraphics[width=16cm]{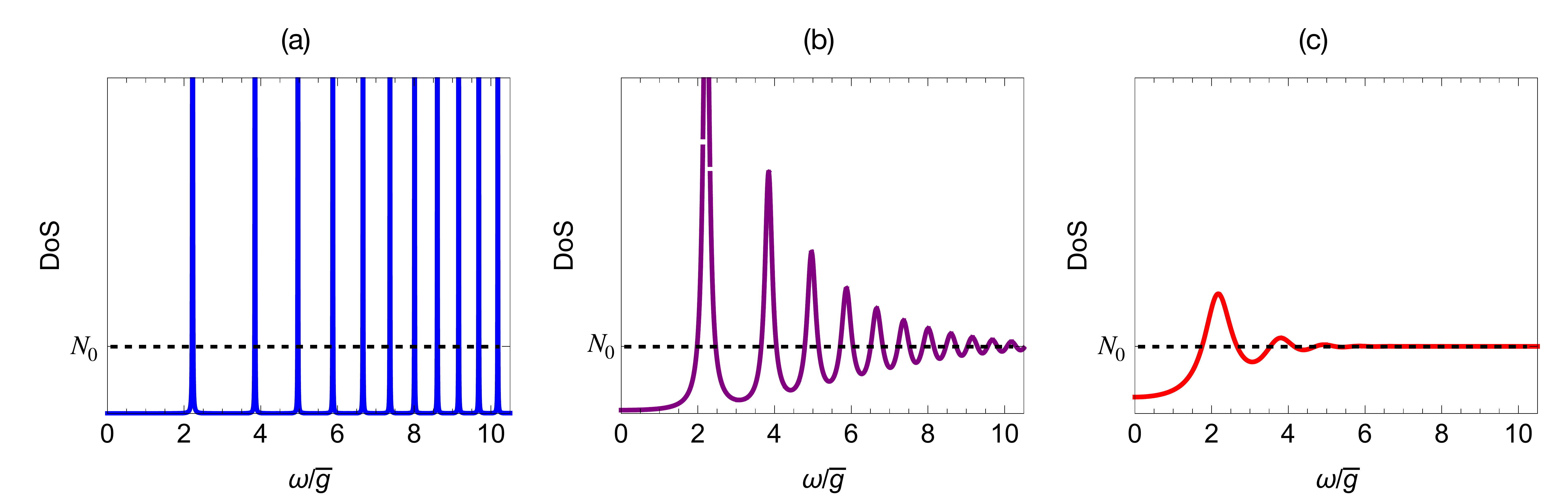}
  \caption{The density of states, $N(\omega)$, at different temperatures, for a toy model with $\Delta_0(\omega)=\omega/\sin(ia+\omega^2(1+ib))$, where $a$ and $b$ are two parameters, which increase with $T$. (a) The $T=0$  limit, $a=b=10^{-4}$. The DoS has a set of $\delta$-functional peaks. (b) A finite but small $T$, $a=b=0.05$. The first few peaks are well defined, but the peaks at large frequencies get overdamped and disappear. (c) A higher temperature, $a=b=0.25$. The peak at the smallest  frequency is still present,  at about the same frequency as at $T=0$, but
    other peaks are washed out, and the spectral weight below the peak increases, i.e., the DoS  at low frequencies fulls in.}\label{fig:dos}
\end{figure}

Let's start with $\gamma =2$. At $T=0$ the DoS is the set of $\delta-$functions, (Fig.\ref{fig:dos}(a))
At a finite $T$, two new features appear.  First,  $\Delta_0 (\omega)$
decreases  with  increasing $\omega$ and displays no oscillations above $\omega_{max}$,
similar to the case with finite $\omega_D$ discussed in \ref{sec:omega_D}.
 As a result, $\delta$-functional peaks in the DoS at larger frequencies get broadened
 and eventually disappear.
Second, other $D_\xi (\omega)$
from the continuum spectrum of condensation energies contribute to the DoS with Boltzmann factors.
  For all these solutions, Im$\Delta_\xi (\omega)$
  remains finite down to $\omega=0$.  As a result, the DoS also becomes non-zero at the smallest  $\omega$ (this phenomenon is often called a gapless superconductivity~\cite{maki,Maki_1973,maki4,maki3})
We model both effects by introducing a phenomenological $\Delta (\omega) = \omega/\sin(ia + (\omega/{\bar g})^2 (1+ib))$, where $a$ and $b$ increase with $T$.  We show the corresponding DoS in  Fig. \ref{fig:dos}(b) and (c).

\subsubsection{gap filling vs gap closing}

\begin{figure}
  \includegraphics[width=15cm]{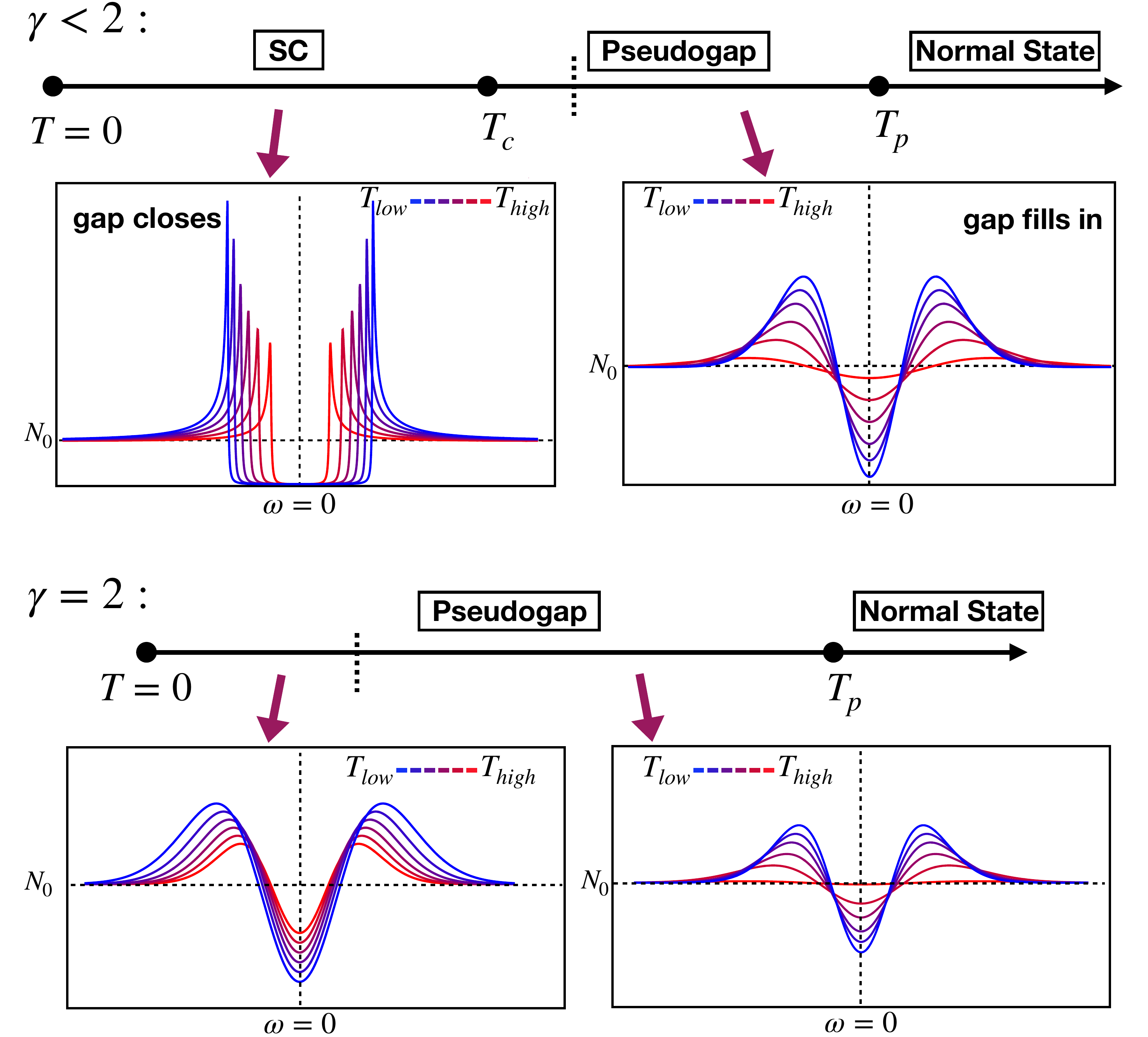}
  \caption{The temperature evolution of the DoS $N(\omega)$. For $\gamma<2$ (upper panel) there is a SC  order at
   $T<T_c$. In this regime and in the pseudogap state at $T \geq T_c$,  the temperature variation of $N(\omega)$  resembles that in a conventional BCS superconductor, i.e. when $T$ increases, the position of the maximum in $N(\omega)$ moves to a smaller frequency. At larger $T$ within the pseudogap phase,
 $N(\omega)$ displays gap filling behavior when   the peak position increases with increasing $T$ and $N(\omega=0)$ increases towards its normal state value.
   For $\gamma=2$ (lower panel), $T_c=0$, but the two different regimes of pseudogap behavior are present. }\label{fig:gap_closing_filling}
\end{figure}

We argue, based on earlier works\cite{Abanov_19,Wu_19_1}), that  there are two different regimes of system behavior within the pseudogap phase. At low $T$, the position of the peak in the DoS scales with $\Delta_0 (0)$ and decreases as $T$ increases (the gap ``closes''
 with increasing $T$). At  higher $T$, the peak in the DOS shifts to higher frequencies and the spectral weight below the peak increases (gap ``fills in" with increasing $T$).  We illustrate this in  Fig.\ref{fig:gap_closing_filling}.
  This last behavior is at least partly related to the fact that in some finite range of $T$ below $T_p$, the gap function on the Matsubara axis is strongly peaked at the first Matsubara frequencies $\pm \pi T$ (Refs. \cite{Abanov_19,Wu_19_1}). On the real axis the
  corresponding $\Delta (\omega)$ displays $\omega/T$ scaling.  For such $\Delta (\omega)$,
  the peak  frequency in the DoS increases linearly with $T$.

\begin{figure}
  \includegraphics[width=15cm]{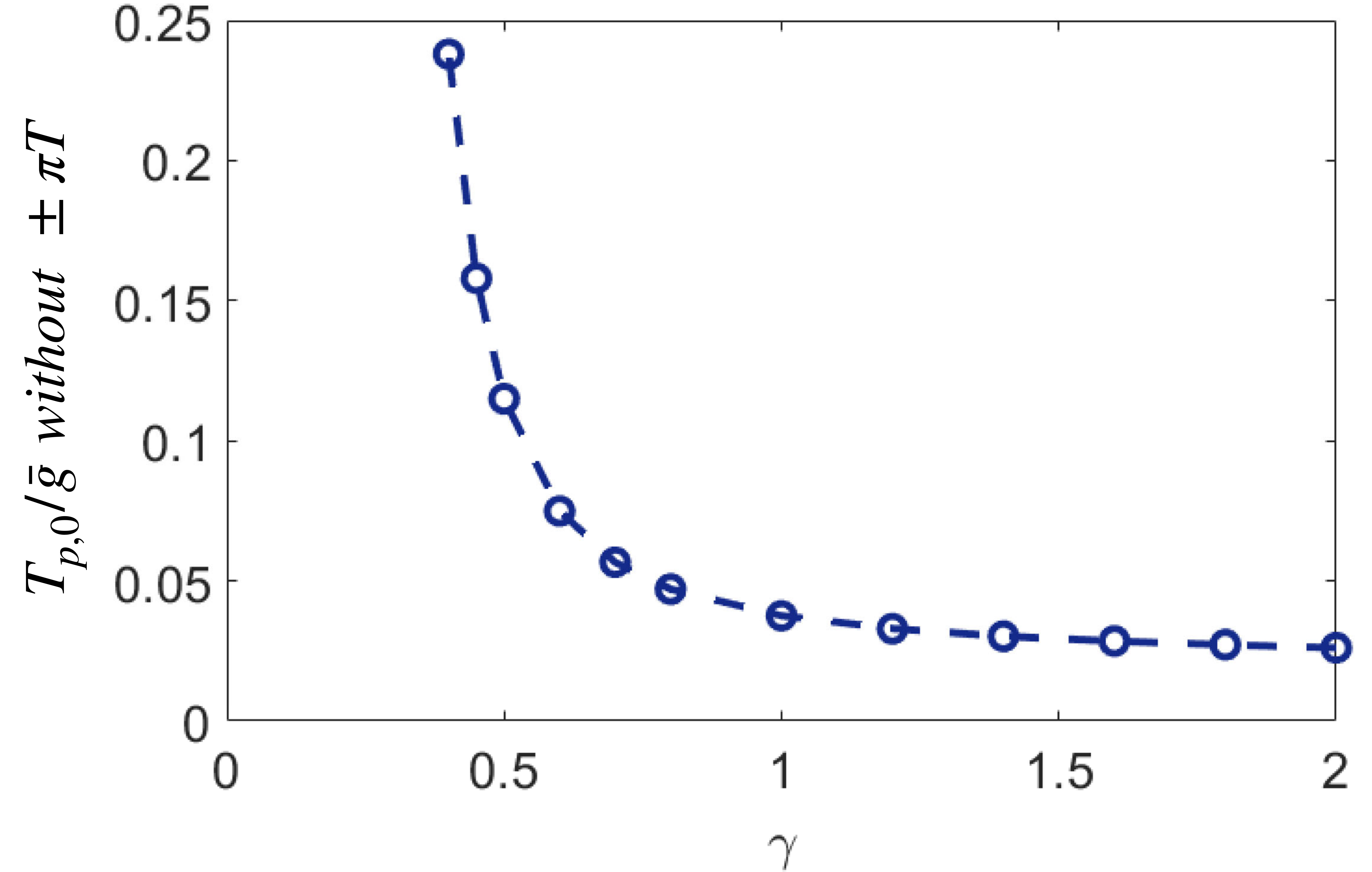}
  \caption{
The onset temperature of the pairing, obtained without including Matsubara frequencies $\omega_m = \pm \pi T$.  For $\gamma=2$, this temperature is roughly
$1/7$ of the actual $T_{p,0}$.}\label{fig:no_first_Matsubara}
\end{figure}

  At a finite $\omega_D$ and/or $2-\gamma$,  the ``gap filling'' behavior
    holds in some range between the onset temperature of the pairing  $T_p$ and a finite superconducting $T_c$ (Fig.\ref{fig:gap_closing_filling}).
  To estimate  the crossover  temperature between the two regimes, we compare the actual $T_p$ with
   the one obtained by neglecting the contributions from fermions with $\omega_m = \pm \pi T$.  We show the results in Fig. \ref{fig:no_first_Matsubara}.   We see that for $\gamma =2$ the onset temperature without $\pm \pi T$ fermions is strongly reduced -- it is  about  $1/7$ of the actual  $T_{p} \sim 0.18 {\bar g}$.  This implies that the ``gap filling'' behavior  holds in a wide range below $T_p$ and crosses over to ``gap closing'' behavior only near $T_c$.

\section{Conclusions}
\label{sec:conclusions}

In this paper, we  extended our earlier analysis of the $\gamma-$model to $\gamma =2$.  The $\gamma =2$ model describes, among other cases,  the pairing, mediated by an Einstein boson, in the limit
    when the bosonic mass $\omega_D$ tends to zero.  On the real axis, the effective interaction in this limit
    $V (\Omega) = - {\bar g}^2/\Omega^2$  is repulsive, and, at a first glance, should not give rise to pairing. However, the same interaction on the Matsubara axis, $V (\Omega_m) =  {\bar g}^2/\Omega^2_m$, is attractive, and earlier calculations on the  Matsubara axis  found that the onset temperature of the pairing,
    $T_p$,  tends to a finite value  $T_p = 0.1827{\bar g}$ at $\omega_D \to 0$ ($T_p = 0.1827 \omega_D \sqrt{\lambda}$ in terminology of Ref.\cite{ad},
    which is the same expression because $\lambda = {\bar g}^2/\omega^2_D$).  The issue we discussed in this paper is whether this $T_p$ is close to the actual superconducting $T_c$, or $T_c$ is smaller, and there is a range of  pseudogap behavior between $T_c$ and $T_p$.  We argued that  the actual $T_c$
    scales with $\omega_D$  and is much smaller than $T_p$ when
    $\omega_D/{\bar g}$ is small.

         To prove this, we solved the non-linear gap equation at $T=0$ and $\omega_D =0$  and found a continuum of solutions, governed by a single parameter $\xi$ ($0\leq \xi \leq \infty$).
         This in turn gives rise to a continuum spectrum of condensation energy, $E_{c, \xi}$, which can be viewed as a continuum gapless spectrum of ``longitudinal" gap fluctuations.   An infinite  set of the gap functions and the condensation energies exists already for $\gamma <2$, but is a discrete one.  For $\gamma =2$, this spectrum becomes continuous in a manner similar to how a discrete set of energy levels in a finite size crystal becomes a continuous vibration spectrum when system size becomes infinite.  In our case, $1/(2-\gamma)$ plays the role of a system size.

          Without the contribution from the gapless longitudinal branch, superfluid stiffness $\rho_s (T=0)$ is larger than $T_p$, and thermal
           corrections to  superconducting order parameter scale approximately as $T/\rho_s (0)$ and remain small  at all $T < T_p$. However,
            upon including contributions from the longitudinal branch, we found that thermal corrections become of order one already at much smaller $T\sim \omega_D$. We identified this temperature with the actual superconducting $T_c$.  We emphasize that $T_c$ vanishes at $\omega_D =0$, and the behavior of the stiffness depends on the order in which the double limit $\omega_D \to 0$ and $T \to 0$ is taken. This strongly suggests that the $\gamma =2$ model is critical at $T=0$.  At smaller $\gamma$, the ground state is not critical at $\omega_D =0$,   and  $T_c \sim {\bar g} (2-\gamma)$.  It is finite but at $\gamma \leq 2$ is still much smaller than $T_p \sim {\bar g}$.

            We presented collaborative evidence that the $\gamma =2$ model is critical, from the analysis of the continuum set of  gap functions along real frequency axis and in the upper half-plane of frequency.  We found that for each solution, there is an  infinite array of $2\pi$ vortices in the upper frequency half-plane. The array of vortices stretches up to an infinite frequency, where each gap function from the continuous set has an essential singularity. We speculated that different gap functions from the continuous set are different extensions from the array of vortices, onto the upper half-plane of frequency.

             At a finite $\omega_D$, the set of gap functions becomes discrete and contains only a finite number of solutions, all of which
               behave regularly in the high-frequency limit.    The
             the number of vortices also becomes finite. Still, at  small $\omega_D/{\bar g}$, the system behavior over a wide frequency range mimics that at $\omega_D =0$.

             We showed the phase diagram of the $\gamma =2$ model in variables $T$ and $\omega_D$ in Fig. \ref{fig:phasediagram1}
               and the phase diagram of the $\gamma$ model at $\omega_D =0$ in in variables $T$ and $\gamma$ in Fig. \ref{fig:phasediagram2}
              In both cases, there is range of pseudogap behavior between the onset temperature of the pairing $T_p$ and the actual $T_c$.
              In the pseudogap region, the bound pairs are formed, but there is no macroscopic phase coherence.
             We argued that in most of the pseudogap regime, the DoS and other observables display ``gap filling" behavior, in which the peak position remains at a finite frequency up to $T_p$,
              while  the states below the peak gradually fill in.

 In the next (last) paper in the series we consider the behavior of the $\gamma$ model for $\gamma >2$ and show that the new physics
  emerges at $T=0$, which gives rise to a reduction  and eventual vanishing of the superfluid stiffness in the ground state.

  \acknowledgements
  We thank   I. Aleiner, B. Altshuler, E. Berg, D. Chowdhury, L. Classen,  K. Efetov, R. Fernandes,  A. Finkelstein, E. Fradkin, A. Georges, S. Hartnol, S. Karchu, S. Kivelson, I. Klebanov, A. Klein, R. Laughlin, S-S. Lee, G. Lonzarich, I. Esterlis, D. Maslov, F. Marsiglio, I. Mazin, M. Metlitski, W. Metzner, A. Millis, D. Mozyrsky, C. Pepan, V. Pokrovsky,  N. Prokofiev,  S. Raghu,  S. Sachdev,  T. Senthil, D. Scalapino, Y. Schattner, J. Schmalian, D. Son, G. Tarnopolsky, A-M Tremblay, A. Tsvelik,  G. Torroba,  E. Yuzbashyan,  J. Zaanen,
   and particularly R. Combescot and  Y. Wang  for useful discussions.
   The work by  A.V.C. and Y.M.W. was supported by the NSF DMR-1834856.  Y.-M.W, S.-S.Z.,and A.V.C  acknowledge the hospitality of KITP at UCSB, where part of the work has been conducted. The research at KITP is supported by the National Science Foundation under Grant No. NSF PHY-1748958.
   A.V.C. also acknowledges the hospitality of  Stanford University, where some results of this work have been obtained. His stay at Stanford has been supported through the Gordon and Betty Moore Foundation's EPiQS Initiative, Grants GBMF4302 and GBMF8686.

\appendix

\section{Expansion in $D^2 (\omega_m)$ for $\gamma =2$ and $\gamma <2$}
\label{app:no_omega}

In this Appendix  we present some details of the analysis of the non-linear gap equation for $\gamma =2$ and elaborate on the claim in the main text that a continuous set of gap functions
 exists only for $\gamma =2$, while for smaller $\gamma$, the set is a discrete one.

\subsection{$\gamma =2$}

We begin with $\gamma =2$.
 Consider first the  limit of small
   frequencies $\omega_m \ll {\bar g}$.  For such $\omega_m$,  $\Delta (\omega_m)$ in the l.h.s. of the gap equation (\ref{ss_11}) can be neglected,
    as its inclusion leads to terms with extra  $(\omega_m/{\bar g})^2$.  This approximation is equivalent to  neglecting $\omega_m$ compared to the self-energy $\Sigma (\omega_m)$ and is
      similar to the ``no $\omega_m$''  approximation, used in the studies of SYK-type models~\cite{gu-2020,patel-sachdev-2019,Wang_2020}.
    The non-linear gap equation at $T=0$ without $\Delta (\omega_m)$ in the l.h.s reduces to
    \beq
    \int d \omega_{m'} \frac{D (\omega_{m'}) - D (\omega_m)}{\sqrt{1 +D^2 (\omega_{m'})}}
    ~\frac{\sgn\omega'_m}{|\omega_m - \omega_{m'}|^2} =0
     \label{ap_1}
  \eeq
  where, we recall,  $D(\omega_m) = \Delta (\omega_m)/\omega_m$.

 The  linearized gap equation is obtained from (\ref{ap_1}) by neglecting $D^2 (\omega_{m'})$ in the denominator. The exact solution of the
 linearized gap equation is Eq. (\ref{5_1}):
\beq
 D (\omega_m) = 2\epsilon \cos{\left(\beta \log{\left(\frac{|\omega_m|}{{\bar g}}\right)^2} + \phi_\epsilon\right)} ~\sgn\omega_m
 \label{ap_2}
 \eeq
  where
  $\epsilon$ is an arbitrary  overall factor, $\phi_\epsilon$ is yet undetermined constant, and  $\beta = 0.38187$ satisfies $\pi \beta \tanh(\pi \beta) =1$.

We now expand Eq.  (\ref{ap_1}) in powers of $D^2$.
We will be searching for the solution in the form
\beq
 D_\epsilon (\omega_m) = 2 \sum_{n=0}^\infty \epsilon^{2n+1} Q_{2n+1} \cos{\left( (2n+1) \left(\beta_\epsilon \log{\left(\frac{|\omega_m|}{{\bar g}}\right)^2}\right) + \phi_\epsilon\right)}
 \label{ap_3}
 \eeq
 Substituting into (\ref{ap_1}) and collecting contributions at each order in $\epsilon^{2n+1}$,
  we find that $D_\epsilon(\omega_m)$ given by  (\ref{ap_3}) does satisfy Eq. (\ref{ap_1}),
    and that all integrals are ultra-violet convergent, i.e., there is no need for regularization.  The calculations are lengthy, but straightforward. We checked explicitly that $\beta_\epsilon$ is the same in all terms in (\ref{ap_3}) and is related to the original $\beta$ by
 \beq
 \beta_\epsilon =\beta \left(1 -\frac{\epsilon^2}{2} + 0.806 \epsilon^4  + ...\right)
  \label{ap_4}
  \eeq
The numerical coefficients are $Q_3 = 0.222 + O(\epsilon^2)$, $Q_5 = 0.043 + O(\epsilon^2)$. We cited this result  and Eqs. (\ref{ap_3}) and (\ref{ap_4}) in
 Sec. \ref{sec:Mats_expansion}.

  At larger frequencies,  we need to keep $\Delta (\omega_m)$  in the l.h.s. of (\ref{ss_11}).  In the opposite limit $\omega_m \gg {\bar g}$, the leading term in $D_\epsilon (\omega_m)$ is obtained by pulling $1/\omega_m^2$ from the integrand in the r.h.s..  Then we obtain
  \beq
  D_\epsilon(\omega_m) = \frac{a_\epsilon}{\omega^3_m}
  \eeq
   where
   \beq
   a_\epsilon =
    \frac{{\bar g}^2}{2} \int d \omega_{m'} \frac{D (\omega_{m'})}{\sqrt{1 +D^2 (\omega_{m'})}}
    ~\sgn \omega_m
     \label{ap_1_a}
  \eeq
 Substituting this form of $D(\omega_m)$ into the integrand in the r.h.s, we find that the integral is ultra-violet convergent, i.e., the solution (\ref{ap_1_a}) is self-consistent.

The two solutions have to merge at $\omega_m \sim {\bar g}$.
For the linearized gap equation (the limit $\epsilon \to 0$) we verified that this does happen for a certain value of $\phi_\epsilon$ in
(\ref{ap_3}). We conjecture that the same holds for other $\epsilon$, i.e.,  for a certain $\phi_\epsilon$,  $D_\epsilon (\omega_m)$ smoothly evolves between (\ref{ap_3}) and (\ref{ap_1_a}).  We did similar analysis in Paper I. There, we demonstrated that for arbitrary $\phi_\epsilon$, $D_\epsilon(\omega)$ of Eq. (\ref{ap_3}) approaches the constant at $\omega_m \to \infty$, while the desired term
($D_\epsilon (\omega_m) \propto 1/|\omega_m|^{\gamma +1}$ for a generic $\gamma$) is the subleading one.   For a particular $\phi_\epsilon$, a constant vanishes, and the high-frequency behavior becomes the expected one.

  We see from (\ref{ap_4}) that $\beta_\epsilon$ decreases with increasing $\epsilon$, while the overall magnitude of $\Delta (\omega_m)$ increases.  It is natural to expect that $\beta_\epsilon =0$ at  some critical $\epsilon= \epsilon_{cr}$. We explored this in the main text.

Another way to argue for the existence of $\epsilon_{cr}$ is to
 depart from the opposite limit $\epsilon \gg 1$, where $D(\omega_m)$ is supposed to be large.  In this case, we introduce $\Xi (\omega_m) =1/D(\omega_m)$ and re-express the gap equation as
    \beq
    \int d \omega_{m'} \frac{\Xi (\omega_{m'}) - \Xi (\omega_m)}{\sqrt{1 +\Xi^2 (\omega_{m'})}}
    ~\frac{1}{|\omega_m - \omega_{m'}|^2} =0
     \label{ap_5}
  \eeq
Note the absence of $\sgn \omega'_m$ in the integrand.
 At small $\Xi$, we neglect the $\Xi^2 (\omega_{m'})$ in the denominator and search for the solution in the form $\Xi (\omega_m) =  \mbox{sign} \omega_m |\omega_m/{\bar g}|^b$. Substituting into (\ref{ap_5}) we find
   $b = \pm 1$, i.e.,
 \beq
 \Xi (\omega_m) = \left(A_1 \frac{{\bar g}}{|\omega_m|} + A_2 \frac{|\omega_m|}{{\bar g}}\right)
\label{ap_6}
\eeq
The first term does not satisfy the normalization condition and has to be discarded (see Paper I for the details on this). This leaves  no parameter to adjust in order to match with the behavior at high frequencies.  This implies that there is no solution for the gap at large $\epsilon$.

\begin{figure}
  \includegraphics[width=15cm]{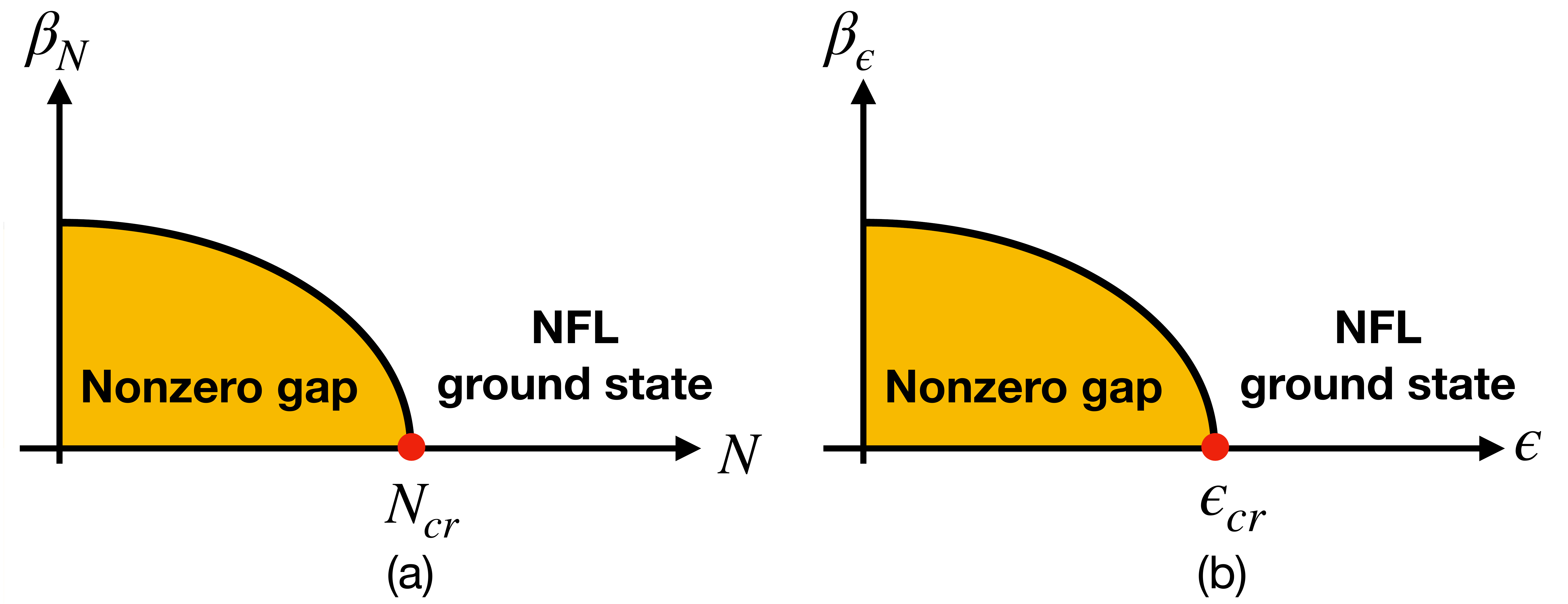}
  \caption{The comparison between the behavior of $\beta_\epsilon$ in the  $\gamma =2$ model and $\beta_N$ in the model with $\gamma <1$,  extended to $N >1$. }\label{fig:Ncr_Cmax}
\end{figure}

There is a similarity between this analysis and the analysis in Paper I, where we considered the $\gamma$ model for $\gamma <1$ and extended it using a continuous variable $N$ to make interactions in the particle-hole and particle-particle channels non-equivalent.
There, we  found that there exists $N_{cr}$, which separates oscillating and non-oscillating solutions, and only
 oscillating solutions are compatible with high-frequency behavior.  Here, $\epsilon$ plays the same role as $N$.
We illustrate this in Fig. \ref{fig:Ncr_Cmax}

\subsection{$\gamma <2$}

We now extend this approach to $\gamma <2$.  The gap equation for $D(\omega_m)$ at $\omega_m \ll {\bar g}$  has the same form as in (\ref{ap_1}), only $|\omega_m - \omega_{m'}|^2$ in the denominator is replaced by $|\omega_m - \omega_{m'}|^\gamma$.  The solution of the linearized equation for $D(\omega_m)$ is
\beq
 D (\omega_m) = 2\epsilon \left(\frac{{\bar g}}{|\omega_m|}\right)^{1-\gamma/2}  \cos{\left(\beta_\gamma \log{\left(\frac{|\omega_m|}{{\bar g}}\right)^\gamma} + \phi\right)} ~\mbox{sign}\omega
 \label{ap_7}
 \eeq
 where  $\beta_\gamma$ is some regular function of $\gamma$.
 As before, we search for the solutions in the form
 \beq
 D_\epsilon (\omega_m) = 2 \sum_{n=1}^\infty \left(\epsilon \left(\frac{{\bar g}}{|\omega_m|}\right)^{1-\gamma/2}\right)^{2n+1} Q_{2n+1}  \cos\left[(2n+1) \left(\beta_{\gamma,\epsilon} \log{\left(\frac{|\omega_m|}{{\bar g}}\right)^\gamma} + \phi_\epsilon\right)\right]
 \label{ap_8}
 \eeq
 Substituting into (\ref{ap_7}), we find that the integrals that determine $Q_{2n+1}$
 now contain  infra-red divergencies. The only way to eliminate the divergencies is to assume that $\Delta_\epsilon$ tends to a finite
  value at $\omega_m \to 0$.  But this is only possible for a discrete set of finite $\epsilon$.
    We also note in passing that because the
 actual expansion parameter is  $\epsilon ({\bar g}/|\omega_m|)^{1-\gamma/2}$, the expansion of $\beta_{\gamma, \epsilon}$  in powers of $\epsilon$ yields
 $\beta_{\gamma,\epsilon}  = \beta_\gamma(1 + a (\epsilon({\bar g}/|\omega_m|)^{1-\gamma/2})^2)$.  For $a \neq 0$, this gives rise to additional terms, which are not matched by the terms in Eq. (\ref{ap_8}).  The only option then is to set $a=0$, i.e., leave $\beta_{\gamma,\epsilon}$ equal to its bare value $\beta_\gamma$.

\begin{figure}
  \includegraphics[width=13cm]{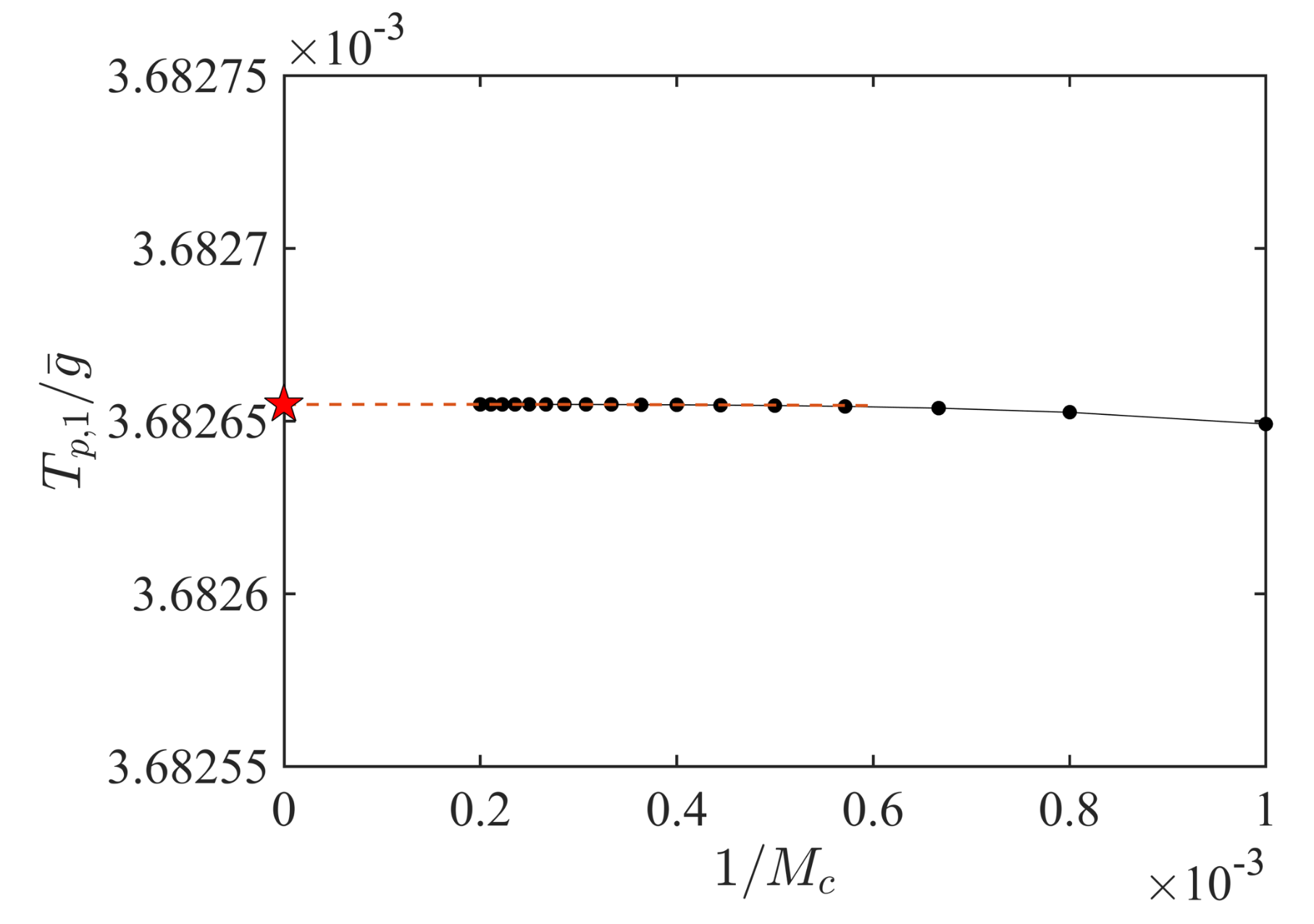}
  \caption{Determination of the temperature $T_{p,1}$, at which the $n=1$ solution develops in the $\gamma =2$ model. $M_c \sim 10^3$  is the largest value of the Matsubara number, used in this numerical calculation. Extrapolating $M_c$ to $\infty$ yields a finite value $T_{p,1}\simeq3.6827\times10^{-3}{\bar g}$.}\label{fig:Tp1}
\end{figure}

 The outcome is that the continuous set of gap functions exists only for $\gamma =2$. For smaller $\gamma$, this set is discrete.
 We also emphasize that the distinction between $\gamma =2$ and $\gamma <2$ holds only at $T=0$. At any finite $T$, the set of gap functions is a discrete one for all $\gamma \leq 2$, and the solutions with different
  $n$ from the set vanish at different temperatures $T_{p,n}$.  In  Fig.~\ref{fig:Tp1} we show the results of high-accuracy numerical calculation of $T_{p,1}$ for $\gamma =2$.  We see that  $T_{p,1}\simeq 3.6827\times10^{-3}{\bar g}$ is finite.

\section{The gap function $\Delta_\infty$}
\label{sec:app_exact}

The exact solution of the linearized gap equation at zero temperature has been derived for $0<\gamma<1$ in Paper I and $1<\gamma<2$ in Paper IV. Here we extend the analysis of Paper IV to $\gamma =2$.

  We first solve for the gap function $ \Delta_{\infty}(\omega_m)$  along the Matsubara axis.
   Following the same computational steps as in the analysis for $\gamma <2$, we
    obtain $D_{\infty} (\omega_m) = \Delta_{\infty}(\omega_m) /\omega_m$ in the form
 \beq
 D_\infty (\omega_m) = \epsilon
  \frac{{\bar g}}{\omega_m} \int_{-\infty}^\infty dk b_k e^{-ik \log{(\omega_m/{\bar g})^2}},
 \label{nn_2_app}
 \eeq
where  $\epsilon$ is an infinitesimal number,
  \beq
  b_k = \frac{e^{-i I_k }}{\left[\cosh(\pi (k-\beta))\cosh(\pi (k+\beta))\right]^{1/2}},
  \label{nn_2_1_a}
  \eeq
  and 
  \beq
  I_k = \frac{1}{2} \int_{-\infty}^\infty dk' \log{|\epsilon_{k'} -1|} \tanh{\pi (k'-k + i \delta)},
  \label{nn_2_2_a}
  \eeq
 Here $\epsilon_{k'} = \pi k' \tanh (\pi k')$ and $\beta\simeq 0.38187$ is the solution of $\pi \beta \tanh{(\pi \beta)}=1$.
    We cited these results in Eqs.  (\ref{nn_2}), (\ref{nn_2_1}) and (\ref{nn_2_2}) in the main text.

The integrals (\ref{nn_2_2_a}) and (\ref{nn_2_app}) can be computed numerically.  We showed the result for
$D_\infty (\omega_m)$ in Fig.\ref{fig:Dinfty} in the main text.  The function $D_\infty (\omega_m)$ oscillates at
 $\omega_m < {\bar g}$ and decays as $1/|\omega_m|^3$ at $\omega_m > {\bar g}$.

\subsection{Series expansion}

The integral in Eq.~(\ref{nn_2_2_a})  can be evaluated by closing the integration contour along an infinite arc in the complex  plane of frequency.  For $\rvert \omega_m \rvert < {\bar g}$, the arc must be in the upper half-plane, and for $|\omega_m| > {\bar g}$, in the lower half-plane.
 The integral is equal to the sum of the contributions from each pole of the function $b_k$ in the upper or lower half-plane. The position of these poles are obtained from the representation of $b_k$ as an infinite product of the Gamma-functions: (see Papers I and IV for details on this)
 \beq
 b_k = \frac{\Gamma(1-ik)}{\Gamma(1+ik)} \Gamma\left(\frac{1}{2} + i (k+ \beta)\right)\Gamma\left(\frac{1}{2} + i (k- \beta)\right) \prod_{m=1}^{\infty} \frac{\Gamma\left(\frac{1}{2} + i (k -i \beta_m)\right) \Gamma\left(1+ m - ik\right)}{\Gamma\left(\frac{1}{2} - i (k +i \beta_m)\right) \Gamma\left(1+ m + ik\right)}
\label{nn_4}
 \eeq
 Here $\beta_m >0$ are the solutions of $\pi \beta_m \tan ({\pi\beta_m}) =-1$.  There is  an infinite set of such $\beta_m$, specified by an integer $m =0,1,2..$. Each $\beta_m$  is located  within the interval  $1/2 + m <\beta_m < 3/2 + m$.
 Viewed as a function of complex $k$, $b_k$ has poles from individual $\Gamma$-functions in the upper  half-plane at $k = \pm \beta + i (n+1/2)$ and
 $k = i \beta_m + i(n+1/2)$ where $n,m=0,1,2,…$, and in the lower half-plane, at $k = -i(n+1)$ and $k = -i(1+ m +n)$,
 where $n=0,1,2,...$ and $m=1,2,...$.

 \subsection{$\rvert \omega_m \rvert < {\bar g}$}

  For $\rvert \omega_m \rvert < {\bar g}$, the relevant poles are at $k = \pm \beta + i (n+1/2)$ and at $k = i \beta_m + i(n+1/2)$, $n=0,1,2,...$.
  This yields series expansion for $D_\infty (y)$ with $y=(\rvert \omega_m \rvert /{\bar g})^{2}$ in the form
 \beq
   D_\infty (y) = {\text Re}~ \sum_{n=0}^{\infty } e^{(i \beta \log{y} + \phi)} C^{<}_n~ y^{n}
+  \sum_{n,m=0}^{\infty }D^{<}_{n,m} y^{(n + \beta_m)}
\label{app_9}
\eeq
 The leading term in (\ref{app_9}) at small $\omega_m$ comes from the contribution of the poles at $k=\pm \beta + i/2$
 \beq
   D_\infty (y) = C^{<}_0 \cos{\left(\beta \log{y} + \phi\right)}
\label{app_10}
\eeq
The first subleading term comes from the contribution of the pole at $k=i (1/2+\beta_0)$ and scales as $y^{ \beta_0}$, where $\beta_0 \simeq 0.89$.

 In the direct perturbation expansion in $y$, the series in $y^n$ (the first term in (\ref{app_9}))
  come from fermions with internal $y' \sim y$ and form the ``local'' series.  The second term in (\ref{app_9})) is the sum of  contributions from fermions with  $y' = O(1)$, which for $y \ll 1$ can we regarded as  ``non-local''. The total $D_\infty (y) = D_{\infty,L} (y) + D_{\infty, NL} (y)$.

 The coefficients $C^{<}_n$ in (\ref{app_9})
  can be obtained analytically, as we already found in Papers I and III for $\gamma \leq 1$ and Paper IV for $1<\gamma <2$. At $\gamma=2$, the result takes a very simple form
 \beq
  C^{<}_n = C^{<}_0 {i^n \over \beta^n n! }
  \label{dd_10a}
  \eeq
 Substituting this into Eq.~(\ref{app_9}),
  we find that the first term (the local contribution) becomes
  \beq
\label{eq:x<1_3_1}
D_{\infty,L} (y) \propto
 \cos{\left[\beta( \log {y} - y) + \phi \right]}.
  \eeq
  It oscillates with the periodicity set by $\beta \log y$ for $y  \ll 1$, i.e., $\rvert \omega_m \rvert \ll {\bar g}$,
  which is the right behavior of the gap function at small frequencies, see (\ref{app_10}).

    We note, however, that the first
   subleading term in (\ref{eq:x<1_3_1}) scales as $y~\sin({\beta \log {y}})$.
   This contribution is smaller than the
    actual subleading term, which scales as  $y^{0.89}$ and does not oscillate.  This implies that, besides the leading term, the form of $\Delta_{\infty} (y)$ is determined by non-local  corrections.

 \subsection{$\rvert \omega_m \rvert > {\bar g}$ and logarithmic correction}

 For $y>1$, i.e.,  $\rvert \omega_m \rvert > {\bar g}$, relevant poles are in the lower half-plane. According to Eq.~(\ref{nn_4}), a pole at $k = -i (n+1)$ ($n=0,1,2,...$) is of order $n+1$, namely
 a simple pole at $n=0$, a double pole at $n=1$, etc.  The leading  term in the limit of $\rvert \omega_m \rvert  \to \infty$ is the contribution from a simple pole at $k=-i$  ($n=0$), This contribution accounts for
  $1/y$ behavior of $\Delta_\infty (y)$ at large $y$.   However, the subleading terms from the rest poles
   contain extra logarithms on top of powers of $1/y$:
\begin{equation}
 \Delta_\infty (y)= \sum_{n=0}^\infty {\tilde C}^{>}_n~ y^{-2(1+n)} \left(\log{y}\right)^n,
 \label{app_11}
\end{equation}

To demonstrate the presence of the logs,
 consider as an example the contribution from the double pole at $k=-2 i$. We shift $\gamma$ to
  $2-\delta$, $\delta >0$  and then take  the limit  $\delta \to 0$. The expression of $b_k$ for $\gamma \leq 2$ is presented in Paper IV. Using it, we find that a double pole splits into two simple poles at $z_1 = -2 i$ and $z_2 = -(2+\delta/2)i$.  In the neighborhood of the two poles, the  function $b_k$  takes the form $\sim 1/(z-z_1)/(z-z_2)$. The contribution from the these two poles is obtained by  circling out  a loop ${\cal C}$ enclosing $z_1$ and $z_2$. Evaluating the integral and taking the limit $\delta \to 0$, we obtain
\begin{align}
\sqrt{y} \lim_{\delta \to 0^+}\frac{{\bar g}}{\omega_m}  \oint_{{\cal C}} d z { 1 \over (z+2 i) (z+(2+\delta/2)i)}
e^{-i z \log(y^{1-\delta/2})} = 2 \pi \frac{\log{y}}{y^{2}}.
\end{align}
Similarly, the triple pole at $k=-3i$ gives rise to $(\log{y})^2/y^3$, etc.
 Collecting the contributions from every pole on the lower-half plane, we obtain (\ref{app_11}).

\subsection{The universal oscillating term at large $y$}

We now show that the high-frequency form of $D_\infty (\omega_m)$ contains an additional oscillating  contribution. This contribution is exponentially small on the Matsubara axis, but, as we will see, it becomes the dominant one on the real axis.  To extract this contribution, we note that for large $\rvert\omega_m\rvert/{\bar g}$, the
 argument of the cosine function, $I_k + k \log{y}$, passes through extremum at $k\sim k_* = y/\pi$.
 Expanding around this point and evaluating the Gaussian integral, we
  obtain the universal piece $D_{\infty;u} (y)$ in the form
\beq
 D_{\infty;u} (y) = 2\sqrt{2}  \epsilon e^{-y }  \cos{\left[ \frac{(\pi^2-2)}{2\pi} y
   +\frac{\pi}{4}\right]}.
\label{app_11_a}
 \eeq
We see that $D_{\infty;u} (y)$ is exponentially small, yet this oscillating term is present.
The total $ D_{\infty} (y)$ is the sum of (\ref{app_11}) and (\ref{app_11_a}).

\subsection{$D_\infty (y)$ along real axis}

Let's now  transform from Matsubara to real axis.  We use $\omega$ instead of $y$ for better transparency.
 By construction, the gap function along the real axis is obtained by replacing $i\omega_m \to \omega+i0^+$ in the integrand in the r.h.s. of Eq. (\ref{nn_2_app}).
 Under this transformation, $\log (|\omega_m|/{\bar g})^{2}$ transforms into $ \log (\rvert \omega \rvert/{\bar g})^{2} - i \pi$. The integral in Eq.~(\ref{nn_2_app}) splits into two parts:
 \beq
 D_\infty ( \omega ) = \epsilon
  \frac{{\bar g}}{\omega}  \int_{0}^\infty dk \frac{e^{-\pi k} e^{-i I_k - i k \log (\rvert \omega\rvert/{\bar g})^{2} } + e^{\pi k} e^{i I_k + i k \log (\rvert \omega\rvert/{\bar g})^{2}} }{ \sqrt{ \cosh(\pi (k-\beta ))\cosh(\pi (k+\beta)) } } .
\label{nn_3_b}
 \eeq
Evaluating each integral by expanding near the point where $I_k \pm k \log (\rvert \omega\rvert/{\bar g})^{2}$ passes through extremum and approximating the denominator in (\ref{nn_3_b}) by its form at large $k$, we find
 that the first term is  small in $e^{-2\pi k}$, while in the second term the exponential factor cancels out.
 Ignoring the first term, we obtain
  \beq
 D_{\infty;u}(\omega) \approx \sqrt{2} \epsilon  e^{{i \over \pi} \left[ \left({\omega \over {\bar g}}  \right)^2 + \log \left({\omega \over {\bar g}}  \right)^2  \right]}.
\eeq
Other contributions to $D_{\infty}(\omega)$ contain powers of ${\bar g}/\omega$ and are smaller. As a result, on the real axis,  $D_{\infty}(\omega) \approx D_{\infty;u}(\omega)$ at $\omega \gg {\bar g}$.

The same calculation can be carried out for an arbitrary complex frequency $z = \omega' + i \omega^{''}$  in the upper frequency plane. For this, one has to replace $i\omega_m$ by  $z \equiv \rvert z \rvert e^{i \psi}$ ($0<\psi<\pi$) in the integrand in the r.h.s. of (\ref{nn_2_app}).
This  changes  $\log (|\omega_m|/{\bar g})^{2}$  to $ \log (\rvert z \rvert/{\bar g})^{2} + i (2 \psi - \pi) $ and gives
$\Delta_\infty (z)$, which we presented in  Eq.~(\ref{delta_z}) in the main text.

 \section{Extended $\gamma-$ model}
\label{sec:AppA}

In Papers I-III and other  works~\cite{raghu_15,Wang2016,Wang_H_17,Wang_H_18,Abanov_19,Wu_19_1,Torroba_19,Chubukov_2020a,Torroba_20},
we and others extended  the $\gamma$ model to in-equal interactions
 in the particle-particle and particle-hole channel.
  This was done by adding a factor $1/N$ to the interaction in the particle-particle channel and leaving the interaction in the particle-hole channel intact.  The advantage of extending the model to $N \neq 1$ is that
  superconducting order in the ground state exists for $N < N_{cr}$, while for larger $N$ the ground state is a non-Fermi liquid.  By analyzing the gap equation near this point, one can  obtain useful information about how a  discrete set of  solutions
 emerges.  In Paper III we argued that the extension to $N \neq 1$ makes sense for $\gamma <1$, while
  for $\gamma \geq 1 $, the model with $N \neq 1$ possesses singularities, not present in the original $\gamma$ model.  We proposed another way to extend the model with $\gamma >1$,  which is free from singularities. The idea is to first explicitly cancel out singularities in the original $\gamma$ model with $\gamma >1$, and only  then extend the model to $M \neq 1$ by making interactions in the particle-particle and particle-hole channel in-equivalent.
   The extended model is then free from singularities, and one can obtain critical $M_{cr}$, where superconducting order disappears at $T=0$ (by our construction, it exists at $M > M_{cr}$).
  In this Appendix, we analyze the extended $\gamma$ model for $\gamma =2$.   We show that a continuous set of solutions for the gap equation emerges at $M_{cr} +0$.

We first briefly describe the extension procedure.
The two coupled Eliashberg equations are for the pairing vertex $\Phi (\omega_m)$ and the
 self-energy $\Sigma (\omega_m)$. For $\gamma =2$, the equations are
  \bea \label{eq:gapeq_aa}
    &&\Phi (\omega_m)  =
  {\bar g}^2 \pi T \sum_{m' \neq m} \frac{\Phi (\omega_{m'})}{\sqrt{{\tilde \Sigma}^2 (\omega_{m'}) +\Phi^2 (\omega_{m'})}}
    ~\frac{1}{|\omega_m - \omega_{m'}|^2}, \nonumber \\
    && {\tilde \Sigma} (\omega_m)  =\omega_m +
    {\bar g}^2 \pi T \sum_{m' \neq m} \frac{{\tilde \Sigma}(\omega_{m'})}{\sqrt{{\tilde \Sigma}^2 (\omega_{m'})  +\Phi^2 (\omega_{m'})}}
    ~\frac{1}{|\omega_m - \omega_{m'}|^2}
\eea
where ${\tilde \Sigma} (\omega_m) = \omega_m + \Sigma (\omega_m)$.
At $T=0$, the r.h.s. of each of the two equations contains a divergent integral $\int dx/x^2$. To regularize the divergencies, we keep the temperature small but finite and set $T=0$ at the end of calculations.
At a finite $T$, the sum over  $m'$ is non-singular as singular  self-action term with $m'=m$   cancels out by the same reason as the contributions from non-magnetic impurities.

We then introduce
\bea
{\bar \Phi} (\omega_m) &=& \Phi (\omega_m) \left(1- {\bar g}^2 \frac{\zeta (2)}{(2\pi T)^{}}\frac{1}{\sqrt{\tilde{\Sigma}^2(\omega_m)+\Phi^2(\omega_m)}}\right) \nonumber \\
{\bar {\tilde \Sigma}} (\omega_m) &=& {\tilde \Sigma} (\omega_m) \left(1- {\bar g}^2 \frac{\zeta (2)}{(2\pi T)^{}}\frac{1}{\sqrt{\tilde{\Sigma}^2(\omega_m)+\Phi^2(\omega_m)}}\right)
\label{3_9}
\eea
where $\zeta (2) = \pi^2/6 = \sum_{n=1}^\infty 1/n^2$.
Because $\Phi (\omega_m)/{\tilde \Sigma} (\omega_m) = {\bar \Phi} (\omega_m)/{\bar{\tilde \Sigma}} (\omega_m)$, Eqs. (\ref{eq:gapeq_aa}) can be re-expressed solely in terms of ${\bar \Phi} (\omega_m)$ and ${\bar{\tilde \Sigma}} (\omega_m)$:
  \bea \label{3_10}
   && {\bar \Phi} (\omega_m)=
     {\bar g}^2 \pi T \sum_{m' \neq m} \left(\frac{{\bar \Phi} (\omega_{m'})}{\sqrt{{\bar {\tilde \Sigma}}^2 (\omega_{m'}) +{\bar \Phi}^2 (\omega_{m'})}} - \frac{{\bar \Phi} (\omega_{m})}{\sqrt{{\bar {\tilde \Sigma}}^2 (\omega_{m}) +{\bar \Phi}^2 (\omega_{m})}}\right)
    ~\frac{1}{|\omega_m - \omega_{m'}|^2}, \nonumber \\
    && {\bar {\tilde \Sigma}} (\omega_m)  = \omega_m \nonumber \\
     && +  {\bar g}^2 \pi T \sum_{m' \neq m}\left(  \frac{{\bar {\tilde \Sigma}}(\omega_{m'})}{\sqrt{{\bar {\tilde \Sigma}}^2 (\omega_{m'})  +{\bar \Phi}^2 (\omega_{m'})}} -  \frac{{\bar {\tilde \Sigma}}(\omega_{m})}{\sqrt{{\bar {\tilde \Sigma}}^2 (\omega_{m})  +{\bar \Phi}^2 (\omega_{m})}}\right)
    ~\frac{1}{|\omega_m - \omega_{m'}|^2}
\eea
These equations are now free from singularities at $T=0$,  when
the summation over Matsubara numbers is replaced by the integration over $\omega_m$.

We now extend the modified  Eliashberg equations (\ref{3_10}) by multiplying the interaction in the particle-particle channel by a factor $1/M$:
  \beq \label{3_12}
    {\bar \Phi} (\omega_m)=
     \frac{{\bar g}^2}{2M} \int d \omega'_{m} \left(\frac{{\bar \Phi} (\omega'_{m})}{\sqrt{{\bar {\tilde \Sigma}}^2 (\omega'_{m}) +{\bar \Phi}^2 (\omega'_{m})}} - \frac{{\bar \Phi} (\omega_{m})}{\sqrt{{\bar {\tilde \Sigma}}^2 (\omega_{m}) +{\bar \Phi}^2 (\omega_{m})}}\right)
    ~\frac{1}{|\omega_m - \omega'_{m}|^2}
\eeq
The gap function $\Delta (\omega_m)$ is expressed via ${\bar \Phi} (\omega_m)$ and ${\bar {\tilde \Sigma}} (\omega_m)$ in the same way as via the original $\Phi (\omega_m)$ and ${\tilde \Sigma} (\omega_m)$:
 $\Delta (\omega_m) = \omega_m {\bar \Phi} (\omega_m)/{\bar {\tilde \Sigma}} (\omega_m)$. The equation on
 $\Delta (\omega_m)$ is
\beq
\Delta (\omega_m) =\frac{{\bar g}^2}{2M}  \int ~\frac{d \omega'_{m}}{|\omega_m - \omega'_{m}|^2} \left(\frac{\Delta (\omega'_{m}) - M \frac{\Delta (\omega_{m})}{\omega_m} \omega'_{m}} {\sqrt{\Delta^2 (\omega'_{m}) + (\omega'_{m})^2}} - \frac{\Delta (\omega_{m}) (1-M)}{\sqrt{\Delta^2 (\omega_{m}) + \omega^2_{m}}}\right)
\label{3_12a}
\eeq
Absorbing $1/M$ into ${\bar g}^2_M = {\bar g}^2/M$, introducing a dimensionless $\bar{\omega}_m = \omega_m/{\bar g}_M$, $D(\bo_m) = \Delta (\bo_m)/\bo_m$ and re-arranging, we obtain from (\ref{3_12a})
\bea
&&D (\bo_m) \left(\bo + \frac{1-M}{2} \int ~\frac{d \bo'_{m}}{|\bo_m - \bo'_{m}|^2}  \left(\frac{\sgn\bo_m}{\sqrt{1 + D^2 (\bo_{m})}} - \frac{\sgn \bo'_{m}}{\sqrt{1 + D^2 (\bo'_{m})}}
\right)\right) = \nonumber \\
&& \frac{1}{2} \int ~\frac{d \bo'_{m}}{|\bo_m - \bo'_{m}|^2}  \frac{D(\bo'_{m})-D(\bo_m)}{\sqrt{1 + D^2 (\bo'_{m})}} \sgn \bo'_{m}
\label{3_12b}
\eea
Both integrals in (\ref{3_12b}) are free from singularities and infra-red convergent.

For infinitesimally small $D(\bo_m)$, Eq. (\ref{3_12b}) becomes
\beq
 D(\bo_m) \left(\bo_m + \frac{1-M}{\bo_m}\right) = \frac{1}{2} \int d \bo'_m \frac{D(\bo'_m)-D (\bo_m)}{|\bo_m-\bo'_m|^2} \sgn\omega'_m, \label{3_19}
 \eeq
At small $\bo_m$, the solution of the gap equation is
\beq
D(\bo_m) = 2\epsilon \cos{\left(\beta_M \log{\bo^2_m} + \phi\right)} \sgn\omega_m,
\label{3_19a}
\eeq
It has the same form as Eq. (\ref{5_1}),
but now $\beta^2_M = M/\pi^2$. This form implies that $M_{cr} =0$.

We now assume that $M$ is small and solve the non-linear gap equation.  Our key intension is to check whether we still have a continuum of solutions.   For this purpose, it is sufficient to  focus on small $\bo_m$, when we
  can
 neglect bare $\bo_m$ in the l.h.s. of (\ref{3_12b}).

 As in  Sec. \ref{sec:Mats_expansion}, we search for the solution of (\ref{3_12b}) in the series in
$\epsilon^2$ for both $D(\bo_m)$ and $\beta$.
 To leading order in $M$, we  obtain
\beq
D(\bo_m) = \frac{2\epsilon}{\pi} M^{1/2} \log{\bo^2_m} \left(1 - 3 \epsilon^2 + ....\right)^{1/2}
\label{ap_3_3}
\eeq
where dots stand for $\epsilon^4$ and higher order terms. The $M^{1/2}$ dependence (same as $(M-M_{cr})^{1/2}$ as $M_{cr} =0$) is an expected one.
 The  logarithmic dependence on frequency is consistent with the result in Paper I, where we obtained
 $\log{\bo^\gamma_m}$ dependence at $N=N_{cr}$.  However, there such dependence exists only for $N = N_{cr}$, while here we have an infinite set of solutions with the same frequency dependence, but different amplitudes, parametrized by $\epsilon$.   All solutions appear simultaneously at $M =0+$

 A complimentary piece of evidence for multiple solutions  comes about if
  we simplify the l.h.s. of (\ref{3_12b}) by dropping
  $D^2$ terms in the denominator of the $(1-M)$ term. The gap equation then reduces to
 \beq
D (\bo_m) \left(\bo + \frac{1-M}{2 \bo}\right)  =
 \frac{1}{2} \int ~\frac{d \bo'_{m}}{|\bo_m - \bo'_{m}|^2}  \frac{D(\bo'_{m})-D(\bo_m)}{\sqrt{1 + D^2 (\bo'_{m})}} \mbox{\sign} \bo'_{m}
\label{3_12c}
\eeq
The full gap equation for the original $\gamma$ model is reproduced if we set $M=1$, so
 Eq. (\ref{3_12c}) can be viewed as another extension of the original model.
 The solution of the linearized gap equation is the same as before, with $\beta^2_M =M/\pi$, hence $M_{cr} =0$.
 At $M \to 0$,  the expansion in $\epsilon^2$ now yields, to leading order in $M \equiv M -M_{cr}$:
 \beq
 D(\bo_m) = 2\epsilon \left(\cos{f_M (\bo_m)} - \frac{\epsilon^2}{16M} \cos{3 f_M (\bo_m)} + ...\right)
 \label{ap_3_4}
 \eeq
  where
  \beq
 f_M (\bo_m) = \beta_{M,\epsilon} \log{\bo^2_m} + \phi
 \label{ap_3_5}
 \eeq
 and
 \beq
 \beta^2_{M,\epsilon} = \beta^2_M \left(1 - \frac{3\epsilon^2}{M} + ...\right)
 \label{ap_3_6}
 \eeq
We see that the expansion holds in powers of $\epsilon^2/M$ and is valid up to $\epsilon \sim (M)^{1/2}$, at which
$\beta^2_{M,\epsilon}$ vanishes. At larger $\epsilon$, $\beta^2_{M,\epsilon}$ becomes negative, and the solution disappears (there is no normalized solution of the linearized gap equation).
 We see that there is again an infinite set of solutions, specified by
  $\epsilon$, which runs between $0$ and $\epsilon_{cr} = O(\sqrt{M})$.

\bibliography{gamma_2}

\end{document}